# Enduring Quests
# Daring Visions

## NASA Astrophysics in the Next Three Decades

<>***On the cover***: *At top left, an artist's concept of Kepler-69c, a super-Earth-size planet in the habitable zone of a star like our sun. In the background, an illustration of a stage in the predicted collision between our Milky Way galaxy and the neighboring Andromeda galaxy, representing Earth's night sky 3.75 billion years hence. In the foreground, an artist's impression of gravitational waves emitted by two orbiting black holes.* **Credits: NASA/Ames/JPL-Caltech; NASA/ESA/ Z. Levay and R. van der Marel (STScI), T. Hallas and A. Mellinger; K. Thorne (Caltech) and T. Carnahan (NASA GSFC)**

# Table of Contents

# Table of Contents

# Executive Summary

The past three decades have seen prodigious advances in astronomy and astrophysics. We have discovered thousands of planets around other stars, revealing a startling diversity of planetary systems and demonstrating that our own solar system is one of many billions across the galaxy. We have mapped the life cycle of stars from their formation in nearby stellar nurseries to their sometimes explosive deaths. We have tallied supermassive black holes weighing millions to billions of times the Sun's mass, finding them in the centers of most big galaxies, where they gorge on gas and stars and blast superheated material into their environments. We have traced the growth and transformation of galaxies from young chaotic shreds a billion years after the Big Bang to the majestic forms they display today. We have established, using precision cosmological measurements, a well-tested account of the Big Bang model's underlying physics. The current picture points to a universe where an early epoch of inflation established uniformity over a vast horizon, where dark matter dominates the gravitational forces that provide the underlying framework of cosmic structure, and where dark energy drives expansion at an ever-accelerating rate.

NASA missions have made essential contributions to all of these dramatic developments.

This roadmap presents a science-driven 30-year vision for the future of NASA Astrophysics that builds on these remarkable discoveries to address three defining questions:

- *Are we alone?*
- *How did we get here?*
- *How does the universe work?*

Seeking answers to these age-old questions are Enduring Quests of humankind. The coming decades will see giant strides forward in finding earthlike habitable worlds (exoEarths), in understanding the history of star and galaxy formation and evolution, and in teasing out the fundamental physics of the cosmos.

Different astronomical phenomena are typically most prominent in distinct regions of the electromagnetic spectrum: young, dust-enshrouded stars in the far-infrared; older stars and galaxies in visible and near-optical wavelengths; star-forming regions in the ultraviolet, and the million-degree gas of galaxy clusters and black hole accretion disks in X-rays. In addition to spanning the electromagnetic spectrum, the roadmap missions will open a revolutionary new window on the cosmos by detecting gravitational waves, ripples of space-time that emanate from colossal events like the merging of two black holes. These scientific investigations and their enabling space missions are described here in three broad categories of time: the Near-Term Era, defined by missions that are currently flying or planned for the coming decade; the Formative Era, with notional missions (referred to here as "Surveyors") that could be designed and built in the 2020s; and the Visionary Era of technologically advanced missions (referred to as "Mappers") for the 2030s and beyond.

For studies of exoplanets, the Near-Term Era challenge is to develop a comprehensive understanding of the demographics of planetary systems. By compiling a census of "cold" and free-floating planets using gravitational microlensing, the WFIRST-AFTA mission will complement the spectacular trove of "hot" and "warm" planets orbiting close to their parent stars already identified by the Kepler mission. The Formative Era goal is to characterize the surfaces and atmospheres of planets around nearby stars through direct imaging and spectroscopy. The Large UV-Optical-IR (LUVOIR) Surveyor will enable ultra-high-contrast spectroscopic studies to directly measure oxygen, water vapor, and other molecules in the atmospheres of exoEarths, and thereby extend the work of the James Webb Space Telescope (JWST) and the coronagraphic observations from WFIRST-AFTA. In the Visionary Era, the ExoEarth Mapper will combine signals from an armada of large optical telescopes orbiting hundreds of kilometers apart, producing the first resolved images of earthlike planets around other stars. For the first time, we will identify continents and oceans—and perhaps the signatures of life—on distant worlds.

Tracing cosmic origins from the first galaxies and quasars to stars and planets forming today requires observations spanning an enormous range of time and wavelength. In the Near-Term Era, JWST will supply unprecedented views of protostars and nascent star clusters, and together with WFIRST-AFTA will resolve nearby star-forming regions and populations of stars across the full range of local galaxies. In the Formative Era, the Far-IR Surveyor will pioneer space interferometry to resolve protoplanetary disks, and the LUVOIR Surveyor will decode the galaxy assembly histories through detailed archeology of their present structure. The high-resolution X-ray Surveyor observations of supernova remnants will trace supernova-driven feedback mechanisms that affect the chemical and dynamical evolution of galaxies. These missions will also directly trace the history of galaxies and intergalactic gas through cosmic time, peering nearly 14 billion years into the past; in concert with the Gravitational Wave Surveyor, they will untangle the complex symbiosis between galaxies and the supermassive black holes at their centers. In the Visionary Era, radio telescope arrays in the shadow of our moon will map intergalactic hydrogen through the epoch of re-ionization, when ultraviolet photons from the earliest stars broke through the opaque curtain of absorbing cosmic gas to fill space with free electrons.

In the first nanoseconds of cosmic time, in the interiors of neutron stars, and in the severely warped space-time around black holes, nature created extreme conditions that can never be achieved in terrestrial laboratories. In the Near-Term and Formative Eras, WFIRST-AFTA followed by the CMB Polarization Surveyor will measure the cosmos with exquisite precision, probing deeply into the physics of cosmic inflation and pinning down the mechanisms that drive today's accelerating expansion. The X-ray and Gravitational-Wave Surveyors will test the striking predictions of Einstein's general relativity near the event horizons of spinning black holes, and they will determine the properties of nuclear matter at the highest densities and pressures realized in nature. In the Visionary Era, the Black Hole Mapper may resolve the shadow cast by a black hole's event horizon onto its own disk of accreting hot gas, and the high-resolution Gravitational Wave Mapper may detect space-time ripples produced during the epoch of cosmic inflation.

Achieving the goals of the Formative and Visionary Eras will require a spectrum of technological advances. Some of the key challenges are: the construction of large, lightweight optical elements; the wavefront control required to achieve ultra-high-contrast performance with large mirrors; the execution of controlled formation flying, precision metrology, and beam combination needed for interferometry; large-area X-ray optics; detector technologies that achieve orders-of-magnitude improvements in sensitivity, field of view, and spectral resolution. Ongoing technological revolutions in communications and computing, as well as novel construction methods such as robotic assembly and 3D printing, may enable a diversity of small missions. Observatories that seem impossibly daunting today may be feasible a decade or two from now.

These quests and missions will inspire the nation and the world, as NASA has done since its inception. The excitement of astrophysics and the challenges of space draw students at every level into science, mathematics, and engineering, helping to build the expert workforce required for success in the global economy. Solving the technical challenges of space astrophysics has direct economic benefit of its own, through spin-off technologies and by propelling technical innovation and performance to the highest levels. From the Formative Era to the Visionary Era, the notional missions of this roadmap build upon the legacy of NASA's most impressive achievements and boldest aspirations.

While future plans are necessarily built on current knowledge, the most important results to come from these missions are likely beyond the edge of present imagination. Thirty years from now, these investigations may show that life arises frequently on planets orbiting other stars, or that fundamentally new states of matter arise in the cores of neutron stars, or that the first stars in the cosmos were unlike any that exist today, or that there are dimensions of space-time beyond the four of our everyday experience. In their compelling science objectives and their extraordinary technical ambitions, the research programs and missions of this roadmap are Daring Visions for the future of NASA Astrophysics.

# Executive Summary

# Preface

The U.S. astrophysics community is widely recognized for employing a large fraction of its members in scientific surveys to develop a prioritized path forward in decadal increments, the so-called Decadal Surveys. These surveys provide input to federal funding agencies (e.g., NASA, NSF, DOE) for the development and implementation of the highest-ranked recommendations. Based on the current and projected budgets, NASA assembles a roadmap toward the realization of these top recommendations, with the most recent being the Origins and Beyond Einstein Roadmaps in 2003 and the Astrophysics Roadmap in 2006.

In 2010, the community produced the "New Worlds, New Horizons" Astro2010 Decadal Survey with concrete recommendations for future missions and technology developments. The Wide-Field Infrared Survey Telescope—Astrophysics Focused Telescope Assets, WFIRST-AFTA, is emerging as the first large space mission after the launch of JWST to be realized in response to this survey. In March 2013, the Astrophysics Subcommittee of the NASA Advisory Council/Science Committee assembled a Task Force to develop a vision for NASA's Astrophysics Division spanning the next three decades. The Charter given to this team (see Appendix A) was to produce a compelling long-term vision for space astrophysics building upon the Astro2010 baseline.

The team was asked to identify the enduring scientific questions over the coming decades and the required technology developments to address these questions. Given the long time span under consideration, they were asked to identify notional missions—generic concepts with capabilities that fulfill the visionary science goals.

The planning was divided into three Eras: the "Near-Term Era," the "Formative Era," and the "Visionary Era." Each represents roughly ten years of mission development in a given field, although the scientific investigations that exploit these missions will typically extend into the following decade. The first of these Eras covers ongoing NASA-led activities and planned missions such as JWST, NICER, TESS, and WFIRST-AFTA, as well as international partnership missions such as Astro-H, Euclid, and Athena+. The missions of the Formative Era will build on the technological developments and scientific discoveries of Near-Term Era missions, with remarkable capabilities that will enable breakthroughs across the landscape of astrophysics. They will also lay the foundations for the missions and explorations of the Visionary Era, which will take us deep into unchartered scientific and technological terrain.

After considering many possible ways to frame the scientific challenges of the next 30 years, the task force returned in the end to the three fundamental questions that have long defined the goals of the NASA Astrophysics program: Are we alone? How did we get here? How does the universe work? These questions link us to our forebears across many civilizations and several millennia. We are poised to make dramatic progress through observations of other planets, stars, and galaxies, using capabilities that go far beyond those available today. The search for answers constitutes the Enduring Quests of this roadmap. Our modes of investigation demand the skills and ingenuity of scientists and engineers. The discoveries that emerge will inspire the public and engage citizen scientists young and old.

The topics covered in this report are not all-inclusive. To support our efforts, we invited community input in the form of talks, written abstracts, and a town hall meeting. We received many excellent contributions, and we are grateful to our colleagues for their investment in this process. From the pool of suggestions, we distilled the leading themes of the remote future, in some cases extending even beyond the 30-year horizon. In both technological and scientific terms, these are the Daring Visions that arise throughout our chapters and are reviewed in our conclusions.

For the Near-Term Era, the team adopted the priorities already identified in "New Worlds, New Horizons." With regard to the following two Eras, it is important to stress here that this report does not rank either their scientific themes or their missions. Priorities will be set by future Decadal Surveys, in response to scientific discoveries, technological opportunities, and budgetary constraints. Many of these missions will be realized through international partnerships. The long-term vision presented here will hopefully provide useful input for the next Decadal Surveys, for NASA, and for the community as a whole.

In the remainder of the document "we" is used in its broadest sense, referring most directly to the U.S. astronomical community and more generally to the much larger community of humans who seek to understand our universe.

This team hopes that 30 years from now, some of us, and all of our students, will be navigating the uncharted space-time of our vision with a cutting-edge flotilla. Toward the end of this 30-year period, we will be designing and building extraordinary missions: arrays of telescopes that work in concert to map the continents and biospheres of planets around other stars and to image the event horizons of black holes; radio telescopes on the far side of the moon that map the epoch 13 billion years in the past when the first stars filled the universe with light; and gravitational wave telescopes so sensitive that they can detect ripples of space-time echoing from the Big Bang itself. Yet experience tells us that the most important discoveries are the ones we cannot anticipate—perhaps not even imagine—because every time we significantly improve our technical capabilities or open a new window on the universe, we take a leap into the scientific unknown.

# 1 Enduring Quests

From ancient times, we have looked into the cosmos and struggled to understand our place in the universe. From the first efforts to comprehend the nature of the physical world by the ancient Greeks, through the development of modern science by Galileo, Kepler, and Newton, and up to the construction of the most ambitious scientific observatories and experiments of today, we have tirelessly sought to discern our origins.

Through the 20th century and into the 21st, humankind has made extraordinary progress toward answering profound and enduring questions about the nature of the universe and our place within it. Efforts led by NASA began with the exploration of our solar system and, through the pioneering Explorers and Great Observatories of today, have opened new eyes on the universe back to its earliest times. We have discovered a wealth of planets around nearby stars, revealing an astonishing diversity of planetary systems. Some of these planets might be hospitable to life as we find it on Earth, while others are scorching hot or icy cold. We have pieced together the lifecycles of stars, beginning with their births in nearby stellar nurseries and ending with their deaths, either as fading embers or in explosive fireworks. We have taken advantage of natural laboratories in space, such as white dwarfs and neutron stars, to explore the properties of nuclear matter and subatomic forces at densities that can never be produced on Earth. We have mapped the formation and evolution of galaxies from their chaotic youth a billion years after the Big Bang to the majestic forms of today. We have discovered that giant black holes millions of times the mass of our Sun are ubiquitous at the centers of the most massive galaxies. We have learned that the expansion of the universe is accelerating rather than slowing, and we have measured tiny irregularities in the cosmos that were imparted by quantum fluctuations in the first trillionth of a trillionth of a trillionth second of cosmic history.

The discoveries of the past 60 years have set the stage for rigorous scientific investigations of some of our most ancient and fundamental questions:

## Are we alone?

## How did we get here?

## How does the universe work?

This roadmap outlines a path to revolutionize our understanding of planets, stars, and the cosmos and to make great strides toward answering these profound questions. The process will require the development of new technologies, and the vision and wherewithal to undertake highly ambitious programs over the next 30 years. These three questions set defining goals for the three primary fields within NASA's Astrophysics Division: Exoplanet Exploration, Cosmic Origins, and the Physics of the Cosmos.

The extraordinary suite of missions described in this roadmap will enable us to:

- Probe the exponential early growth of the universe, when quantum fluctuations grew into the seeds for large-scale structure.
- Probe the origin and ultimate fate of the universe, and determine the forms of matter and energy that govern it, by mapping the growth of cosmic structure through its history.
- Directly chart the warped space around distant black holes by detecting the ripples of space-time produced when they merge.
- Unveil the chaotic flows of superheated gas swirling around black holes that fuel the most powerful engines in the universe.
- Use telescopes as time machines to map the full history of galaxy formation and assembly, from the birth of the first stars through the turbulent epoch of rapid growth to the galaxies we see today.
- Make star-by-star maps of nearby galaxies across the full range of observed galaxy types to decode their histories and understand how and when they acquired their present-day forms.
- Characterize the evolution of planetary systems like our solar system by understanding the nature of newborn stars, the process of planet formation around them, and the crucial transport of water to inner planets.
- Complete the reconnaissance of planets and planetary systems, including gas giants, rocky planets like Earth and Mars, ocean-covered water worlds, planets close to and far from their parent stars, and even free-floating planets that have been ejected to interstellar space by gravitational interactions with their siblings.
- Directly image the planets around nearby stars and search their atmospheres for signs of habitability, and perhaps even life.

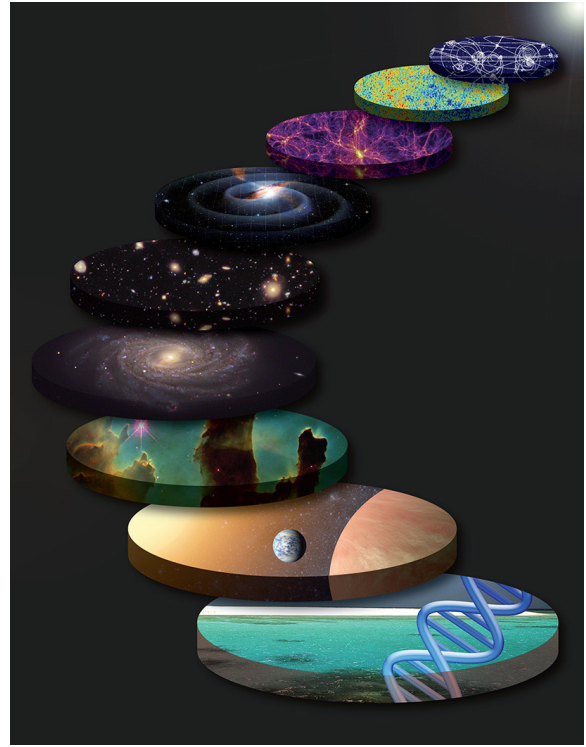

*Figure 1.1 The roadmap science themes are pictured here as a series of iconic stepping stones. Starting at top from the Big Bang and the cosmic microwave background and moving downward through the evolution of galaxies to the formation of stars and planets, these steps from the distant universe to nearby life-bearing worlds trace out the history of the cosmos. Credit: NASA*

The next three chapters consider each of the science themes in turn, beginning with the quest to understand the populations of planets around other stars and search for signatures of life, continuing with the quest to map the life cycles of stars and the history of galaxies, and then the quest to understand the basic physical laws that govern our universe. Throughout these chapters we include examples of how NASA Astrophysics can inspire the public and educate the next generations of scientists and engineers, and we devote Chapter 5 specifically to strategies for education and public engagement, which are essential to achieving NASA's mission. In Chapter 6 we describe the notional missions of the Formative and Visionary Eras and the technological developments required to make them possible. We also identify emerging technological trends that could lead to radical new opportunities in the design and realization of space missions. We conclude, in Chapter 7, with a vision of discoveries that may come from the next three decades of NASA Astrophysics missions, though history suggests that even our most daring visions will fall short of reality.

As reference for the science chapters, **Figure 1.2** provides a chart of the notional missions for the Formative and Visionary Eras and of those Near-Term Era missions that are most important to the discussions in this report. Each mission is placed on the chart within a broad wavelength range indicated at the y-axis, which covers the entire electromagnetic spectrum; the gravitational wave and cosmic ray missions are placed at the top of the figure as representatives of multi-messenger astronomy. The Eras are separated horizontally in equal segments that encompass roughly 10 years each.

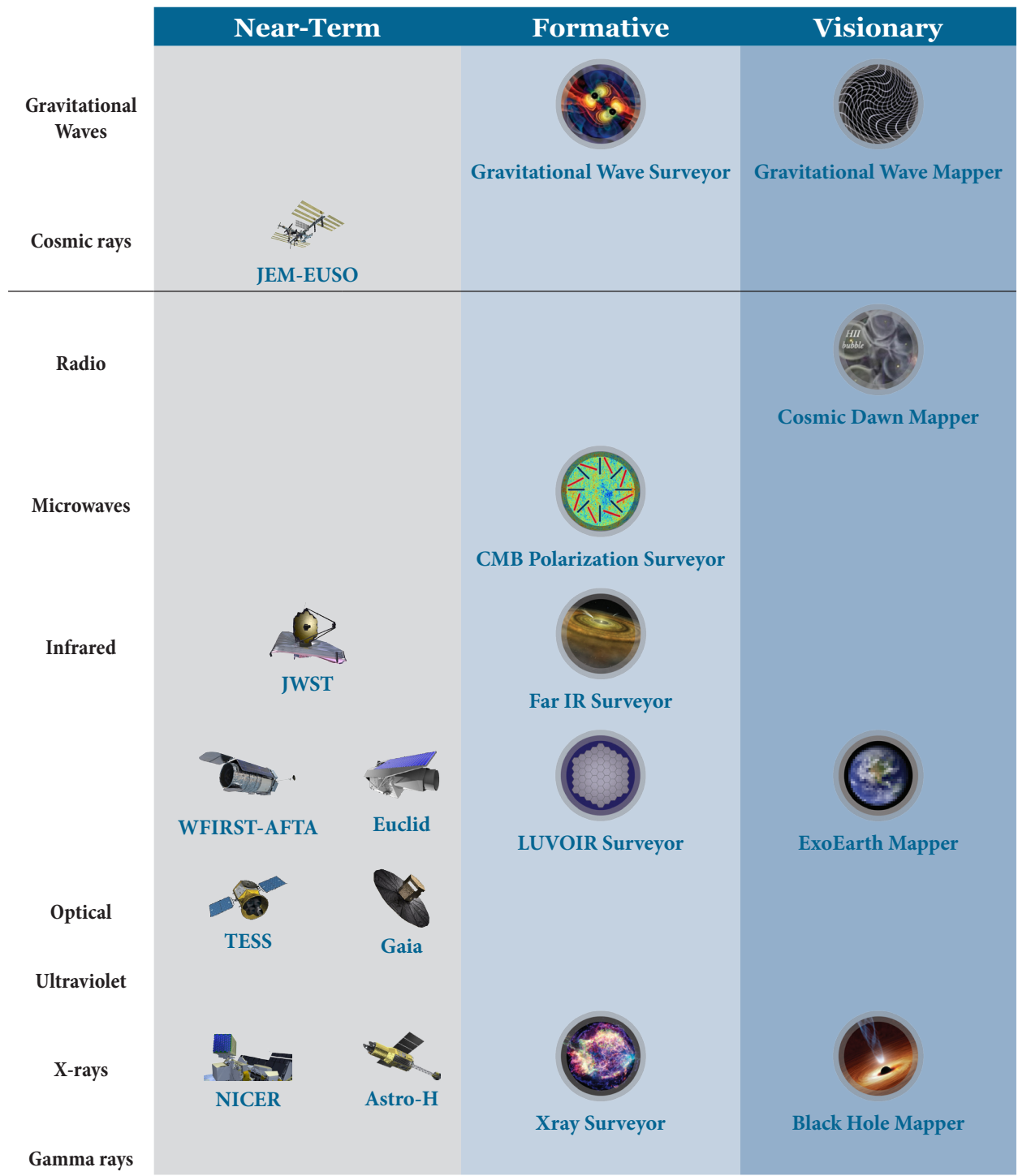

*Figure 1.2 Chart of the missions currently planned for launch during the Near-Term Era and of the notional missions of this roadmap for the Formative and Visionary Eras.*

For the Near-Term Era, **Figure 1.2** displays only missions that have not yet been launched, though (as discussed in the subsequent chapters) important Near-Term science will also come from currently operating missions such as HST, Chandra, and Kepler. This section of the figure also includes several missions led by the European or the Japanese space agencies (ESA, JAXA), with NASA participation, that are especially relevant to discussions in the chapters that follow. ESA recently selected two concepts for its L2 and L3 programs, the Hot and Energetic Universe and Gravitational Waves, respectively. The missions for these concepts have not been selected yet; they are scheduled to launch in the late 2020s—early 2030s. Both are of great relevance to this roadmap because their expected capabilities overlap those of the Formative Era X-ray Surveyor and Gravity Wave Surveyor. These and other roadmap missions may ultimately be realized through international collaborations, given their broad scientific interest and demanding technical challenges.

Finally this chart does not call out any of the ground-based facilities addressed in this report; such a listing is beyond the scope of this roadmap. However, as discussed throughout the following chapters, observations from these powerful ground-based electromagnetic, cosmic ray, and gravitational wave observatories will make crucial contributions to achieving the scientific goals of the Roadmap, in concert with capabilities that can be realized uniquely from space.

We have somewhat arbitrarily adopted the terms "Surveyor" for all of our Formative Era missions and "Mapper" for all of our Visionary Era missions. This nomenclature does not mean that the former are necessarily wide-field survey facilities or the latter high-resolution imagers, but we do intend to convey the ability of Visionary Era missions to reveal extraordinary details, delving deeply into phenomena discovered and characterized by the Surveyors. Whether mapping the surfaces of exoplanets, the event horizons of black holes, the dawn of the first stars in the cosmos, or the space-time reverberations of the Big Bang, these powerful missions represent a Daring Vision of NASA's long-term future.

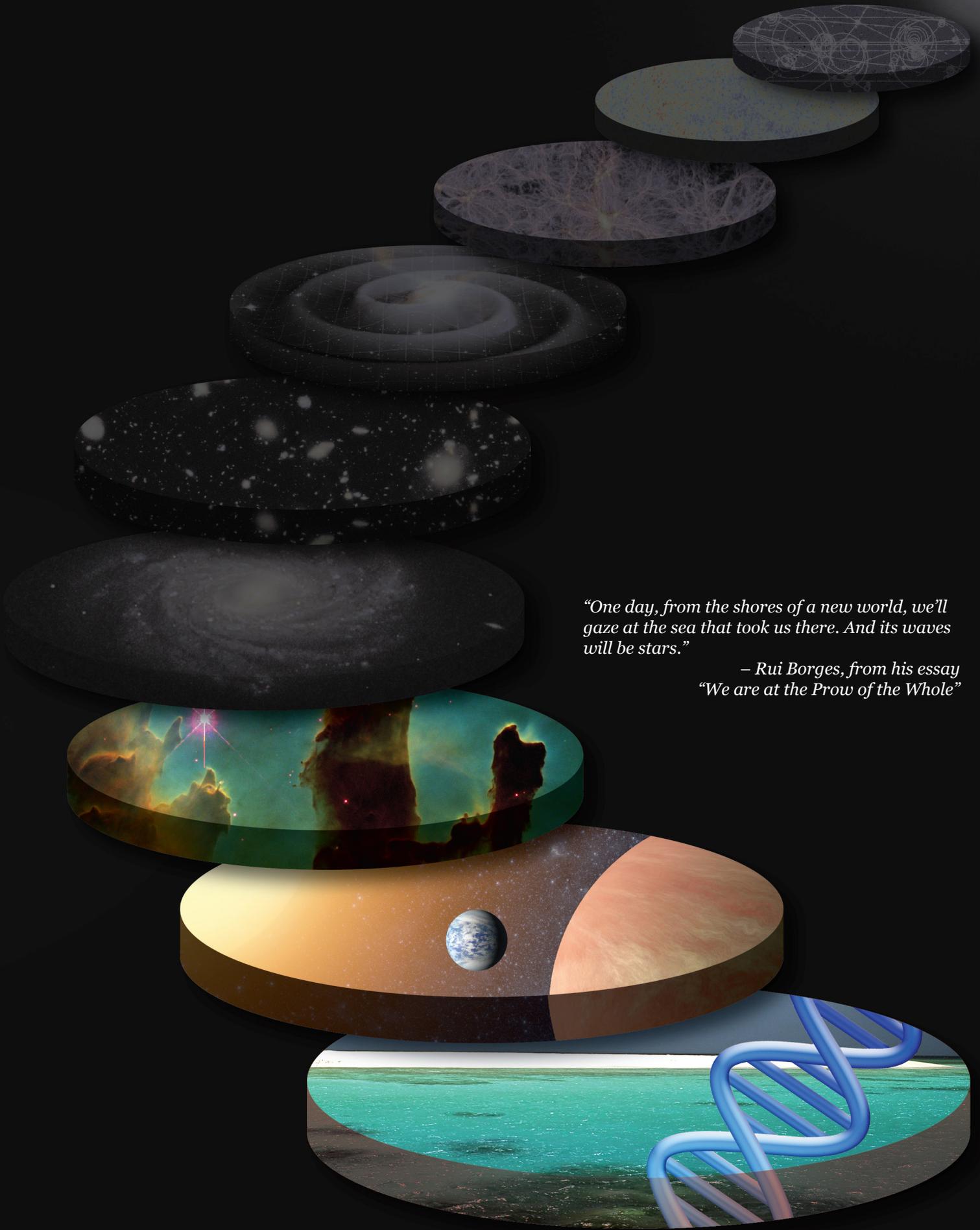

"One day, from the shores of a new world, we'll gaze at the sea that took us there. And its waves will be stars."

— Rui Borges, from his essay "We are at the Prow of the Whole"

# 2  Are We Alone?

Throughout human history, we have yearned to comprehend our universe and understand our place within it. These ambitions have continually driven human exploration, compelling us to search our world, seek out new territories, and spread out over its entire surface. These same forces lure us beyond our cosmic home to discover and explore new worlds, and seek out the existence of life different from ourselves. Our drive to explore carries with it promise, costs, and perils. Exploration brings tangible benefits from new territories, new resources, unforeseen opportunities, and ultimately a deeper understanding of ourselves. However, it also carries demands of societal commitment, resource allocation, and—for the most ambitious undertakings—global cooperation. Furthermore, the risks of human exploration are significant, particularly as we move ever further from our familiar habitat. For many decades, NASA has led humanity's greatest and most ambitious expeditions, sending humans to our moon and robotic probes to many planets in the solar system (**Figure 2.1**).

The next stage in this Quest—and possibly the most exciting one—opened with the discovery of planets around other stars in the mid-1990s. These exoplanets have already turned out to hold surprises beyond the expectations of scientists and closer to the dreams of science fiction. Exotic worlds unlike any in the solar system are abundant, and yet we find planetary systems reminiscent of our own as well. Over the next several decades, we could conceivably answer these fundamental questions: What is our place in this cosmic landscape? How common—and how close—are Earthlike worlds? Is there life beyond the solar system? In this chapter, we lay out a possible path toward finding those answers, calling on the talents of a wide range of scientists and the complementary capabilities of space- and ground-based facilities.

We begin our vision in the exciting present, with the burst of exoplanet discovery ushered in by NASA's 2009 Kepler mission. Over the next few years, Kepler will finish its statistical survey and determine the abundance of exoplanets as small as Earth in orbits as large as one astronomical unit (AU, the average Earth-Sun distance). Additional space- and ground-based studies will find the prevalence of other types of planets, locate planets around the stars closest to the Sun, begin determining their characteristics,

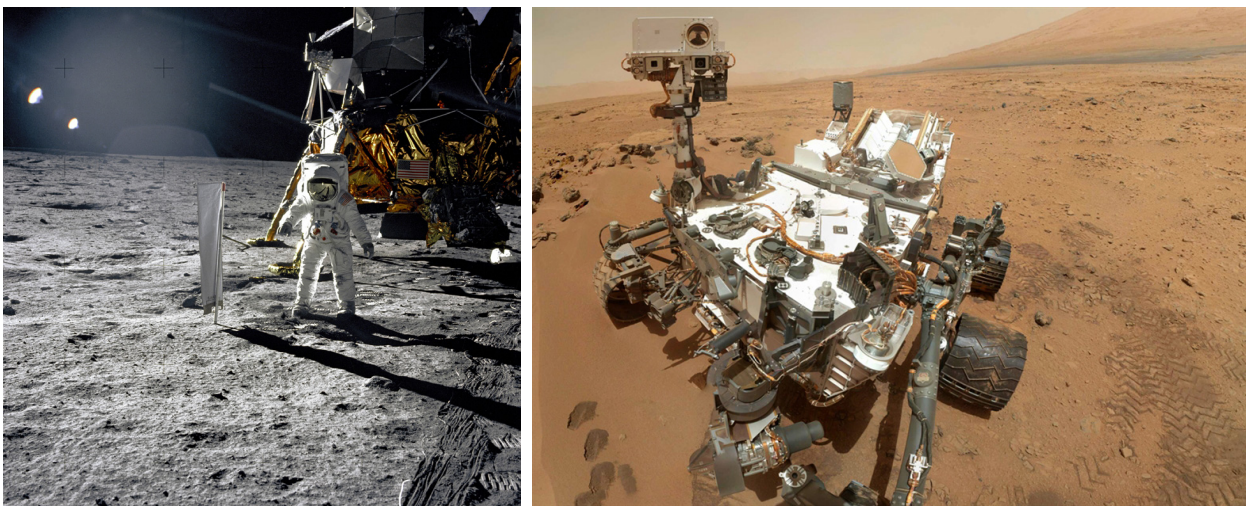

*Figure 2.1* Left: Edwin "Buzz" Aldrin and the Lunar Module "Eagle" at the moon's Sea of Tranquility during Apollo 11, July 20, 1969. Right: Self-portrait of NASA's Curiosity rover on the surface of Mars, Oct. 31, 2012. **Credits: NASA/N. Armstrong and NASA/JPL-Caltech/Malin Space Science Systems**

### The Habitable Zone

All life on Earth requires liquid water, which guides our search for other life-bearing worlds. The "habitable zone" around a star is the region where an Earth-like exoplanet has a surface temperature in the right range for liquid water. Earth-like planets closer to the star have steam (too hot) and those farther away have ice (too cold). Earth-like life also requires planets with enough gravity to retain an atmosphere (larger than Mars) but not so much as to accrete a thick layer of dense gas (smaller than Neptune). However, even if these criteria are met, there is no guarantee that the planets actually possess water on their surfaces. If they do, then we refer to them as "exoEarths."

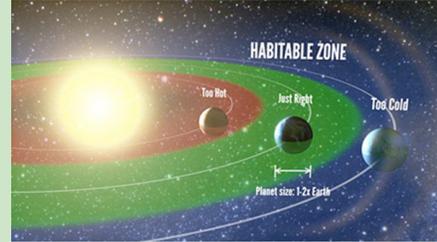

*Figure 2.2* Habitable planets must orbit at the right distance from their host stars to retain liquid water on their surfaces. This so-called habitable zone is shaded green. **Credit: E. Petigura and G. Marcy (Berkeley) and A. Howard (UH-Manoa)**

and measure interplanetary dust in the habitable zones (**Figure 2.2**) of nearby Sun-like stars. All these intertwined efforts pave the way for ambitious future missions to explore habitable exoplanet environments and, eventually, search for life in our solar neighborhood.

## 2.1 The Exoplanet Zoo

What are other planetary systems like? Before the first exoplanet discoveries, theories originally developed to explain the solar system naturally predicted that other systems would look much like our own. Nature turned out to be much more imaginative than we were. To date, most confirmed exoplanets look extremely different from those in our solar system and from the majority of planets predicted by planet formation models (**Figure 2.3**). It has become clear that there exists an unanticipated and incredible diversity in the properties and architectures of exoplanetary systems. A major stumbling block for scientists struggling to understand the origins of this exoplanet diversity is that, even after 20 years and more than 1,000 confirmed discoveries, our statistical census remains woefully incomplete. NASA's Kepler mission has begun the process of obtaining this census. The effort will continue throughout this decade, first with ground-based direct imaging and radial velocity (RV) surveys, and then with the Transiting Exoplanet Survey Satellite (TESS) mission. The statistical census will be completed in the next decade with surveys for outer planets using direct imaging and gravitational microlensing using the WFIRST-AFTA mission. Within a few decades, we will have a statistically complete survey of planets with masses greater than or equal to Earth's. It will include a robust estimate of the prevalence of Earth-size exoplanets in stellar habitable zones. This estimate is crucial for the development of missions in the Formative and Visionary Eras that will determine if these are truly exoEarths—"pale blue dots," in the words of Carl Sagan—and search them for signs of life.

### *Abundant Hot and Warm Planets: Kepler Begins the Exoplanet Census*

In a triumph of technology, inspiration, and perseverance, the existence of planets orbiting other stars was finally established as we entered the new millennium. These first exoplanets looked nothing like the worlds in our solar system. They were gas giant exoplanets orbiting very close to their host stars—so-called "hot Jupiters"—and they confounded the conventional wisdom of planet formation. As more and more worlds were discovered, it became clear that planets of nearly every conceivable type might be found in almost any environment. A host of new questions arose: What is the full diversity of planetary systems? Is our solar system unusual? Are there other planets like Earth? In other words, are there rocky planets with thin atmospheres capable of hosting liquid water, and perhaps life?

As the data volume increased, giant planets orbiting further from their host stars were discovered. As the detection sensitivity improved, systems with smaller planets were added to the sample. However, the RV

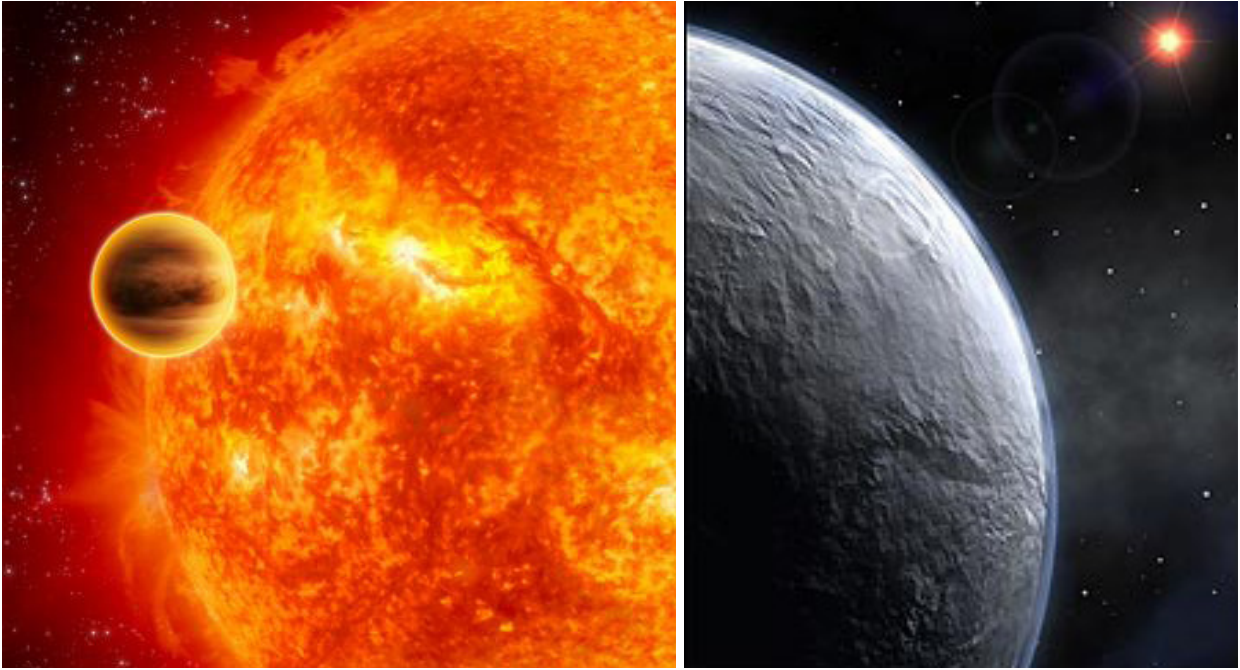

*Figure 2.3* Exotic types of exoplanets not seen in the solar system have been discovered. At left is an artist's concept of the gas giant planet HD 189733b, which orbits its host star every two days. At right is an artist's concept of OGLE-2005-BLG-390Lb, a solid icy "super-Earth" exoplanet about six times Earth's mass. **Credits: ESA/C. Carreau and ESO**

technique (**Figure 2.4**), used to discover the vast majority of these planets was reaching its technical limits. The goal of detecting Earth-mass planets or indeed analogs to most of the planets in our solar system was out of reach. By 2009, twenty years after the first exoplanet was found, the number of known exoplanets exceeded 400, with only 25% of them smaller than Jupiter and only ~6% smaller than Neptune. While these breakthroughs were occurring, NASA was preparing its first mission capable of detecting Earth-size planets in abundance: Kepler.

As NASA's 10th Discovery-class mission, Kepler was built to measure the prevalence of planets in the inner regions of planetary systems, with a particular focus on detecting small, rocky planets similar to Earth. By using the transit method (**Figure 2.4**), combined with unprecedented photometric precision and pointing stability, Kepler broke open the discovery space of planets down to the size of Earth or smaller. Kepler's primary objective was to measure the fraction of stars that host Earth-size planets orbiting in the habitable zone—the range of distances from the star where it is warm enough to have liquid water on the surface given the right atmospheric conditions (**Figure 2.2**).

With the analysis of three of the four years of data acquired during the primary mission, Kepler has already revolutionized our understanding of planetary systems. Over 3,500 planets have been discovered to date, 86% of which are smaller than Neptune (**Figure 2.5**). The early results from Kepler's census are profound: nature makes small planets efficiently. One of six stars is expected to have an Earth-size planet within a Mercury-like orbit. Moreover, approximately half of the low-mass red dwarf stars in our galaxy are expected to have an Earth- or super-Earth-size planet in the habitable zone. Since 70% of stars in the galaxy are red dwarfs, it seems that the nearest potentially habitable exoplanet is not far from the Sun (see **Figure 2.14** for more discussion of habitable planets around red dwarf stars). Analysis of the remaining year of Kepler data will tell us the prevalence of Earth-size planets around stars like our own Sun.

> ### Exoplanet Detection
>
> Today, there are four major means of detecting planets. The first is the radial velocity technique (RV), which measures changes in a star's velocity as an orbiting planet pushes and pulls on it. This method is used on ground-based telescopes and gives the planet's mass.
>
> The second is the transit technique, used by the Kepler and TESS missions. It measures the slight dimming as a planet passes in front of its host star and blocks a tiny portion of its surface. This method gives the planet's size.
>
> The third method is gravitational microlensing, which measures the brightening of a distant background star as another planetary system passes in front of it, acting like a lens. This will be used by WFIRST-AFTA's wide-field instrument. This method also gives the planet's mass.
>
> Finally, the direct-imaging technique uses a variety of sophisticated methods to suppress the light from the bright host star in order to directly collect light from a nearby faint planet. Direct imaging is currently being employed by ground-based telescopes and Hubble Space Telescope (HST) to discover and characterize young planets. In the future, it will be used with ground-based Extremely Large Telescopes (ELTs), WFIRST-AFTA, the LUVOIR Surveyor, and ultimately the ExoEarth Mapper, to enable a wide variety of studies of planets and their atmospheres.
>
> 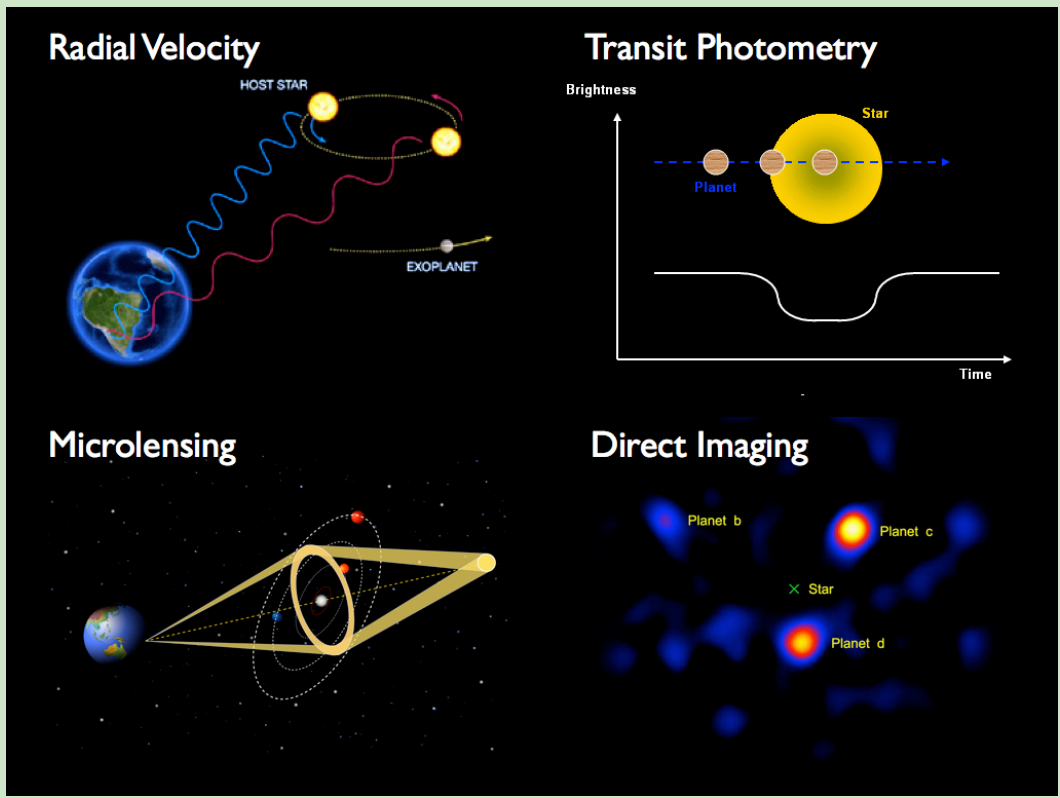
>
> ***Figure 2.4*** *A cartoon illustrating four exoplanet detection techniques.* **Credit: NASA /N. Batalha (NASA Ames)**

### *Zooming in on our nearest neighbors*

By design, Kepler is sensitive to planetary systems orbiting stars that are relatively distant and faint. However, the planets we can learn the most about are the ones orbiting the nearest stars, within a distance to the Sun of roughly 30 light-years. These systems will be studied for decades to come. Ultimately, they are the ones we are most likely to visit first in some distant future that includes interstellar travel.

The RV technique was foundational for the field of exoplanets, and it will continue to make many of the significant exoplanet discoveries anticipated over the next two decades. It is currently the best way to find

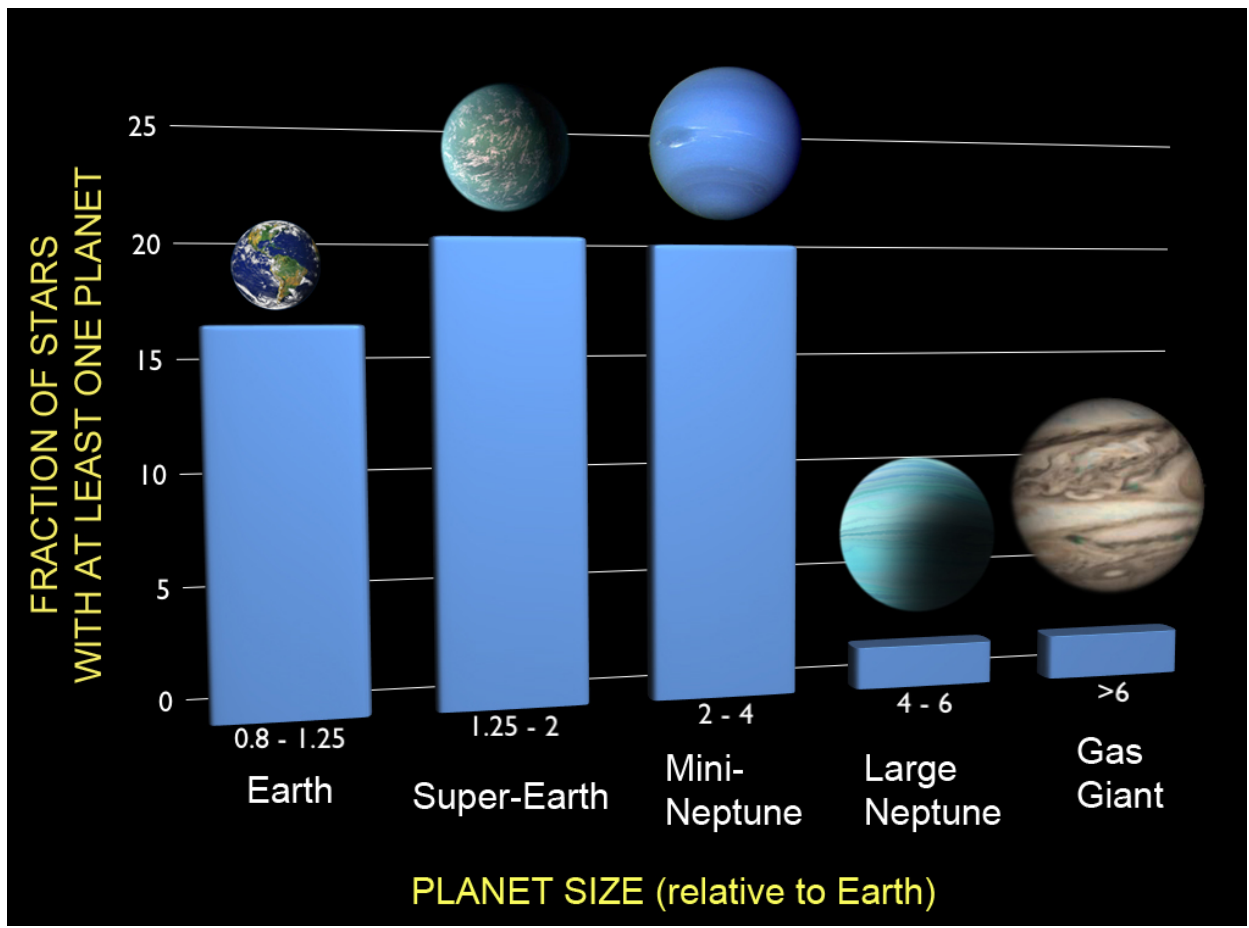

*Figure 2.5* Kepler exoplanet discoveries inform us about the fraction of stars with at least one planet, as depicted here for planets in orbits shorter than 85 days. These statistics were derived from the analysis of the first two of the four years of Kepler data. The size boundaries for different types of planets are still the focus of active research and are likely to change with more knowledge. **Credit: F. Fressin (Harvard CfA)**

giant planets in long period orbits around stars, and such ground-based campaigns should be continued with diligence. Not only will they provide a census of outer giant planets, they will also provide targets for subsequent characterization studies.

The precision of the RV technique has steadily improved to the point where the detection of Earth-mass planets in short-period orbits around bright stars is now also possible (e.g., the rocky Earth-size planet Kepler-78b). Efforts are underway to improve this precision even further, potentially allowing routine detections of short-period Earth-mass planets around a range of host stars. Developing this capability is not only important for surveying nearby systems down to low masses, but also for providing crucial mass measurements for planets detected by transit surveys, such as the TESS mission (discussed in Section 2.2).

In principle, it may be possible to improve RV precision to the point where the detection of Earth-mass planets in the habitable zones of nearby Sun-like stars is achievable and thus confirm Kepler's estimate of their frequency. RV measurements might then yield critical mass measurements for exoEarths detected by the direct imaging missions of the Formative and Visionary Eras (see Section 2.3). Recognizing the importance of these ground-based efforts, NASA has invested in this line of research, with recent examples being the Eta-Earth survey and the Kepler follow-up program on the Keck I ground-based telescope.

### Citizen Scientists

Harnessing the curiosity of the public to do science is possible in the age of digital astronomy. The website planethunters.org has used the incredible pattern-recognition capability of the human brain to find transiting planets in Kepler data that were missed by automated software due to complex stellar variability. The WFIRST-AFTA search for exoplanets with gravitational microlensing will also allow the public to participate, since the signal detection is difficult to automate but may be easy to detect by trained people.

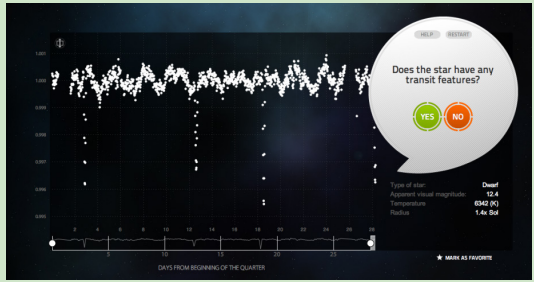

*Figure 2.6 Snapshot from the* **planethunters.org** *website showing a planet transit (the dips) and a prompt for classification.*

Complementing the RV surveys described above, which are generally only sensitive to old, stable stars, ground-based direct-imaging surveys over the next decade will provide crucial constraints on the outer regions of young exoplanetary systems. Direct imaging of exoplanetary systems will be performed with increasingly capable Extreme Adaptive Optics instruments being added to today's largest ground-based telescopes. These instruments will provide a statistically significant sample of Jupiter-mass or larger planets on wide orbits around stars less than a billion years old, thus contributing to a more complete understanding of planetary formation and evolution. Spectroscopy of the brightest exoplanets revealed by these surveys will provide unprecedented insight into the physical and chemical properties of giant planets. While not capable of directly imaging Earth-size planets, the current 8- to 10-meter ground-based telescopes, together with the new ground-based Atacama Large Millimeter/Submillimeter Array (ALMA) radio observatory, will study young planets and planet-forming disks, providing a link between the first juvenile stages of star formation and mature planetary systems.

### *The Exoplanetary Suburbs: Surveys for Outer Planets*

Over the next decade, the transit, RV, and direct-imaging surveys mentioned above will provide a census of "hot" and "warm" planets down to the Earth's mass in orbits close to stars near the Sun. Nonetheless, we will still have a very incomplete picture of the "cold" outer parts of planetary systems beyond the "snow line." The snow line is the location in a planet-forming disk where the temperature is cold enough for water vapor to turn into ice. The region beyond the snow line is where the most massive gas and ice giant planets in our solar system reside, and this is also where the currently favored models of planet formation predict that the majority of exoplanets form, including all the gas giants. Furthermore, these suburbs are the source of surface water—a key ingredient for life on Earth—for rocky planets in the inner regions, through impacts of icy asteroids and comets just before the end of planet formation.

The Near-Term Era mission WFIRST-AFTA will complement the Kepler census of exoplanets by surveying the outer regions of planetary systems using two complementary techniques. First, taking advantage of its prime camera's wide field of view, WFIRST-AFTA will use the gravitational microlensing exoplanet detection technique (**Figure 2.4**) to find thousands of exoplanets with masses greater than Mars and orbits larger than Earth's. It can even detect thousands of "rogue" free-floating planets kicked loose from the stars they formed around. This census will also overlap with Kepler for planets near the habitable zone, thus providing a valuable second estimate of the prevalence of potentially habitable planets (**Figure 2.7**).

Second, WFIRST-AFTA will be equipped with a specialized camera with a narrow field of view to directly image and acquire spectra of circumstellar disks and companion giant planets orbiting the closest, brightest stars. This camera will employ an internal coronagraph instrument to suppress light from the bright host stars, enabling direct detection of the much fainter light from planets and disks. In addition, this instrument

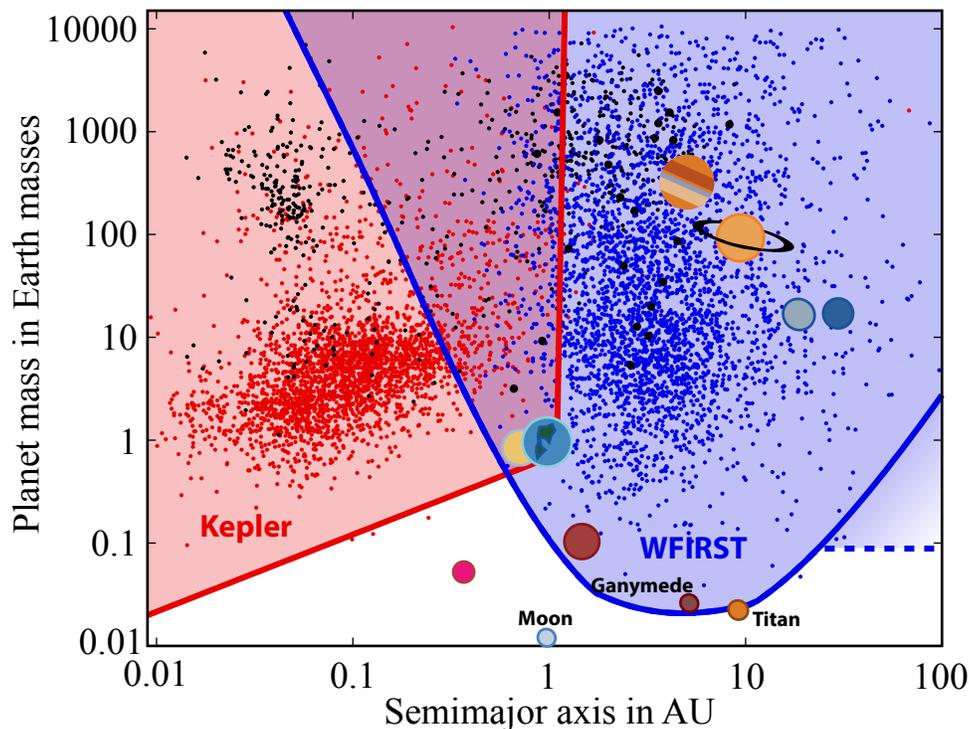

*Figure 2.7 Kepler (red) and WIFRST-AFTA (blue) provide complementary sensitivities and together cover nearly the entire exoplanet discovery space, enabling a complete statistical census of exoplanets. Kepler detections are shown as red dots and simulated WFIRST-AFTA microlensing detections are shown as blue dots. The mass limit for detecting free-floating exoplanets with WFIRST microlensing is indicated by the horizontal dashed line. Solar system planets, as well as the moon, Ganymede, and Titan, are also shown.* **Credit: M. Penny (OSU) and J.D. Myers (NASA)**

will serve as a foundation for the development of technologies and techniques needed for the envisaged Formative Era mission to directly image exoEarths in the habitable zones of nearby stars.

## 2.2 What Are Exoplanets Like?

What are the properties of exoplanets? Through centuries of ground-based observations and decades of in-situ probes, we have been able to study the planets in our solar system in great detail. These studies have allowed us to construct sophisticated models of the interiors, surfaces, and atmospheres of these worlds, which in turn have provided profound insights into their origin and evolution. On the other hand, exoplanets are a diverse collection of worlds with very different properties and environments compared to those within the solar system. The collection includes rocky "super-Earths," gas- and ice-rich "mini-Neptunes," "super-Jupiters" with masses up to the boundary between planets and brown dwarf stars, and the previously mentioned "hot Jupiters."

The best hope of understanding the diverse properties of these exoplanets is to perform comparative planetology: make detailed measurements of the properties of a wide range of planets and compare them to our models and to each other. The Hubble and Spitzer Space Telescopes have started us down this path by providing constraints on the sizes, atmospheres, temperatures, and winds of transiting hot Jupiters. Current and future ground-based telescopes will extend these studies to smaller and cooler planets that might more closely resemble the planets in the solar system. The Near-Term Era TESS mission will identify a set of transiting planets that will be studied in great detail with JWST. In addition, WFIRST-AFTA and JWST will directly image many giant planets, providing complementary information about their atmospheres and compositions. In the following Formative Era, a large ultraviolet, optical, and near-infrared space telescope (LUVOIR Surveyor) will directly observe a sizable sample of diverse planets and provide detailed measurements of their atmospheres. These data will allow us to better understand the properties that dictate habitability and to put our own habitable world in context.

### *Comparative Planetology of Earth-size Exoplanets with TESS*

The Explorer class mission TESS, scheduled for launch in 2017, will continue the census of nearby systems by providing a large sample of transiting planets orbiting close to nearby stars. In addition, because the host stars are bright, these transiting planets will be suitable for follow-up RV measurements to determine planet masses. These systems can also be followed up with, e.g., JWST using transit spectroscopy observations to probe planetary atmospheres. TESS perfectly complements Kepler because it is sensitive to bright, nearby systems, rather than the fainter, more distant systems probed by Kepler.

The TESS mission complements RV and direct imaging surveys by providing precise measurements of planet sizes. With masses provided by RV and sizes from transit observations, the mean planetary densities can be calculated. Density is indicative of a planet's bulk composition; for example, a rocky planet has a higher mean density than one with a large proportion of ice or gas. A key goal is to populate the mass-radius diagram with a statistically significant sample of small ($M_{planet} < 10\ M_{Earth}$) planets. Results from the Kepler mission suggest that these planets exhibit a wide range of compositional diversity—from low-density mini-Neptunes to dense iron-rich super-Mercuries. Furthering these studies is critical for our understanding of the physical properties, habitability, and formation of exoplanets.

An extremely important question that TESS plus RV will answer an extremely important question: What is the fraction of Earth-size planets that are truly rocky? This is a vital parameter, because future direct imaging and spectroscopy of potentially Earth-like planets will only provide indirect information about their bulk compositions. Therefore, the statistical information provided by TESS and RV together on the diversity of bulk compositions of small planets will be needed to put these future observations in context.

### *Understanding the Diversity of Exoplanets with JWST Spectroscopy*

The next flagship mission of the NASA Astrophysics program is JWST (**Figure 2.8**), the successor to both HST and Spitzer, scheduled for launch in 2018. The JWST mission is expected to make an enormous impact in the area of exoplanet science by providing a platform for transit spectroscopy observations with unprecedented precision, resolution, and wavelength coverage. Transit spectroscopy is most sensitive to planets in the inner parts of planetary systems. This technique is a natural complement to direct-imaging techniques, which are sensitive to the outer regions of planetary systems.

JWST observations will revolutionize our understanding of hot Jupiters, the subjects of the first forays into exoplanet atmosphere studies using HST and Spitzer. JWST's high-precision time-resolved spectra of these planets will constrain their chemical abundances, temperature structures, and dynamics. Ultimately, this will lead to a deeper understanding of the origins, evolutionary paths, and physics of these mysterious objects.

Remarkably, JWST will also have the capability to study the atmospheres of smaller close-in exoplanets—the mini-Neptunes and super-Earths that Kepler has shown to be extraordinarily abundant in our galaxy (**Figure 2.5**). In favorable cases, JWST may even probe the atmospheres of close-in Earth-size exoplanets. Beyond measuring the mean densities of small exoplanets (as discussed above), the TESS mission will pave the way for JWST by finding planets near enough to probe their atmospheres with transit spectroscopy. Exoplanet discovery with TESS and subsequent JWST characterization will enable unprecedented comparative planetology that is critical for understanding the properties of small planets beyond our solar system.

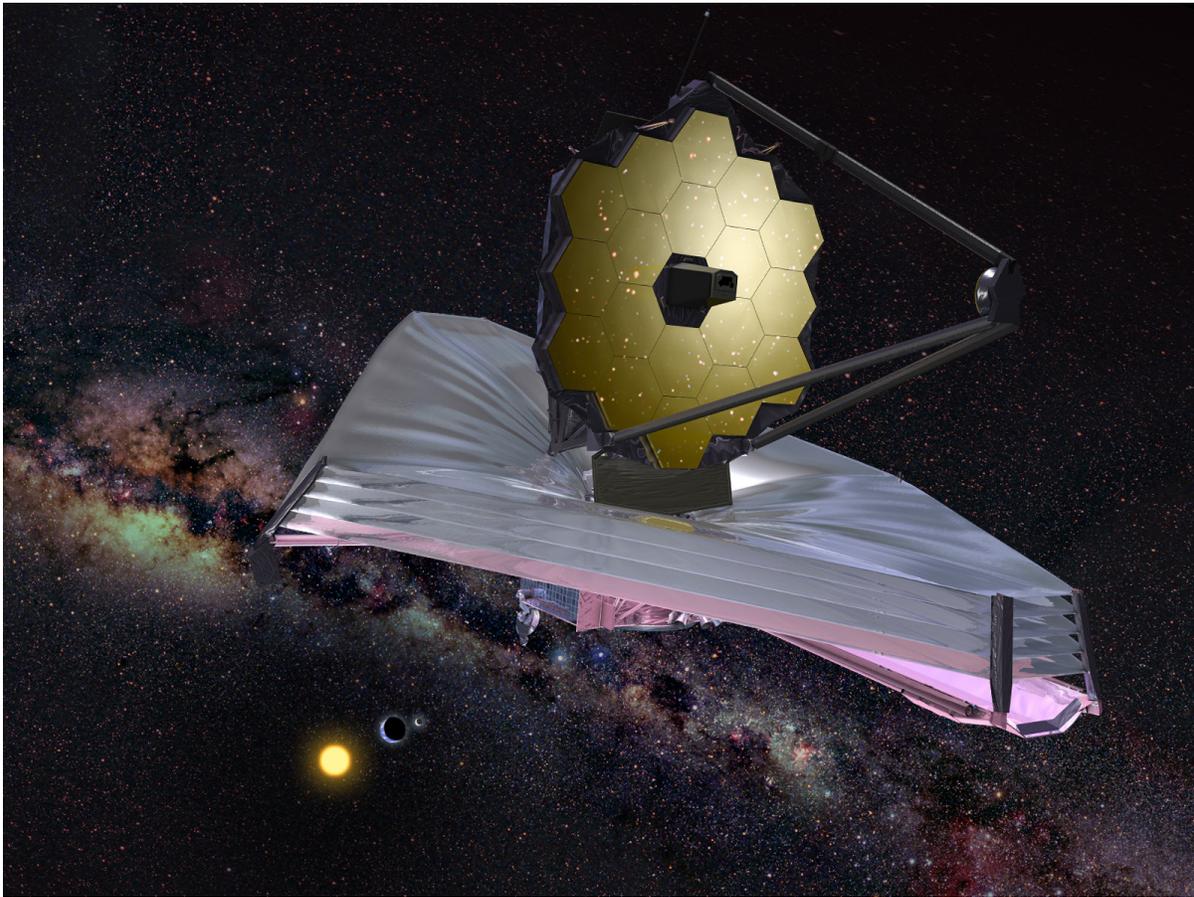

*Figure 2.8* Illustration of JWST, which will usher in the era of exoplanet characterization. **Credit:** NASA

## Studying Giant Exoplanets with the WFIRST-AFTA Coronagraph

DEPLOYMENT of the 2.4-meter optics made available to NASA by the National Reconnaissance Office, equipped with a state-of-the-art high-contrast imaging instrument, will help usher in the long-awaited dream of routine exoplanet direct imaging and spectroscopy from space. The baseline design for WFIRST-AFTA includes such an instrument, whose detailed characteristics and capabilities are being defined as of this writing. It will be capable of direct optical wavelength observations of exoplanets in the outer regions of planetary systems that will far outperform what can be accomplished from the ground due to the blurring effects of Earth's atmosphere.

WFIRST-AFTA will survey many of the nearest stellar systems in order to directly image mature gas giant planets more similar to Jupiter and, most importantly, to begin categorizing them by measuring their reflected light colors. A number of previously unknown Neptune-sized exoplanets also may be discovered. For the best targets, coarse optical spectroscopy sensitive to strong molecular absorption and scattering might be obtained, which will allow further insight into their atmospheric compositions and structures. Using this facility, astronomers will, for the first time, be able to characterize exoplanets orbiting near the snow lines of nearby stars. For our solar system, the snow line is thought to have been somewhere around the location of the asteroid belt, demarcating the boundary between the formation of rocky planets in inner regions and much more massive gas giant planets in outer regions. Therefore, the position and characteristics of exoplanets relative to this boundary provide potent insight to the processes governing planetary system formation and evolution.

*Planetary System Characterization with LUVOIR Surveyor*

Ultimately, only a large space telescope can provide the sensitivity needed to locate the bulk of the planets in the solar neighborhood and reveal the details of their atmospheres. That telescope must be equipped with a next-generation high-contrast system to suppress the blinding light from the host stars and reveal the relatively faint planets nearby. We envision a large ultraviolet, optical, and near-infrared space telescope to achieve this (LUVOIR Surveyor). Such a telescope is the natural next step toward understanding potentially habitable planets (see Section 2.3), but it also will provide a way to understand entire planetary systems. Truly understanding the nature of Earth-like exoplanets requires knowledge of the characteristics of other planetary bodies in the same system, and it demands confronting theories of planetary system formation and evolution with the full complement of diverse exoplanet systems that exist in nature.

The LUVOIR Surveyor will identify Earth-size and larger exoplanets orbiting inside the snow lines of nearby stars. Most importantly, it will perform detailed probes of their atmospheres. This will complete the census of the planetary systems in the solar neighborhood, and extend the extensive comparative planetology work anticipated with JWST, which will be limited to the most close-in planets. Neptunes and super-Earths at wide separations (0.5 to 10s of AUs) will be studied in detail for the first time. This work will provide profound insight into (i) the general processes of planet formation around different types of stars, and (ii) the context of habitable exoplanets, by revealing the complete characteristics of the systems these planets exist in.

## 2.3 The Search for Life

Is there life on other worlds? Although this is one of our oldest and most enduring questions, establishing a definitive answer has long been frustratingly out of reach. However, recently and for the first time in human history, we have finally been able to embark on the systematic, scientific pursuit of an answer. These efforts began near the end of last century with the development of the technology needed to detect planets around other stars, and were followed soon after with the first exoplanet discoveries. In the intervening years, progress has been astonishingly rapid, and we are currently witnessing a great strengthening of the scientific foundation for humanity's speculations about life on other worlds. Upon this foundation, we can now realistically build ambitious and rigorous experiments that will tell us how common Earth-like environments are and whether they harbor life of their own. Here we describe a systematic path toward those transformative goals, starting with near-term work preparing for a search for habitable conditions and signs of extraterrestrial life on nearby worlds.

It begins in the Near-Term Era by determining the prevalence of rocky exoplanets in habitable zones using Kepler and WFIRST-AFTA. At the same time, the ground-based Large Binocular Telescope Interferometer (LBTI) will survey nearby stellar habitable zones for interplanetary dust that can hide small planets in direct images and spectra. In concert with all this, studies of the diversity and characteristics of a wide variety of exoplanets using TESS, JWST, and WFIRST-AFTA will provide context for our understanding of the conditions needed for habitability. Furthermore, astrobiology investigations will help us better understand possible life signs on other worlds.

In the Formative Era, a large space telescope mission (LUVOIR Surveyor) will confirm whether rocky exoplanets in the habitable zones of nearby stars really have habitable conditions on their surfaces. This mission could possibly even find indirect atmospheric signs of surface life. Finally, we describe the ambitious ExoEarth Mapper mission concept, a Visionary Era interferometric space facility that would resolve and map the surfaces of nearby exoEarths, looking for direct evidence of continents, oceans, and surface life.

*Preparing the Way: Reconnaissance of Stellar Habitable Zones*

To effectively design a mission that can probe the atmospheres of potentially habitable planets around Sun-like stars, we first need to know what percentage of the target stars have Earth-size planets in their habitable

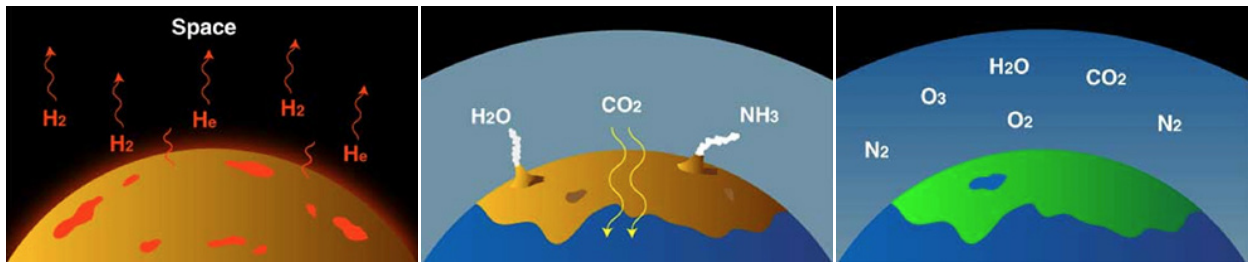

*Figure 2.9* Earth's three atmospheres. Left: The primordial hydrogen-rich atmosphere just after formation of the Earth, about 4.6 billion years ago. Middle: Earth's early atmosphere, about 3.8 billion years ago, contained large amounts of gases released from volcanoes. Right: Earth's present-day atmosphere, containing abundant free oxygen coming from life. **Credit: NASA**

zones. This determines how many target stars we have to observe to find a certain number of exoEarths. As described in Section 2.2, this percentage will be securely determined with data from Kepler, TESS, and WFIRST-AFTA.

> **Is there life on other worlds? For the first time in human history, we have finally been able to embark on the systematic, scientific pursuit of an answer.**

Furthermore, we need to know how "clean" the habitable zones of the target stars are. Dust in the habitable zones (exozodiacal dust) comes from extrasolar asteroids and comets and is thus an expected part of a planetary system. Direct observations of terrestrial planets around nearby stars will likely be limited by light from this dust. The astronomical community has recognized the importance of probing nearby stars to assess typical exozodiacal dust levels (Astro2010 Decadal Survey). NASA has funded construction and operation of a ground-based instrument to perform this task, LBTI on the Large Binocular Telescope at Mount Graham, Arizona, which is now being commissioned. LBTI science operations are expected to begin in 2014.

Finding exoEarths requires synergy between astronomical and biological research. The intersection of these disciplines is the realm of astrobiology, the quest to understand life's origins and recognize its signs. The dynamic contributors to Earth's atmosphere have changed substantially over its long history, dramatically modifying the planet's climate and the bulk composition of its atmosphere (**Figure 2.9**). Before the Great Oxygenation Event, about 2.4 billion years ago, Earth's atmosphere did not contain large amounts of free oxygen although life had long flourished on its surface. The event occurred soon after the rise of photosynthetic cyanobacteria that convert $CO_2$ into $O_2$; the exact timing depended on the balance between oxygen-producing life and oxygen-destroying geologic processes. Other exoEarths will likely have ages different from the present-day Earth. To understand their atmospheres and identify which of them might harbor life, we first need to better understand atmospheric signs of life throughout our Earth's long inhabited history.

### *Pale Blue Dots*

THE next steps are to find rocky exoplanets with liquid water on their surfaces and search for telltale signs of life's influence in the atmospheres of these exoEarths. To accomplish these goals, we must extract light from extremely faint Earth-sized exoplanets out of the blinding glare of their Sun-like host stars. The starlight could be suppressed with an internal coronagraph or external occulter (a free-flying starshade) added to a large optical/near-infrared space telescope (LUVOIR Surveyor). Because the planets are so distant from us, each will appear as a single smudge of light, an "unresolved" image (**Figure 2.10**).

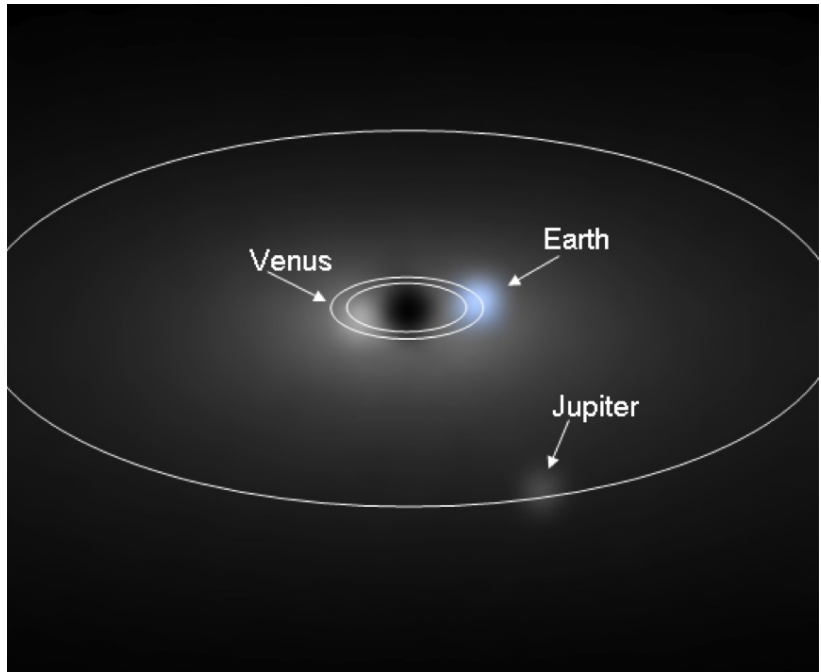

*Figure 2.10 Our "pale blue dot." This is a simulated image of the solar system viewed from interstellar distances. The Sun is blocked out in this image, revealing the relatively faint planets in orbit. The haze of light around the orbits of Venus and Earth comes from exozodiacal dust. **Credit: W. Cash (Univ. of Colorado)***

**The search for life-supporting molecules and biomarkers:** Determining whether a small exoplanet in a star's habitable zone is truly habitable will take more than just measuring its mass/size. We need to determine whether it has the surface materials needed for life—like water—by measuring the planet's spectroscopic properties at optical and adjacent wavelengths. A spectrum of the light collected from the whole planet surface is referred to as a "full disk" spectrum, which also has the signatures of the planet's atmosphere imprinted into it. **Figure 2.11** shows a comparison between the full-disk spectra of Earth, Mars and Venus. Earth's full-disk spectrum is rich in signs of atmospheric biogenic molecules, which are not seen in the others. Water, oxygen, and carbon dioxide—the building blocks of our biosphere—are the key molecules to search for in the atmospheres of rocky exoplanets.

The very presence of these molecules in an exoplanet atmosphere hints at the presence of life. In this case, the molecules are referred to as "biomarkers." A powerful example in Earth's atmosphere is the presence of abundant $O_2$ (coming from photosynthetic plants and bacteria) and methane (coming from animals and bacteria). It will likely be difficult (perhaps impossible) to unambiguously prove that these molecules cannot arise abiotically in an exoplanet atmosphere. However, the mere detection of an Earth-like composition for an exoplanet atmosphere in our solar neighborhood will profoundly change our view of the commonality (or rarity) of the conditions needed for carbon-based life as we know it.

**Characterization of exoEarth systems:** There is still more to learn about a possible exoEarth in addition to its atmospheric composition. Snapshots of the planet taken every few months will show its orbit around the host star and will reveal how much energy it receives, which is vital for understanding the planet's surface temperature.

Next we can look at the finer features, without actually mapping the surface of the planet. Weather and seasonal variations can be measured through changes in total planet brightness as clouds and ice form and dissipate. Given a sufficiently large space telescope (about 8 meters or larger), continents and oceans could be detected and mapped in longitude over the rotation period of the planet (**Figure 2.12**).

A large moon might outshine its planet at some wavelengths, allowing us to detect it. This would enable accurate dynamical measurement of the planet mass. These data will potentially tell us about the planet's surface gravity, the length of its day, its cloud cover, chemical composition, atmospheric pressure, obliquity, and approximate land, ice and ocean fractions, all of which have important implications for habitability.

Other information that will be revealed is the presence of additional planets, both exterior and interior to the habitable zone, which is essential for understanding the architecture of the planetary system: the number, size, and location of all planets and minor body belts. These architectures contain clues to the formation and history of planetary systems, including the possible delivery of water to the habitable zone, and can be used to evaluate leading theories and models.

**Required tools:** Earth is about 10 billion times fainter than the Sun at visible wavelengths. If it were viewed from a distance of 10 parsecs (33 light-years), the separation between Earth and Sun would be excruciatingly tiny (0.1 arcsecond). To isolate light from the faint planet, we must suppress the blinding starlight to an extremely high degree at very small separations from the star. Achieving this high contrast needed to study exoEarths around Sun-like stars necessitates a space-based telescope to eliminate the blurring effect of Earth's atmosphere. Technology development on starlight-suppressing instruments (internal coronagraphs and/or external occulters) will also be required (see Chapter 6).

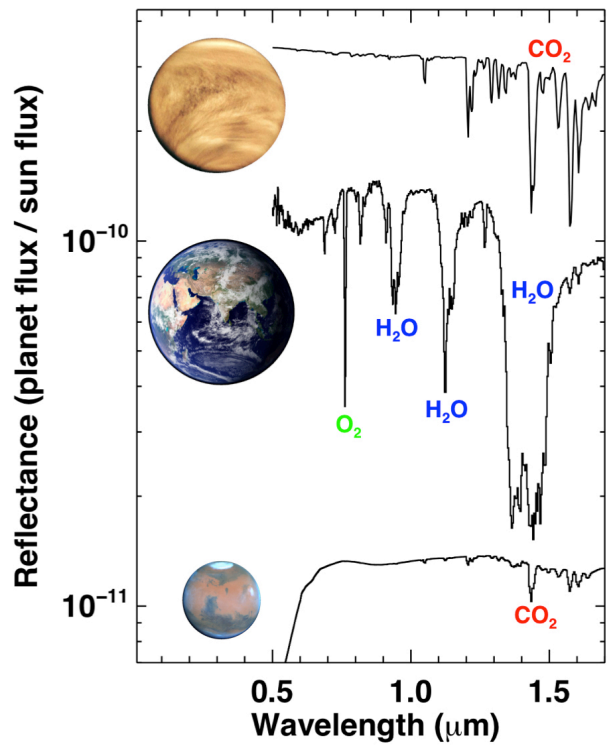

*Figure 2.11 The reflectivity of the solar system's three largest terrestrial planets—Venus (top), Earth (middle), and Mars (bottom)—at visible and near-infrared wavelengths. Each spectrum is imprinted with the signatures of the planet's atmosphere in the form of absorption lines due to molecular gases. Venus and Mars primarily show only lines of carbon dioxide. Earth shows a rich atmosphere filled with oxygen and water vapor.* **Credit: The Virtual Planet Laboratory and V. Meadows (Univ. of Washington) and A. Roberge (NASA GSFC)**

Furthermore, to have sufficient signal from the planet to perform all the observations described above, the telescope mirror must be large enough to collect enough light from the faint planet. We estimate that LUVOIR Surveyor with a diameter between 8—16 meters can achieve our science goals. A large telescope also has the advantage of being less impacted by interplanetary dust enshrouding the planets (exozodiacal dust), concentrating the light from the planet so that it stands out above the "fog" of the dust. Assuming optical observations with this facility, the entire region exterior to the inner edge of the habitable zone for a Sun-like star (>0.5 AU) becomes accessible for stars within 60 light-years (120 light-years for a 16-meter telescope).

*Figure 2.12* Finding oceans and continents on exoEarths. The top panel shows the actual distribution of land and water on Earth. The bottom panel shows the longitudinal map of Earth derived from repeated full-disk images in nine colors, taken with the EPOXI experiment on NASA's Deep Impact spacecraft. This type of observation is similar to what we could do for exoEarths given a sufficiently large space telescope. Full-disk images cannot provide latitudinal (north/south) information on surface features but can recover the longitudinal (east/west) location of continents and oceans as they come in and out of view while the planet rotates.
**Credit: N. Cowan (Northwestern Univ.)**

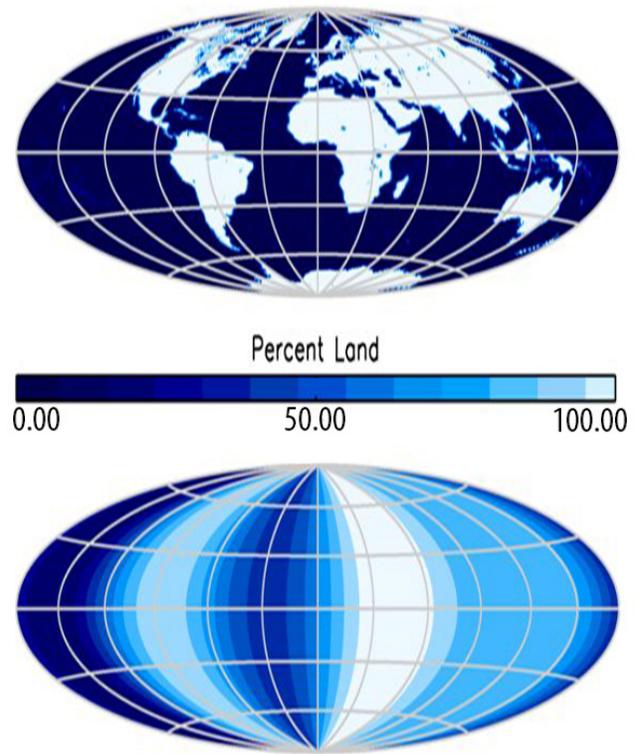

Unless a moon is detected and its orbit measured, high-contrast images alone do not yield the planet mass, an important parameter for interpreting atmospheric spectra and understanding conditions at the surface of the planet. Vigorous development of the RV technique with large ground-based telescopes may enable measurements of the masses of exoEarths after they have been found with space-based imaging. An alternative approach would be to measure the astrometric wobble of the host stars caused by the planets as they orbit, which could possibly be done in conjunction with high-contrast imaging using the LUVOIR Surveyor. Both approaches require technology maturation to reach the sensitivity needed for Earth-mass planets.

In sum, we believe that the first steps toward the detection of truly habitable planets and life will occur with a large optical space telescope equipped for high-contrast imaging and spectroscopy, such as LUVOIR Surveyor. There are two possible devices to suppress the starlight: internal coronagraphs or external occulters (starshades). Each technology has technical and scientific advantages and disadvantages. The occulter approach should be considered if it proves difficult for coronagraph technologies to provide the required contrast and sensitivity, and should therefore remain a topic of further study. With such a large high-contrast space telescope, we will discover how common Earth-sized planets with surface water are within our neighborhood in the galaxy, and begin probing them for life.

### Intelligent Life?

The SETI project is a ground-based attempt to discover technological signals from intelligent life outside the solar system. It contrasts with the program described in Section 2.3 to discover simple life forms, which we think are likely to be much more common. Although there are many reasons we might not be able to detect technological signals (intelligent life is rare, they don't want to talk, etc.), SETI represents a high-risk, high-reward endeavor that complements other scientific efforts to find life.

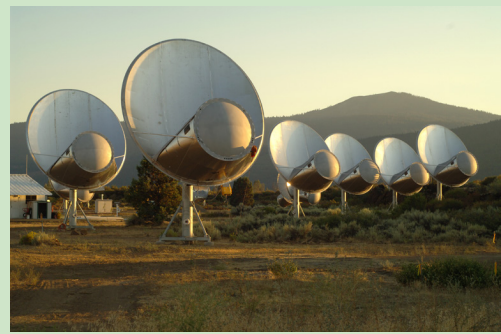

*Figure 2.13* The SETI project's Allen Telescope Array.
**Credit: S. Shostak/SETI Institute**

*Exploring "New Earth": Mapping the Surfaces of Habitable Worlds*

By the end of the Formative Era, we will have a collection of exoEarths around nearby stars, some of which might show tantalizing indirect signs of liquid surface water and possibly life. We desire to learn all we can about these planets—and know them as individual worlds. Somewhere among the collection may be a future home for us; somewhere else we may find the home of otherworldly life forms.

In the Visionary Era, we envision an ambitious "ExoEarth Mapper" that will enable us to resolve the surfaces of these newfound planets. The progression from pale blue dots to maps of fully resolved worlds follows a natural advancement in scientific capability, such as the one that guided the exploration of our solar system over the last centuries. From blurs in Galileo's telescope to Google maps of Mars, our understanding of the character of a planet explodes with the ability to spatially resolve its surface. The ability to create maps of exoEarths around nearby stars will allow exoplanet science to fully take its place alongside Earth and planetary sciences. Following the path laid out by those disciplines, we have an exciting and detailed picture of what we hope to discover.

**The planet contours—oceans and landmasses:** The high-quality, disk-integrated spectra of terrestrial exoplanets obtained in the Formative Era can provide evidence for continents on a planet through changes in its brightness and color over time as the planet rotates. However, this type of observation can only tell us about the longitudinal variation of the surface. For example, modeling of Earth observed as an unresolved dot suggests two major landmasses (Eurasia/Africa and the Western Hemisphere) and only gives their rough relative sizes (**Figure 2.12**). Mapping the surface, on the other hand, will let us measure the land-to-ocean ratio more accurately. Moreover, the shapes of Earth's continents provide evidence for plate tectonics, which is necessary for the long-term stability of Earth's atmosphere and its habitable conditions.

### Earth-size Planets around Red Dwarf Stars

We have long known that red dwarfs—low-mass stars with between a tenth and half of the Sun's mass—are about 10 times more common than Sun-like stars. Because red dwarfs are much less luminous, their habitable zones are located closer in, making planets in these zones more likely to transit their stars. Thanks to these two facts, the Kepler mission has already shown that Earth-size planets in the habitable zones of red dwarfs are quite common. We do not yet know if such planets can really be habitable, given their very different histories and stellar environments. But if they can, these systems provide an opportunity to find habitable planets in systems close to the Sun relatively soon. The upcoming TESS mission will search for transiting Earth-size planets in the habitable zones of red dwarfs near the Sun. These "small black shadows" would be close enough to us that we might begin to study their atmospheres with follow-up JWST transit spectroscopy and, in very favorable cases, potentially find evidence for habitable conditions on such worlds.

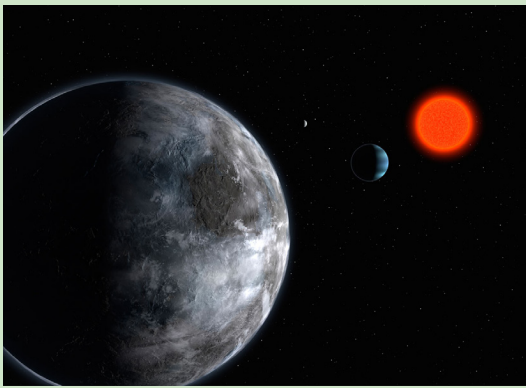

There is another upcoming opportunity to study Earth-size planets in the habitable zones of nearby red dwarfs with the next generation of extremely large telescopes (ELTs) on the ground. These telescopes are planned to begin operations some time in the next decade and will have diameters of 25 to 40 meters, compared to the current generation of ~ 10-meter ground-based telescopes. While Earth-size planets in the habitable zones of Sun-like stars are out of reach from the ground, the ELTs might be able to directly image planets in the habitable zones of the very nearest red dwarfs. This complements space-based efforts to study Earth-like exoplanets around Sun-like stars, where we have greater hopes of identifying and understanding life.

**Figure 2.14** *Artist's concept of a super-Earth and a Neptune-like planet around the red dwarf star Gliese 581.* **Credit: ESO**

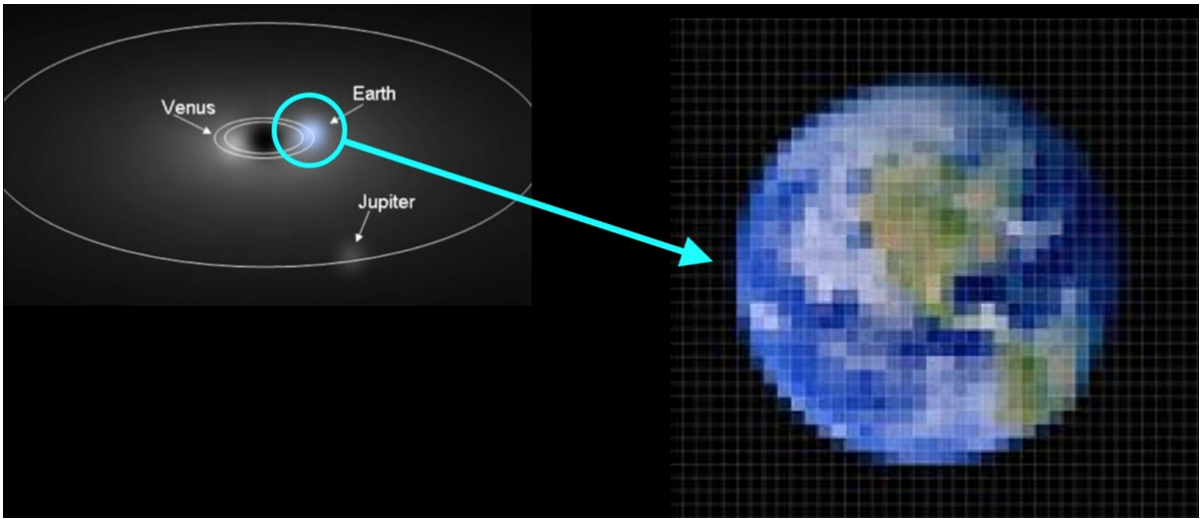

*Figure 2.15 Mapping an exoEarth reveals its character. At left is a simulated image of the solar system viewed from interstellar distances with the Sun blocked out, revealing the relatively faint planets in orbit. At right is a spatially resolved image of Earth, similar to what we hope to accomplish for an exoEarth around another star. **Credit: NASA/W. Cash (Univ. of Colorado)/S. Gaudi (Ohio State Univ.)/A. Roberge (NASA GSFC)***

Similar features might also exist in exoEarths. Finally, obtaining spectra of particular regions of the planet's surface (or even just their colors) can confirm the presence of actual liquid water and allow us to distinguish between land, ice, clouds, and possibly vegetation.

**The planet character—moons, seasons, weather, and climate:** A telescope capable of mapping the surface of an Earth-size planet around a star several light-years away also can routinely discover its moons. Measurement of a moon's orbit will directly provide the planet's mass. Furthermore, the presence and orbit of Earth's moon bear witness to the violent history of the late stages of Earth's formation, as well as stabilize the spin of our planet and enhance ocean tides. Observations of a planet's surface over the entire orbit can reveal whether it has seasons, through growth and shrinkage of any ice caps. We will thus be able to investigate the planet's weather by measuring the time-varying cloud coverage, and looking for patterns associated with atmospheric circulation. Finally, the combination of all these measurements would provide the necessary inputs for realistic three-dimensional modeling of the planet's atmosphere, as is done for Earth using Global Circulation Models (GCMs). Such high-fidelity modeling will reveal the long-term evolution of the planet's atmosphere—its climate—and tell us about the history and future of the planet's surface.

**The planet inhabitants—investigating life:** Hopefully, some of the pale blue dots we discover will show intriguing signs of biomarker gases their atmospheres. Identifying the kind of life that might be responsible for these will be extremely difficult, but might be possible with spatially resolved spectroscopy of the planet's surface. For example, photosynthesis by plant life produces a strong biomarker via a high abundance of oxygen in Earth's atmosphere. This life is visible from space, in the beautiful green shades on parts of our continents. Spectra of these areas obtained from Earth-observing satellites show the so-called "red edge" produced by the photosynthetic molecule (chlorophyll). However, this feature can only be detected when the vegetated region is not covered by clouds, and is rarely detectable even in high-quality disk-integrated spectra of Earth. Mapping the surface of an exoEarth can let us peer through gaps in clouds and probe for light-harvesting vegetation-like life.

**Required tools:** The apparent size of an exoEarth at 10 parsecs (33 light-years) is about 223 million times smaller than the full moon viewed from Earth's surface. To make a map of this blue speck, we must divide its image into a number of smaller pieces, like having more pixels in a digital photo. This demands extremely high imaging resolution and therefore a very large telescope.

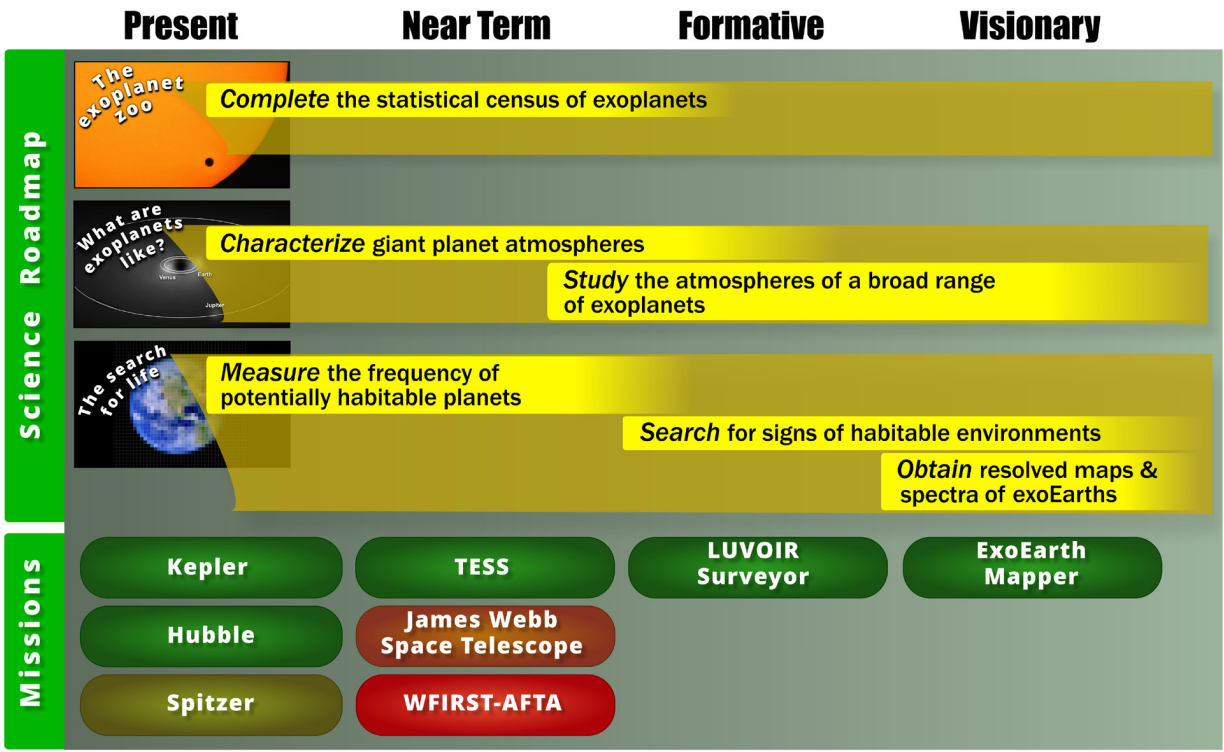

**Figure 2.16** *Schematic of the Exoplanets Roadmap, with science themes along the top and a possible mission sequence across the bottom.* **Credit: F. Reddy (NASA GSFC)**

We would need a mirror with a diameter of about 500 kilometers to make a 30 × 30 element visible-light map of an exoEarth 33 light-years away (as in **Figure 2.15**). That telescope would also have to be outside Earth's atmosphere, in the vacuum of space. Such a large space telescope is far, far beyond our technical capabilities. Even on the ground, today's largest visible telescopes have diameters of only about 10 meters. The planned next generation of large, ground-based visible telescopes will have diameters in the 25- to 40-meter range.

Therefore, we believe that the path forward to an exoEarth mapping mission lies in space interferometry. This technique combines the power of several smaller telescopes widely separated from each other to achieve the image resolution of one huge telescope with a diameter comparable to the separations between the small telescopes. Interferometry has been employed with ground-based telescopes for many years but cannot achieve the image resolution needed for exoEarth maps due to unavoidable limitations imposed by Earth's atmosphere and the curvature of its surface. To someday achieve the goal of mapping exoEarths, we must make progress on a technology development path for space-based interferometry. The gross characteristics of a notional ExoEarth Mapper space interferometer mission appear in Chapter 6.

## 2.4 Activities by Era

The major activities in the exoplanet roadmap are listed below in order of Era.

A graphical summary of the science themes and missions discussed appears in **Figure 2.16**.

A re   W e   A l o n e ?*Near-Term Era*
- **The exoplanet zoo**

  - *Complete Kepler's census of "hot" and "warm" planets and determine the frequency of potentially habitable planets*

  - *Survey nearby planetary systems with ground-based radial velocity and direct imaging, TESS, and finally WFIRST-AFTA*

  - *Complete the statistical census of planetary systems begun by Kepler with the WFIRST-AFTA gravitational microlensing survey*

- **What are exoplanets like?**

  - *Study planet-forming disks and giant planet atmospheres with ground-based imaging, JWST, and WFIRST-AFTA*

- **The search for life**

  - *Measure the frequency of potentially habitable planets with Kepler, ground-based radial velocity surveys, and WFIRST-AFTA*

  - *Quantify dust in habitable zones (exozodiacal dust) with LBTI*

  - *Study habitable zone planets around red dwarf stars with TESS & JWST*

*Formative Era*
- **What are exoplanets like?**

  - *Characterize planet-forming disks and planetary atmospheres with LUVOIR Surveyor*

- **The search for life**

  - *Obtain full-disk images and spectra of pale blue dots with LUVOIR Surveyor*

  - *Make longitudinal maps and detect seasonal variations on exoEarths with LUVOIR Surveyor*

  - *Search for signs of habitability and evidence of biological activity on exoEarths with LUVOIR Surveyor*

*Visionary Era*
- **The search for life**

  - *Resolved maps and spectra of "New Earth" with ExoEarth Mapper*

  - *Confirm surface water and identify possible life with ExoEarth Mapper*

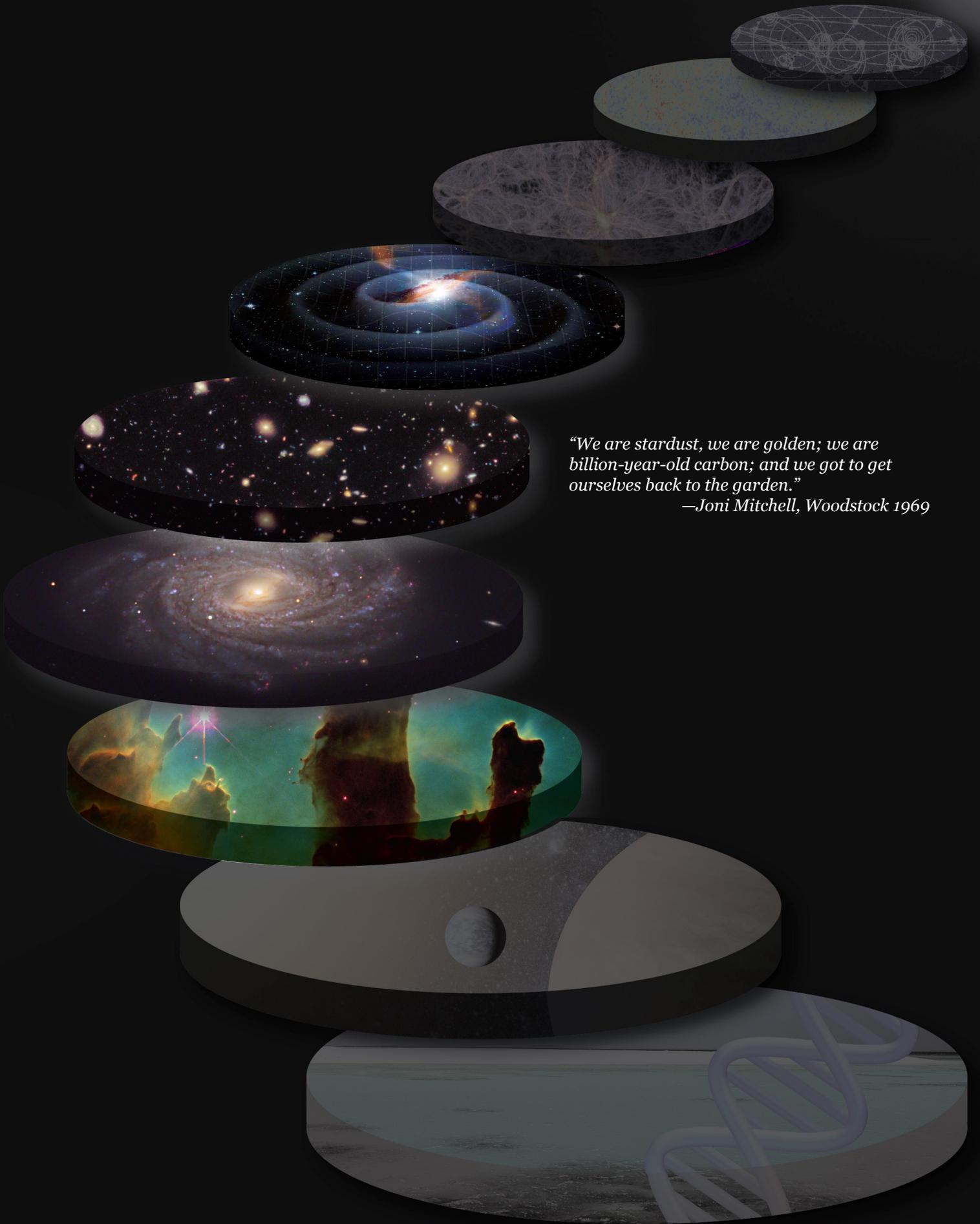

"We are stardust, we are golden; we are billion-year-old carbon; and we got to get ourselves back to the garden."
—Joni Mitchell, Woodstock 1969

# 3 How Did We Get Here?

From the moment our ancestors first looked into the night sky and wondered what was out there, our quest to understand the celestial bodies, from our solar system to the first stars and galaxies and everything in between, has been never-ending. Beyond our own solar system, we seek to explore the great variety of stars, their life cycles, and the generation of elements that ultimately made life on Earth possible. While our Milky Way offers us a unique, close-up perspective of the many different environments within a single galaxy, as well as the many individual components that compose a galaxy, there are billions of galaxies in the universe, each one containing billions of stars and having its own unique life history. We want to understand how these galaxies first began to form, how they were constructed, and how they grew and changed over time. In this chapter we continue our quest to understand our origins by expanding our view beyond the exploration of other planets and out to the components of galaxies (stars, clouds of gas and dust, black holes, and dark matter) and the galaxies themselves, from our own Milky Way to the edge of the universe.

## 3.1 Stellar Life Cycles and the Evolution of the Elements

> One of the most fundamental realizations gleaned by astronomers is that we are made of "star dust." The primordial universe expanded and cooled, allowing the formation of hydrogen, helium, and trace amounts of a few other light elements, but none of the essentials for human life: carbon, nitrogen, oxygen, iron and more. The rich chemistry of nuclei, atoms, and molecules that life is based upon were absent in the young universe. It took the formation of stars, the subsequent nucleosynthesis of elements in their hot cores, and the ejection of these new elements into their environments that allowed life as we know it to emerge. Stellar life cycles, thus, hold the key to the chemical enrichment of the universe, without which no observers would exist to marvel at its beauty.
>
> Extensive astronomical observations are required to understand the processes of star formation and the associated formation of protoplanetary systems, stellar evolution, and, most importantly, the end stages of massive stars, which provide feedback to the interstellar medium (ISM) in the form of outflows (steady winds and explosive events) that chemically enrich, heat, and move the gas between stars. The evolution of gas-star systems is a complex dynamical dance that critically determines the appearances and physical states of galaxies as a whole as well as the specific details of their star-forming regions, where new planetary systems are assembled. We have already sampled a wide variety of nearby stellar nurseries and supernova remnants (the beginning and end of stellar life cycles) and have performed detailed studies of open and globular clusters (stellar midlife stages). In the next 30 years we will extend the census of these environments and obtain their detailed characterizations.

### *Our Solar System: A Home Like No Other?*

We earthlings know and love our neighborhood well. Our affection and curiosity for our solar system and our place within it has been abundantly expressed from our earliest ancestors, who began studying the movements of the planets, the Sun, and the moon and used them to guide their daily activities, to the people around the world who marvel at lunar and solar eclipses. Since the discovery of the first extrasolar planet, we have been on a quest to discover other Earths (Chapter 2) and to understand how star and planetary systems form and evolve. The Hubble and Spitzer Space Telescopes have uncovered the factories where stars and planets are being formed, providing tantalizing clues for what might be happening within them—and we will soon find out! These factories are made of dense molecular gas and dust, hiding within them the rich chemistry and physics involved in the creation of stars and planets. In the next 30 years, we will chart hundreds, possibly thousands, of these star-forming nurseries across the Milky Way and in neighboring galaxies, unraveling the basic processes of how solar systems assemble and the diverse conditions under which different stars and planets can be born.

*Stellar Nurseries*

ALL stars form in immense, dense clouds of molecular gas and dust (**Figure 3.1**). Hidden in these cocoons are clumps of matter that begin to collapse under their own gravity, shedding angular momentum and forming one or multiple stars surrounded by planet-forming accretion disks. Light emerging from these systems is a superposition of direct starlight (mostly in the optical regime) and starlight reprocessed by dust (mostly emerging in the infrared part of the electromagnetic spectrum).

Although we have observed stages of this creation, the precise details, sequence of events, and chemical evolution of star systems remain shrouded in mystery. In the next decade, however, there is immense promise for better understanding these phenomena with ALMA and JWST. ALMA observes at long wavelengths where molecules and dust emit most of their light. In the largest arrangement of its individual antennas, ALMA will be a 10-mile-wide telescope able to map star-forming regions with unprecedented clarity—nearly an order of magnitude better resolution than HST. With its 66 antennas, the collecting power of the telescope will allow it to map the entire variety of molecular clouds and star-forming regions across the Milky Way.

In addition, ALMA will reveal gaps in the protoplanetary dust disks, indicators of planet formation. These exoplanetary systems will be targets also for JWST. Operating at near- to mid-infrared wavelengths, JWST builds on the legacy of the Spitzer Space Telescope with 50 times greater light collecting power, allowing it to map the hotter dust and inner regions of star systems where rocky planets may be forming. Still, it will be limited to the same small number of systems studied with HST because of its relatively small mirror size.

By combining data from ALMA, JWST, and the next generation (> 10 m) of ground-based optical/IR telescopes, over the next three decades we will map hundreds of nurseries of all kinds with different chemical compositions and across all galactic structures and environments (the bar, the molecular ring, the spiral arms, etc.). However, to truly and fully explore the vast array of young star and planetary systems uncovered by ALMA, we will need the resolving power of the LUVOIR Surveyor, to bring together ultraviolet, optical, and near-IR observations at similar resolutions.

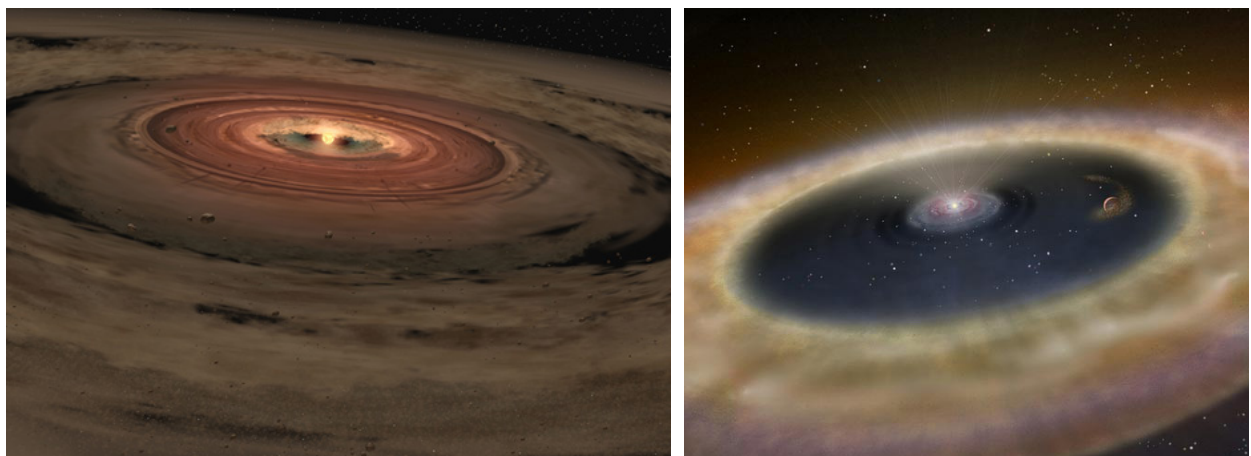

*Figure 3.1* These panels illustrate what we might see if we could peek inside a stellar nursery. On the left is a protoplanetary disk of gas and dust surrounding a young protostar. On the right is a more advanced stage of the same system, in which a planet has formed and cleared out debris near its orbit, creating a gap. In the Near-Term and Formative Eras these gaps can be mapped with telescopes such as ALMA, JWST, and the LUVOIR Surveyor. **Credits: NASA-JPL/Caltech and K.L. Teramura, UH IfA**

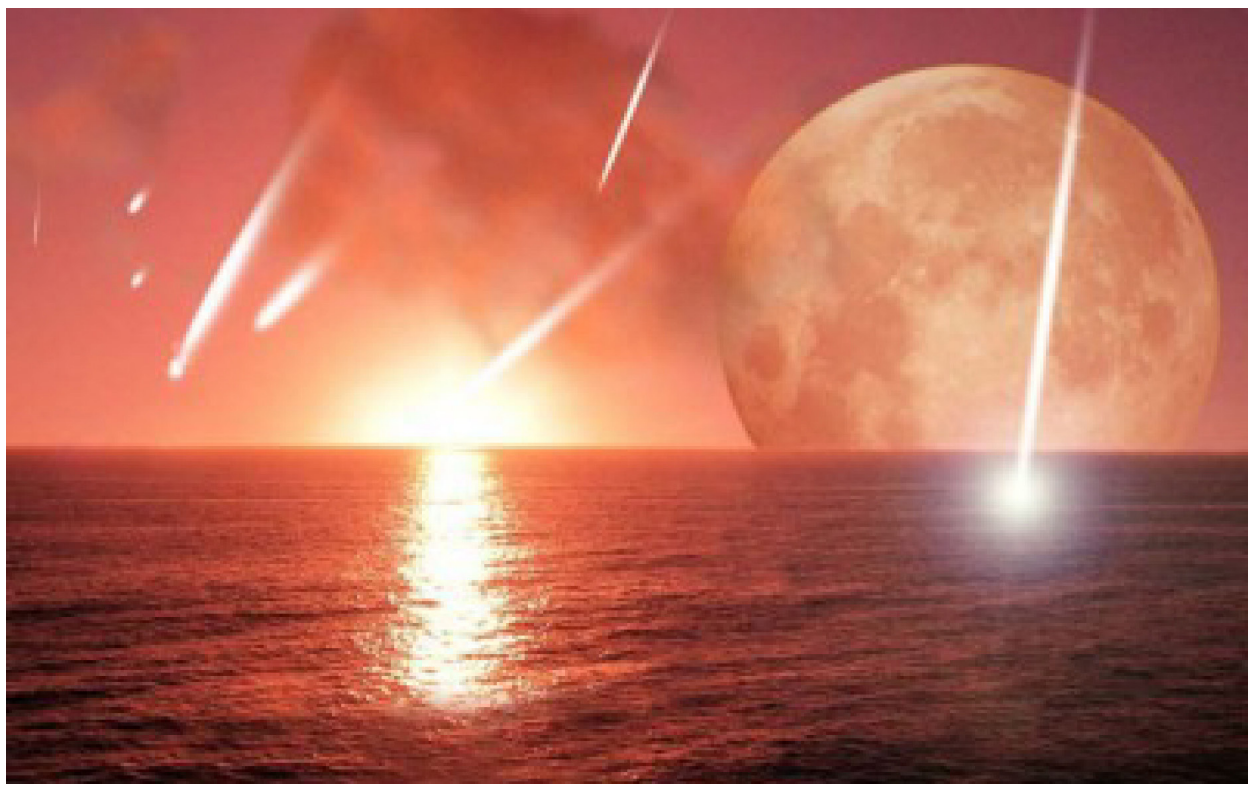

*Figure 3.2 An illustration of how Earth's water may have been deposited from an early bombardment of comets.* **Credit: D.A. Aguilar (CfA)**

## Water: Where does it come from?

Water is fundamental to life on our planet, but where does it come from? The leading theories suggest that water was deposited on Earth through bombardment by asteroids and comets (**Figure 3.2**). As we begin to study the formation of other star systems and look for sister planets, we also need to understand where the most basic ingredients of life, such as water and organic molecules, reside in a stellar nursery before they become incorporated into asteroids and comets. We need to directly detect water in protoplanetary disks and map its locations. Although some by-products of water, such as deuterated species, are visible at millimeter/submillimeter wavelengths, the chemical processes creating them may have multiple pathways leading to inconclusive results about the distribution and evolution of water in other star systems. In the next 30 years, the Far-IR Surveyor will directly measure water emission lines from young stellar systems identified by ALMA and JWST. It would also provide the resolution needed to distinguish whether water is present everywhere in a star system or only in its outskirts. Moreover, by studying star systems at different evolutionary stages, we should be able to determine how the distribution of water evolves with time.

## Debris Disks

Debris disks are the evolutionary link between gas-rich protoplanetary disks and mature planetary systems, representing the late stages of planetary system formation (**Figure 3.3**). These low-mass dusty disks around main-sequence stars are produced by colliding and evaporating asteroids and comets, the building blocks of planets. Younger debris disks (10–100 million years) are the likely sites of ongoing rocky planet formation. In older systems (0.1–1 billion years), impacts by water-rich asteroids from the outer stellar system may be delivering water and other volatiles to young rocky planets. Probing the composition of debris disks will reveal the makeup of planetary building blocks. Initial efforts in this area have been made with HST, but to survey more than just a few systems, the LUVOIR Surveyor's much greater sensitivity is needed. Furthermore, patterns in the dust of debris disks (like gaps and clumps) can reveal the presence of young

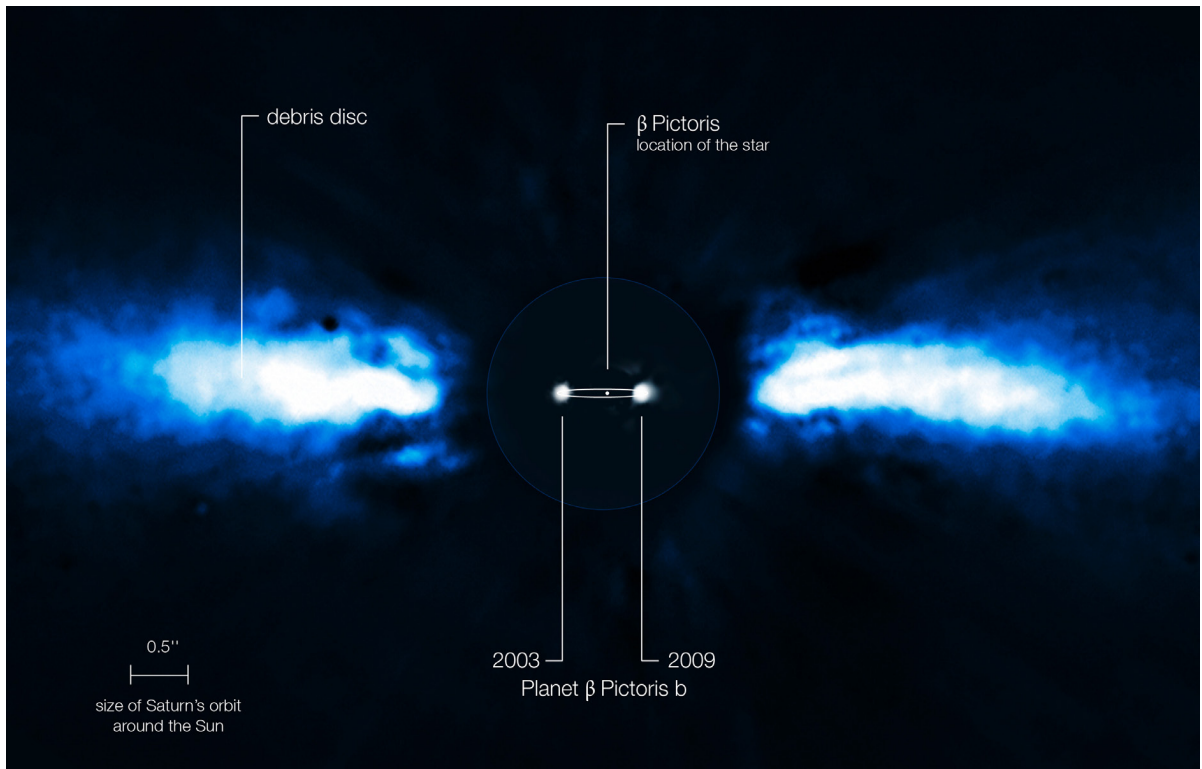

*Figure 3.3* *An image of the 12 million-year-old edge-on Beta Pictoris debris disk and its young gas giant planet. This is one of the few exoplanets that has been directly imaged to date.* **Credit: ESO/A.-M. Lagrange (Grenoble)**

planets that cannot currently be detected in any other way. These young planets are important targets for understanding planetary evolution. Structured debris disks found with ALMA will be the best systems to deeply search for reflected light from young planets using high-contrast imaging and spectroscopy with the LUVOIR Surveyor.

## Stellar Feedback

The cycle of stellar evolution is connected in many ways to the grander schemes of galactic and cosmic evolution. Stars act as cauldrons for element creation, but these newly synthesized elements need to be returned to the interstellar medium (ISM), where they are eventually mixed and incorporated into new star-forming regions. This happens through expulsion of the outer layers of stars during their advanced evolutionary stages (stellar winds, planetary nebulae) as well as the violent final explosions experienced by the most massive stars and the incineration of white dwarfs in Type Ia supernovae. In addition to chemical enrichment, stellar outflows also return energy and momentum to the ISM and produce copious amounts of ionizing radiation that profoundly alters their environments. The combined set of these processes is referred to as "feedback," which constitutes a crucial link between star and galaxy evolution. Together with growing black holes in galactic centers, and accretion via minor and major mergers with other galaxies, the dynamical evolution of galaxies is sensitive to stellar feedback, which couples the gas and stellar components. This, in turn depends on the initial mass function (IMF), which describes the statistical distribution of stellar masses found in a star-forming region. We need to establish whether or not the IMF is "universal" or varies significantly between stellar nurseries. Especially at low stellar masses, we have yet to determine how the mass spectrum is established in the star-planet transition regime. The formation of hydrogen-burning stars, substellar (not hydrogen-burning) objects like brown dwarfs, and their associated planetary systems, as well as free-floating planets, is the outcome of a complex sequence of events. Even the mass boundary

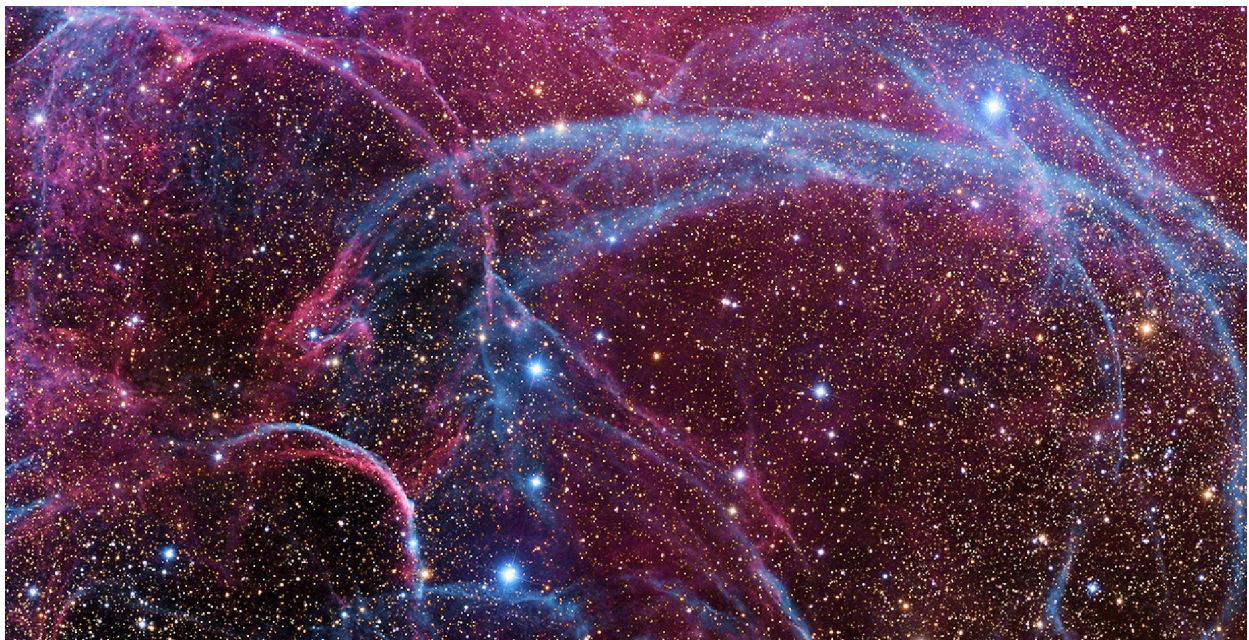

*Figure 3.4* Over 10,000 years ago, a star in the constellation Vela exploded. The expanding shell of ejecta drives shock waves into the surrounding medium, still visible today in X-rays and optical bands. This image captures parts of the filamentary structure created by these shocks, which accelerate subatomic particles to extreme energies. **Credit: A. Lau, Y. Van, S.S. Tong (Jade Scope Observatory)**

between brown dwarfs and planets is not yet fully established. The IMF of objects in this substellar regime, and in particular the transition from stars to planets, remains to be explored with a much more detailed census.

Accreting massive black holes in galactic cores can provide feedback via the outflows they drive, but star formation does so in a more distributed fashion via punctuated supernova (SN) explosions. To study SN-induced feedback, a detailed understanding of their resulting remnants is one of the key goals for the next generation of instruments. Based on what we have learned to date from gamma-X-UV-optical-IR-Radio, breakthroughs are expected from high-angular/energy- resolution imaging spectroscopy. Freshly synthesized elements in a supernova are ejected at very high velocities and fluid dynamic instabilities create inhomogeneous outflows. These mixed ejecta drive shock waves into their external medium (either the ISM or a circumstellar medium composed of material shed via strong stellar winds—a process prevalent in massive stars—before the explosion). Centuries after the supernova, small-scale filamentary shock structures develop as part of the interaction between stellar ejecta and the swept-up surrounding medium (**Figure 3.4**). To fully understand the spatial and temporal structure of outflows from supernovae, and the shock waves that result, we need to obtain X-ray spectroscopic data with high spatial resolution. The X-ray Surveyor would allow us to map many remnants in sufficient detail to obtain position-resolved physical properties, such as temperature, density, and composition profiles, and to separate thermal from non-thermal emission components. Currently, NASA's Nuclear Spectroscopic Telescope Array (NuSTAR) has focused on a few nearby galactic supernova remnants, such as the famous remnant Cassiopeia A, and found evidence that the short-lived (~100 years) radioactive isotope titanium-44 synthesized in the explosion. Another famous supernova, SN 1987A in the Large Magellanic Cloud, is evolving into a spatially resolvable remnant, even at its distance of 163,000 light-years. In the Formative Era, the X-ray Surveyor will enable resolved imaging and spectroscopy of supernova remnants located in galaxies in the Milky Way's neighborhood. Sensitive gamma-ray surveys could directly trace freshly synthesized elements through emission lines from several radioactive isotopes, such as aluminum-26, titanium-44, and iron-60, as well as the diffuse 511 keV line resulting from positron annihilation.

*Particle Accelerators*

THERE is yet another, less apparent component of galaxies: high-energy particles (cosmic rays). A century ago, Victor Hess discovered via high-altitude balloons a background of high-energy charged particles, which we now know pervades the ISM, providing about a third of its energy density. This dynamically important gas of particles requires efficient acceleration, which is now believed to take place in the strong shock waves of supernova remnants. Only recently did gamma-ray observations in the GeV regime (from space) and the TeV regime (from the ground) yield convincing data supporting this paradigm. About half a dozen stars are born every year in the Milky Way; those with sufficiently high mass ultimately end their lives as supernovae, resulting in a few supernova explosions per century. Due to their large energy per explosion, this rate is sufficient to maintain a significant fraction of the ISM in a hot and ionized state, emitting characteristic line radiation in the ultraviolet and X-ray bands. The study of galactic cosmic rays with space based detectors (e.g., Fermi, AMS) and ground-based facilities, such as the Auger array in Argentina (and indirectly through ground-based observations of very high-energy gamma rays), can be advanced with dedicated detectors on the International Space Station such as JEM-EUSO, which will utilize Earth's atmosphere as a particle detector. The study of cosmic ray particle acceleration is an important physics question in its own right, but this high-energy particle component of any actively star-forming galaxy is also responsible for the creation of some crucial elements (lithium, beryllium, and boron). It bathes the galaxy in a diffuse glow of high-energy gamma rays, which represents a significant fraction of its total energy output; the presence of the cosmic ray component can even play a substantial role in the self-regulation of the star-formation rate. The motion of the Sun-Earth system in the galactic potential causes changes to the cosmic ray environment on timescales on the order of 100 million years, which could potentially be linked to extinction cycles on Earth.

## 3.2 THE ARCHAEOLOGY OF THE MILKY WAY AND ITS NEIGHBORS

> Much of what we have learned about galaxies—their assembly, stellar generations, and chemical enrichment—has come from looking at our own Milky Way. For our galaxy, and to a lesser extent its Local Group neighbors, the Magellanic Clouds and the Andromeda galaxy, we can examine—one by one—individual stars, star clusters, and star-forming gas clouds, measuring ages and heavy-element abundances throughout. Such data tell the story of the Milky Way: generations of star formation and their legacy of heavy chemical enrichment, and the growth of stellar mass in various structures—the bulge, disk, and halo. Recently, shallow wide-field surveys of our galaxy's halo have uncovered evidence of the Milky Way's past and ongoing accretion of satellite galaxies, recorded as thin streams of stars. They formed as the Milky Way's gravity induced tides in dwarf galaxies, shearing off trails of stars. These fossils are clues to the archaeology of our galaxy, yet our current census only includes the brightest streams.
>
> In the next 30 years, we will explore the full census of these fossils across the galaxy's halo with sensitive wide-field imagers and characterize each of the substructures, unraveling the properties of the small systems that built up our galaxy. However, the Milky Way is just one galaxy, one whose close neighbors are of a similar type, so we have only a narrow view of the full range of galaxy histories. Astronomers have devoted a large amount of resources using HST to make comparable observations for the Andromeda galaxy. Surprisingly, these results show that the nearest big galaxy to our own—also a spiral—is in many ways different from the Milky Way. Our goal in the next 30 years is to study dozens of galaxies within our neighborhood to the same level of detail we have achieved for Andromeda, thereby expanding our detailed knowledge of how galaxies assemble and evolve to a fully representative range of galaxy types and sizes.

*The Fossil Record of the Milky Way*

ONE of the most beautiful sights in nature is the band of the Milky Way—the midplane of our galaxy—stretched across the dark night sky (**Figure 3.5**). This integrated light, produced from billions of individual stars, is a constant reminder of our place in the cosmos. While we know today that our Milky Way galaxy is a large spiral system like many others in the universe, how much of our galaxy have we truly seen? Since the first large telescopes came online, studies of the Milky Way have mostly included shallow wide-field surveys or deep narrow-field ("pencil beam") probes. A recent example of the former includes the beautiful starcount maps produced by the Sloan Digital Sky Survey (SDSS), which illustrates the changing density of stars across our galaxy, while the latter include, deep-field HST observations of individual Milky Way populations—star clusters, star-forming regions, and other components (**Figure 3.6**). Taken together, these measurements have defined the ensemble structure of our galaxy as a spiral system with a flattened rotating disk and central bulge, all surrounded by a vast and diffuse halo that stretches out hundreds of thousands of light-years. Whereas the Milky Way disk contains a mixed stellar population with both young and intermediate-aged stars—all of which are "second generation"—the Milky Way halo contains stars that are truly ancient.

In the near future, our view of the Milky Way will be transformed by a new generation of state-of-the-art telescopes. Following in the footsteps of SDSS, the Large Synoptic Survey Telescope (LSST) will provide astronomy with its deepest look at the transient sky, mapping 18,000 square degrees every few nights across the entire visible wavelength range in six filters. The imaging depth of LSST will be more than 100 times as sensitive as SDSS and at higher resolution. Complementing LSST, WFIRST-AFTA will overlap the same footprint but will also extend the spectral range of surveys to near-infrared wavelengths at much higher spatial resolution. These new tools will provide an unprecedented census of the Milky Way's structure and detailed stellar populations, and likely will produce a wealth of new discoveries requiring follow-up studies. As one example, LSST will reveal the stellar graveyard of the Milky Way by uncovering 50 million white dwarf stars, the burnt-out cinders of stellar evolution for the earliest generations of Milky Way stars.

One of the most exciting discoveries from the SDSS starcount maps is the detection of coherent streams in the Milky Way's halo. These streams are predicted from simulations of galaxy formation; they represent past accretion events that led to the buildup of mass in the Milky Way. The streams are, therefore, fossils that can be used to understand the archaeology of our galaxy. Their proximity allows us to study our galaxy's assembly in exquisite detail.

Unfortunately, current technology limits our discovery of these systems to just the brightest or nearest streams, and restricts detailed characterization to the most luminous few stars in each one. The huge gains in sensitivity and resolution afforded by LSST and WFIRST-AFTA will increase the contrast of these streams relative to foreground Milky Way stars and background unresolved galaxies, allowing us to establish a complete map of streams out to the furthest realms of the halo.

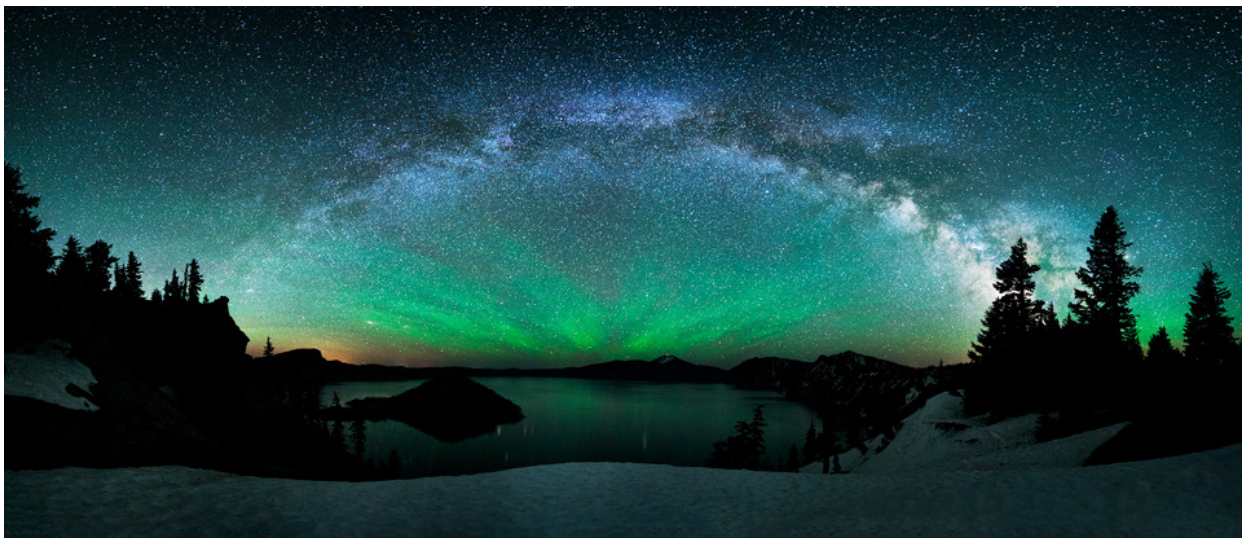

*Figure 3.5* *A beautiful view of the Milky Way, the midplane of our galaxy, over Crater Lake in Oregon.*
***Credit: J.H. Moore***

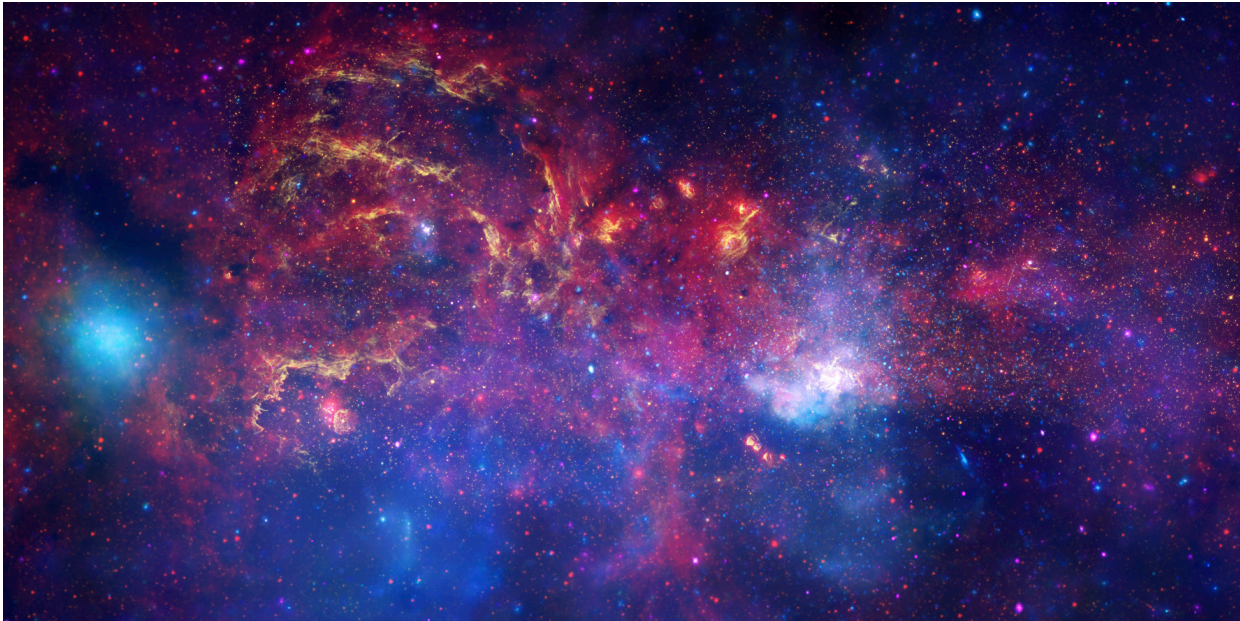

***Figure 3.6*** *A composite of the center of our galaxy from images captured by HST, Spitzer, and Chandra.*
***Credit: NASA***

To unravel the formation process of the Milky Way halo, we must dissect these streams and place their properties within a cosmological context. Not only are LSST and WFIRST-AFTA the discovery vehicles for this exciting science, they will also yield preliminary insights on the global structure and extent of each stream, as well as the colors of their stars, providing a hint at their composition. But a more detailed analysis will be needed in order to use these new discoveries as testbeds for simulations of galaxy formation. We will need to characterize the chemistry, stellar mass, dark matter mass, age, and orbit of each stream around the Milky Way to decipher its role in the overall assembly history. Achieving this will require three new studies:

First, the high-precision astrometry of individual stars from LSST and WFIRST-AFTA will feed the next generation of 30-m ground-based telescopes for detailed spectroscopic analyses. With multi-object spectrographs, these telescopes will establish the velocities and chemical compositions of stars in each stream. The velocities can be used to measure the line-of-sight dispersion and constrain the mass of the stream's progenitor galaxy stream.

Second, the characterization of stellar populations will require the LUVOIR Surveyor, which will enable high-precision color-magnitude diagrams in the Formative Era. Currently, the state-of-the-art instrument for such work is HST, which has a 2.4-m mirror. Although JWST will extend the capabilities of HST in the near IR, achieving sub-billion-year age resolution for all Milky Way substructures will require an even larger telescope to disentangle the most age-sensitive features of composite stellar populations (e.g., the old main-sequence turnoff of hydrogen-burning stars and the white dwarf cooling sequence). This can only be done over a panchromatic baseline that stretches from the UV to IR.

Third, the reconstruction of a stream's dynamical history requires establishing full 3D space velocities of its stellar content. The radial velocity measurements from ground-based 30-m telescopes must, therefore, be coupled with measurements of tangential velocities derived from proper motion studies from space. With 3D velocities in hand, anisotropies in the orbits of these streams around the Milky Way will be measured directly, and the build up of the Milky Way halo will be seen in the form of a "movie" of the accretion events. Proper motions of very distant stars, over wide fields of view, will require a large space telescope operating at high spatial resolution (better than 10 milliarcseconds).

Taken together, a future objective of NASA's astrophysics program will be to fully measure the archaeology of the Milky Way and to chart its evolution over billions of years to its present form. We will first establish a complete inventory of the galaxy's past and active accretion events through sensitive starcount maps across a wide expanse of the halo, and then interpret the detailed nature of these streams through high-resolution photometric, astrometric, and kinematic measurements. This knowledge of halo substructures and their chemistries, velocities, masses, and ages will compose the state-of-the-art data set needed for the next generation of galaxy evolution simulations.

In addition to this characterization of relic accretion events, future tools will enable discovery and age-dating of the ashes of the first generation of Milky Way stars. These objects, white dwarfs and neutron stars, are scattered throughout our galaxy but are typically too faint to be seen. For the cooler white dwarfs, the LUVOIR Surveyors's sensitivity will enable discovery of these remnants out to the edge of the Milky Way galaxy and will also allow complete "cooling sequences" to be characterized in all stellar populations out to tens of thousands of parsecs. The first neutron star cooling sequences may be seen in the nearest stellar populations with LUVOIR, and hotter white dwarfs, neutron stars, and black holes will be explored in detail with the X-ray Surveyor. As these remnants have no nuclear energy sources, their present-day luminosities serve as precise clocks that can be used to date the stars.

Additional insight into the graveyard of the Milky Way will come from observations outside of the electromagnetic spectrum. The Gravitational Wave Surveyor will be able to detect tens of thousands of compact stellar remnant binaries comprised of white dwarfs, neutron stars and black holes, using an entirely new technique. The most numerous sources will be double white dwarfs, which are one of the candidate progenitors of type Ia supernovae and related peculiar supernovae. Detailed knowledge of this ultra-compact binary population opens a new frontier in the study of how stars evolve over cosmic time.

The Milky Way stellar halo is an ideal hunting ground for measuring accretion processes due to the long timescale of orbits at large distances, but it is just one component. Still today, there are major uncertainties in our knowledge of the properties and assembly history of the galactic disk and bulge, and of the formation and dissolution of star clusters and dwarf galaxies. Unlike the halo, all of these populations are seen by ground-based telescopes as "crowded environments," leading to compromised photometric and astrometric precision. Space telescopes have proven to be the best tools to get around this problem because they operate beyond the blurring effect of Earth's atmosphere, have spatial resolutions only limited by the aperture size, and have extremely stable imaging properties.

In the Formative Era, the high spatial resolution of the LUVOIR Surveyor will advance the characterization of these stellar populations to new heights, especially when combined with near-future constraints from the soon-to-launch Gaia mission, which will achieve high-precision distances to Milky Way stars. Taken together, by the end of the Formative Era, we will characterize the stellar initial mass function over all mass scales and in different environments, measure the formation and dissolution of star clusters (where all stars are born), and build a movie of how the galactic disk rotates and how spiral arms work. The detailed spectra of star clusters and individual stars in the disk will tell us the chemical composition of the gas from which they formed over the history of our galaxy. For the disk, this gas has been enriched with heavy chemical elements produced by earlier stars that exploded as supernovae and returned their yield of synthesized elements back into the gas reservoir of the galaxy. For the central bulge of the galaxy, we will test theories of spheroid formation through precise age-metallicity relations and their radial trends, and also study proper motions within the innermost 100 parsecs.

These measurements will establish the physical properties and evolution of all of the Milky Way components and their stellar populations, and place them into an overall context of our galaxy's formation and assembly history.

*Archaeology of Our Galaxy's Neighbors*

A comprehensive study of the Milky Way facilitates major advances in our understanding of galaxy assembly, yet it is still only an example of one kind of galaxy. HST has been able to resolve the brightest stellar populations in nearby galaxies, but only for the nearest large galaxy, Andromeda, can HST reach the depth of the main sequence turnoff that tells the whole story of star formation, evolution, and galaxy assembly. JWST will provide infrared color-magnitude diagrams, which offer a better diagnostic tool than the visible-light diagrams traditionally used. However, only one of the next-nearest groups—the Sculptor Group, analogous to our own galactic neighborhood—is within the reach of JWST. To go farther we need the LUVOIR Surveyor—with a mirror several times that of the 6.5-m aperture of JWST—to extend the reach of such studies to tens of millions of light-years in the Formative Era. This volume contains dozens of major galaxies (**Figure 3.7**), including some types that are very different from the Milky Way, a disk galaxy with several spiral arms. Half of the stars in the universe live in elliptical galaxies (large spheroids with no disk at all) or in spheroidal stellar populations ("bulges") embedded inside disk galaxies. To complement such a grand space telescope, giant ground-based 30-m-aperture telescopes, now under development, will help us separate different populations by measuring stellar velocities and chemistry within each galaxy, parallel to the work on the Milky Way's streams.

The hierarchical model of structure formation—in which large systems grow by merging with others of comparable size and by accreting small satellite galaxies—has become the paradigm for how galaxies grew to their present size, composition, and diversity. By nature, such a model is one of great complexity, where the histories of star formation in individual components are blended. Much of this merging and accretion would have happened in the distant past, and so, from the perspective of the present day, it is challenging to assess the role that such events would have played in galaxy evolution.

Fortunately, for nearby galaxies, and especially our own Milky Way as discussed earlier, some effects of mergers and accretion events can be tracked in detail. Small satellites and very faint streams have now been found through exquisitely detailed starcount maps of the Andromeda galaxy's halo. However, it has taken very large imaging campaigns with HST to actually measure the star-formation history of these streams, and even those studies are confined to tiny regions of the halo due to the small field of view of HST's instruments. In the near future, WFIRST-AFTA will make comparably deep images with ease and at the same spatial resolution as HST, but with a field of view 100 times larger. With this capability, the accretion histories of dozens of nearby galaxies can be studied, including morphological galaxy types very different from the Milky Way and Andromeda. Moreover, we will obtain unprecedented deep and wide images in regions with a high density of galaxies, like the Virgo cluster, where shallower observations have already shown broad streams and plumes, clear evidence of significant encounters between galaxies in this crowded environment.

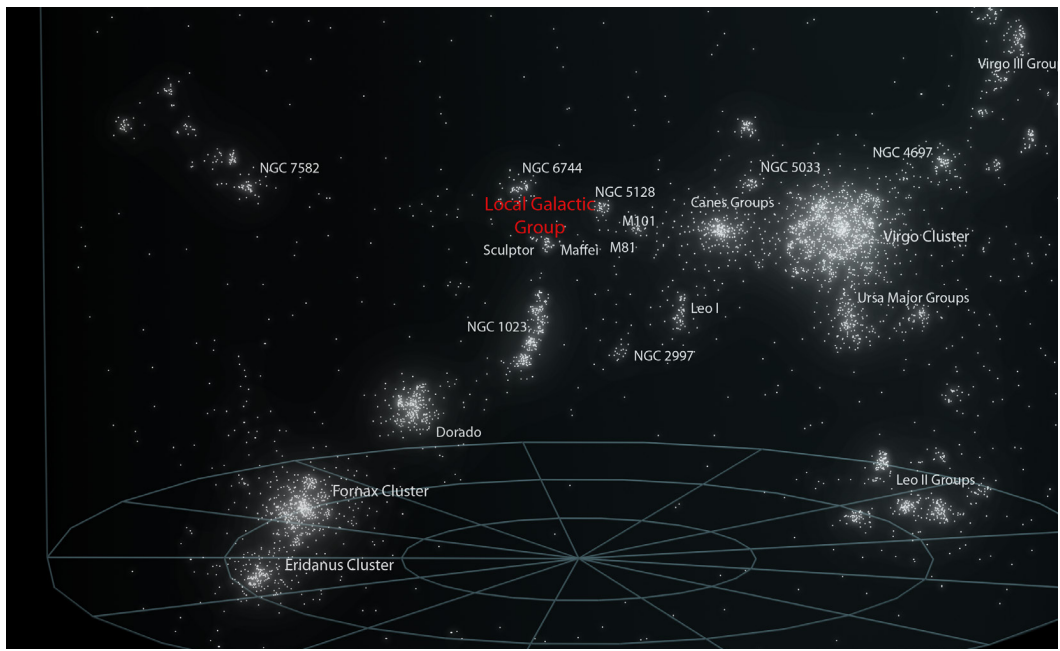

*Figure 3.7* A map of galaxies near the Milky Way, including the Virgo cluster. **Credit: A.Z. Colvin, Wikimedia**

As described above, the identification of such structures is a first step toward analyzing their roles in building the structural elements of the mature galaxies we see today. In the Near-Term and Formative Eras, after the wide-field images from the WFIRST-AFTA have shown us where to look, we will make color-magnitude diagrams from JWST and its successor, the LUVOIR Surveyor, and use ground-based 30-m telescopes to measure the velocities of stars in these structures. Together, these observations will reveal the overall star formation and chemical enrichment histories of a wide range of galaxies and galaxy types, directly reflecting similarities and differences in their evolution.

By the end of the Formative Era, we will have assembled a vast database of color-magnitude diagrams and velocities for representative stellar populations in all the major galaxy types, as well as a comprehensive set of observations of unresolved stellar populations in galaxies over the complete history of the universe. Together with advanced theoretical modeling of galaxy-construction processes, we will reach our goal of describing in detail how our galaxy—and all galaxies—came to be.

## 3.3 The History of Galaxies

> Our solar system formed roughly 4.6 billion years ago in the Milky Way galaxy, about 25,000 light-years from its center. The Milky Way is of average size as spiral galaxies go, and it resides within a small assemblage of galaxies known as the Local Group. Did life appear on Earth because we were just lucky enough to be in a "Goldilocks" zone that had formed after the right sequence of events in the Milky Way's history? And how does this history compare to those of other galaxies? How abundant or rare is life?
>
> Remarkably, the whole of cosmic history is laid out for our inspection. By looking at objects that are far away, we are looking back in time and seeing them as they were in the past. In a way, nature has provided us with a "time viewer" that enables us to address the age-old question, "How did we get here?"

### *Monsters in the Middle: Supermassive Black Holes*

Over the last 30 years, black holes have moved from the realm of science fiction to that of science fact. We now know that supermassive black holes (SMBHs), with millions to billions of times the mass of our Sun, lurk in the centers of most galaxies, including our Milky Way. We also observe that black holes are messy eaters—as they gorge on gas and stars within a galaxy, more material seems to spew out than go in. This outflow of superheated material away from a black hole, generally called "feedback," seems to play a big part in shaping how galaxies grow and change over time and likely played a significant role in the early shaping of the universe and the first galaxies. Most models for galaxy formation and evolution suggest a symbiotic relationship between a galaxy and its central SMBH (**Figure 3.8**). Understanding the details of this relationship is, and will continue to be, a subject of interest to astrophysicists. Furthermore, it may have a profound effect on the locations within a galaxy where life is able to form and evolve.

Both the feeding and the feedback of the central black hole are integral to our understanding of a galaxy's history. While our black hole at the center of the Milky Way appears for the moment to be a sleeping giant, there is tantalizing evidence that it was not always so in the past. The Fermi Gamma-ray Space Telescope has recently found large bubbles of high-energy gamma rays expanding outward from the galactic center, likely a relic signature of the black hole's most recent "meal." However, the current quiet state of our black hole means that the feeding and feedback processes have only been studied in other galaxies. Current technological limitations restrict those studies to a small handful of relatively nearby galaxies with actively feeding black holes.

But to fully understand how black holes grow and influence their galaxies, we need to study this phenomenon in all types of galaxies, from those in which the black hole is barely feeding to those where it is feeding at a maximum rate. We also need to study black holes much farther from us, so we can study them as they were at earlier times in the universe, when galaxies and black holes were young. It is already understood that black hole feedback is likely to be a multifaceted phenomenon, mediated by fast jets of radio-emitting plasma in some circumstances and radiation plus wide-angle winds in others. Correspondingly, a full picture of black hole growth and feedback needs sensitive observations across the electromagnetic and gravitational wave spectrum, as well as sophisticated computer simulations of the coupled galaxy/black-hole growth that can be used to connect physical models to observations (**Figure 3.10**).

Black holes seem very mysterious, but in some ways they are actually very simple objects that can be fully described with two numbers—their mass and their spin (or rate of rotation). Masses are the sum total of their initial formation mass and the amount of matter the black hole swallowed since it was formed. Their spins reveal how matter was swallowed—black hole growth via the addition of small, randomly oriented packets of gas produces a slowly spinning black hole, whereas growth from a large accretion disk produces rapid spin. By extension, measuring the distributions of masses and spins for the whole population of SMBHs as a function of cosmic time will be a powerful probe of the life story of these exotic beasts.

The gravitational force of a black hole on the stars and gas around it yields a kinematic signature that has allowed astronomers to measure masses for more than 100 black holes. But the relatively small mirror of HST and the blurring effects of Earth's atmosphere for larger ground-based telescopes have limited SMBH mass measurements to galaxies close to the Milky Way.

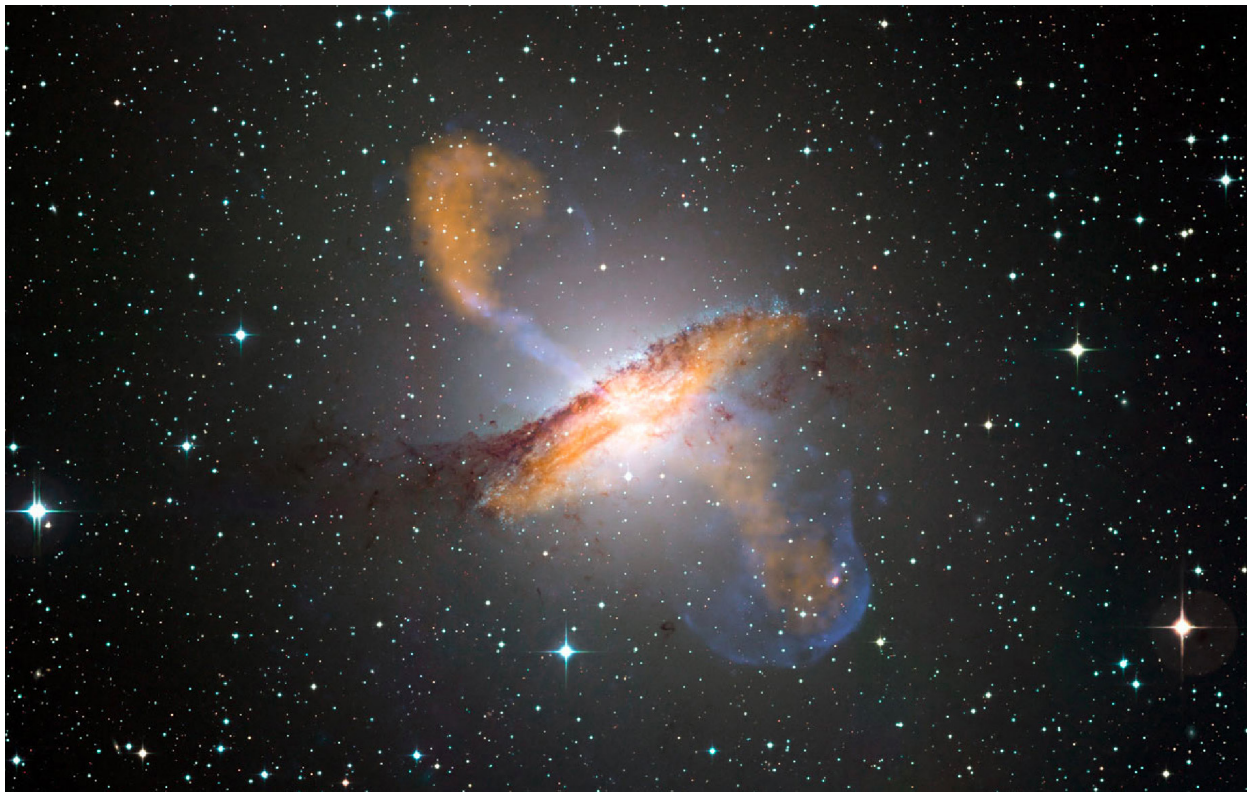

*Figure 3.8* This multiwavelength composite image of the galaxy Centaurus A shows huge jets spewing material, driven by the supermassive black hole in the galaxy's center. **Credit: ESO/WFI (visible); MPIfR/ESO/APEX/A. Weiss et al. (microwave); NASA/CXC/CfA/R. Kraft et al. (X-ray)**

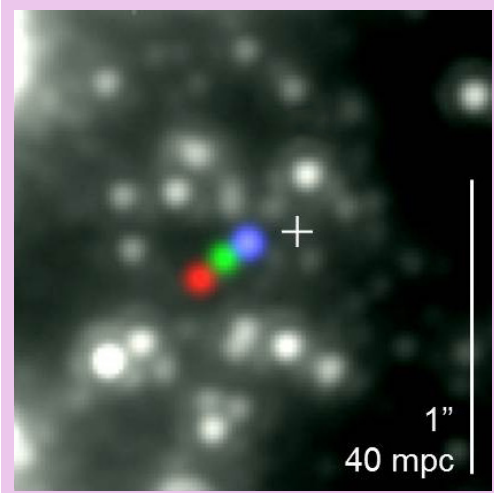

Feedback is usually a process that must be studied in distant galaxies. However, we will be given a rare glimpse of it in the center of the Milky Way over the next few years. A cloud of gas known as G2 has been tracked moving toward the Milky Way's central black hole (**Figure 3.9**), and in late 2013 or early 2014, we will witness the interaction in unprecedented detail. We will watch what happens when G2 gets too close to the monster in the middle—an event that actually happened over 25,000 years ago but that we are just now able to see.

*Figure 3.9 This composite image shows the positions of the gas cloud G2 in 2002 (red), 2007 (green), and 2011 (blue). The cross indicates the position of the black hole in our galaxy's center.* **Credit: MPE**

Measuring spin is even more challenging because it can be determined only very close to the event horizon, a scale that is far too small to be imaged directly with current technology. To make progress today, we must search for the signature of spin in the X-ray spectra of superheated gas close to actively feeding black holes. To date, there are tentative spin measurements for only about 20 black holes with the most recent measurement achieved in 2013 with the combined forces of NASA's NuSTAR and the European Space Agency's XMM-Newton missions. This situation is set to change dramatically with future developments in X-ray and gravitational wave astronomy.

Both mass and spin provide important clues to the past feeding of black holes, as well as to the "seeds" from which they began growing when the universe was very young. Supermassive black holes in average galaxies like the Milky Way tend to have masses 1–10 million times the Sun's. We currently do not know how these black holes initially formed nor how they gained the majority of their mass. A number of theories propose that they started out with approximately 100 solar masses and that their dominant growth occurred through accretion of gas and other black holes during galaxy mergers. Using clues from the galaxy itself or from the superheated gas around a black hole, it is possible to obtain a crude estimate of the monster's mass in most galaxies. When we estimate masses for the most distant sources feeding at the highest rates, active galaxies called quasars, we find that these black holes tip the scales at a billion solar masses when the universe was only 800 million years old. How did they get so massive in such a short amount of time? Other theories suggest that black hole "seeds" were much larger, perhaps 100,000 solar masses, and they are generally better able to account for the rapid growth of quasar black holes. The Gravitational Wave Surveyor will have the sensitivity to detect mergers of supermassive black holes throughout the observable universe, and it will measure the masses and spins of these merging black holes with unprecedented precision.

The hunt for and study of black holes is a multiwavelength and multimessenger endeavor. In the Near-Term Era, ALMA and JWST will provide a major step forward in our understanding of these issues, enabling a hunt for the first seed black holes as well as careful measurements of their masses in a larger range of nearby galaxies than is currently possible. Within the Formative Era, the LUVOIR Surveyor (coupled with supporting radio/submm data from the ground) will push these mass measurements to more distant galaxies, as well as provide unprecedented details on the physics of feedback. The X-ray Surveyor and the Black Hole Mapper in the Formative and Visionary Eras, respectively, will measure the spin distribution of black holes in the local universe and expose the inner workings of quasars and the feeding of SMBHs. Furthermore, moving beyond the electromagnetic spectrum, the future space-based Gravitational Wave Surveyor and Mapper observatories in the Formative and Visionary Eras, respectively, will be able to record the mergers of massive black holes in galactic collisions out to the edge of the universe, and determine their mass and spin back to the epoch of when the seed objects first formed.

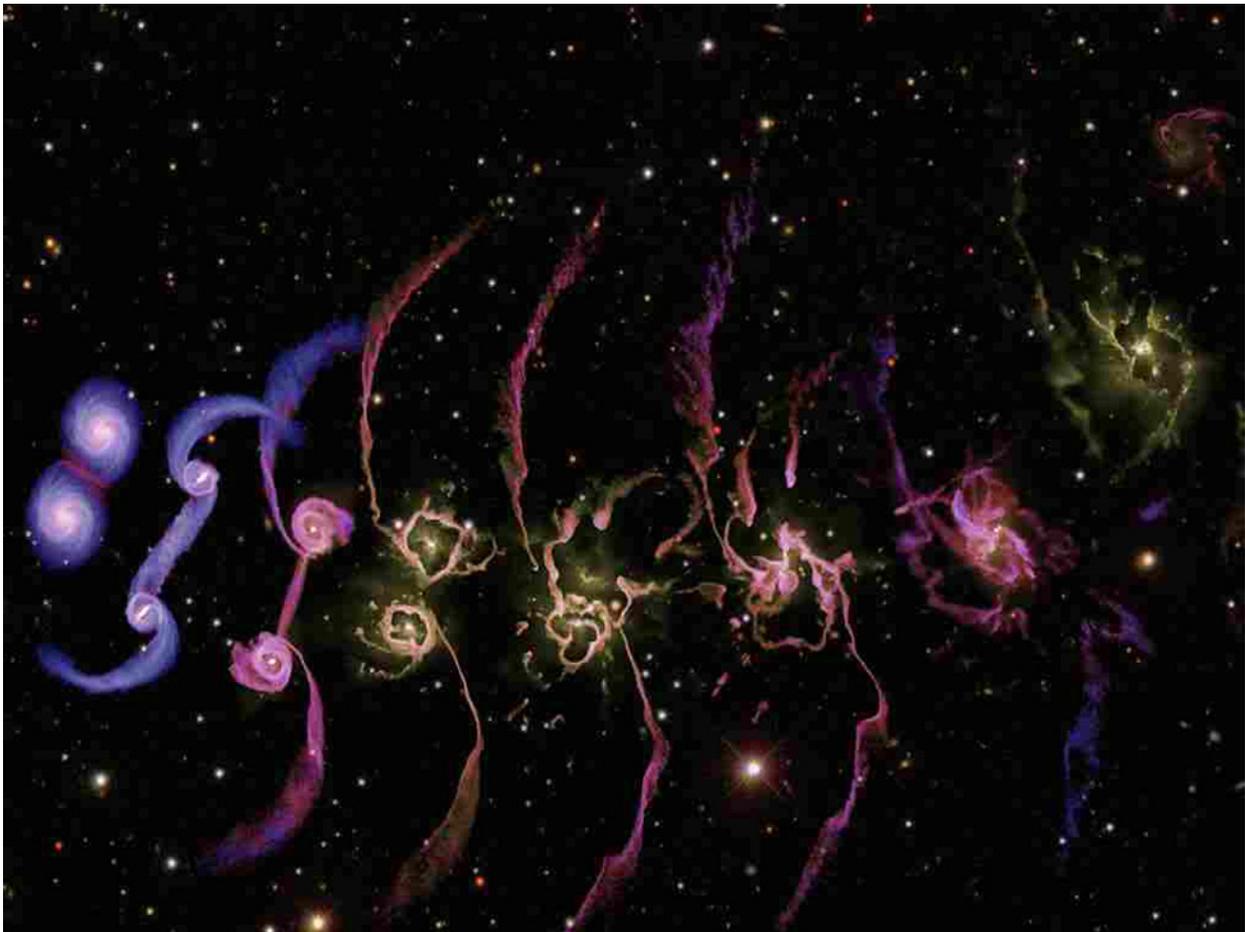

***Figure 3.10*** *Snapshots from the simulated merger of two Milky Way-like galaxies. When gas funnels toward the central supermassive black hole of each galaxy, it initiates a strong feedback that shapes later stages of the merger.* **Credit: V. Springel and T. Di Matteo (Max Planck Institution for Astrophysics) and Lars Hernquist (Harvard Univ.)**

## The Manufacturing and Assembly of Galaxies

Ultra-deep HST observations have revealed a handful of high-redshift galaxies that apparently formed only a few hundred million years after the Big Bang (**Figure 3.11**). Small fluctuations in density in the very early universe grew into larger and larger gravitational potential wells—the likely sites of the first stars and galaxies. In these wells, gaseous matter decoupled from dark matter and sank to the center, condensing into the first stars. Deeper wells may have contained many such condensations, thus leading to the formation of the first galaxies. These were likely simple systems made of immense clouds of hydrogen and trace amounts of helium and lithium. The first baby galaxies likely transformed rapidly accreting additional gas and dark matter from the cosmic web—the filamentary structure that dominates our universe. Although we think this is how the first stars and galaxies were created, we have very little information about the physics and chemistry of these early times. And yet, incredibly, we stand at the threshold of being able to create movies and portraits of the infant universe. Within the next 10 to 20 years, we will directly capture light from the very first stars and the first clouds of gas and dust with ALMA and JWST. Future missions, such as the LUVOIR and Far-IR Surveyors, will

> **We stand at the threshold of being able to create movies and portraits of the infant universe.**

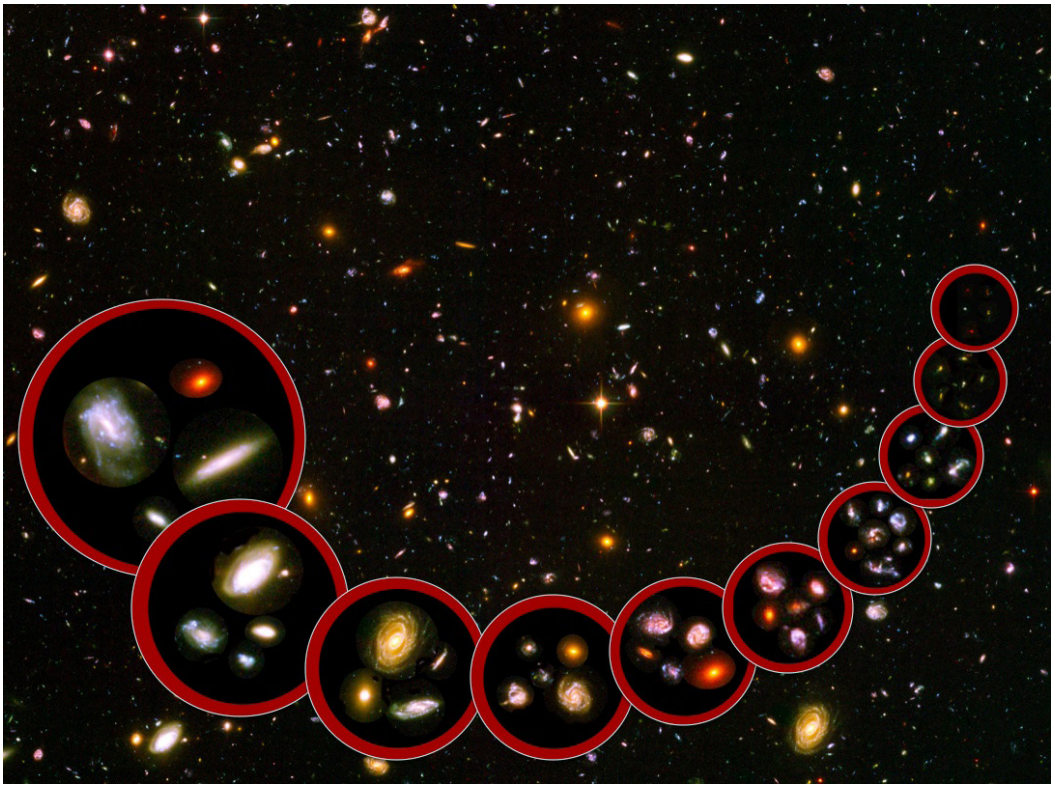

*Figure 3.11* *The panels show galaxies selected from the Hubble Ultra Deep Field at "lookback times" (left to right) of 1, 3, 5, 6, 8, 9, 10, 11, 12, and 13 billion years, compared to the universe's full age of 13.8 billion years. Galaxies evolve strongly in size and shape, growing both from the inside through star formation and from outside by mergers of smaller galaxies. The smooth symmetry of today's galaxies arose only in the last 8 billion years (five leftmost images). Earlier panels show galaxies in full assembly, chaotic in form, with stellar birthrates 10 to 100 times higher than today's galaxies.* **Credit: NASA/A. Dressler (Carnegie Institution of Science)**

provide critical information on the chemical and dynamical evolution of these sources. In the Visionary Era, we will understand the detailed physics behind the separation of dark matter from baryonic matter, the process that led to the gravitational collapse of gas, formation of the first stars, and the initial rapid growth of the first galaxies.

Between 3 and 7 billion years after the Big Bang, the universe underwent an incredible growth spurt, forming more than half of all of the stars we see today. The rate at which the universe formed stars during this period was at least 20 times higher than it is today. SMBHs at the centers of galaxies also grew, and galaxies continued to interact and merge with other galaxies. What caused this growth spurt? Was it influenced by internal mechanisms or was it dominated by interactions between galaxies? In the next two decades existing and planned facilities will allow us to map the detailed mass assembly history of galaxies. We will be able to determine how fast the gas reservoirs in galaxies were consumed and measure how and why star formation varied, both from galaxy to galaxy and within galaxies. Since much of the star formation activity occurs deep inside dense clouds of molecular gas and dust, studying galaxies at far-infrared wavelengths, where the majority of their photons are emitted, will be extremely important. To peer into young star-forming regions on scales of a few thousand light-years will require high-spatial-resolution observations from a space-based far-infrared interferometer.

Compared to their energetic first years and tumultuous middle years, the last 7 billion years of cosmic time has been relatively quiet for most galaxies. During this most recent era, the number of mergers and interactions has steadily decreased and galaxies have gradually taken on the shapes we see today (**Figure 3.12**). But these changes were not uniform across the galaxy population. Studies show that the most massive galaxies changed and acquired their present day forms first, whereas the lowest mass galaxies matured last. Why did they evolve this way? Some combination of their gas fraction, dark matter halo properties, and environment likely played a role. Current investigations of these properties, e.g., the gas fraction, are extremely limited and based on aggregating single background quasar sightlines passing through a foreground galaxy's halo. Future high-resolution spectroscopic observations by the LUVOIR Surveyor will provide capabilities to study sightlines that are 10 times fainter, and therefore to map the physical properties, kinematics, and distribution of gas throughout a large number of galaxy halos.

During this epoch of rapid growth, when galaxies grew from youngsters to adults, the "cosmic web"—an intricate structure of dark matter and gas—emerged as well (**Figure 3.13**). We describe the peak of cosmic hierarchy as "large-scale structure." Broad, high-density plateaus grew with accompanying superclusters of galaxies, studded with crowded and rich clusters, the most extreme galaxy environments. Mechanisms that quench star formation in galaxies, and processes that can even transform galaxies from spirals to

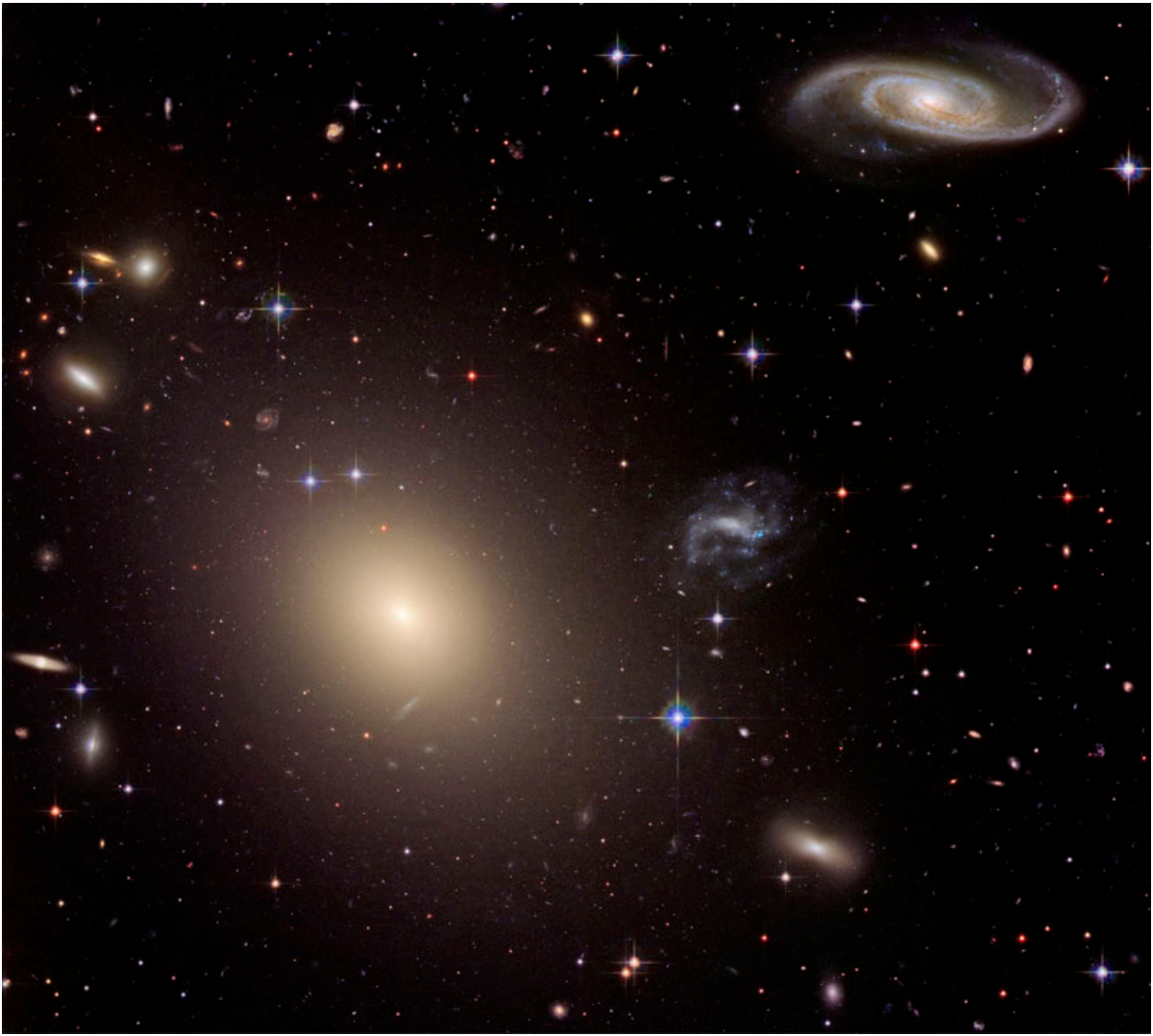

*Figure 3.12 The photogenic cluster Abell S0740 contains galaxies of many different shapes and sizes, illustrating the various galaxy types we can find today: ellipticals—both large and small—spirals, and lenticulars.* **Credit: NASA/ESA/Hubble Heritage Team**

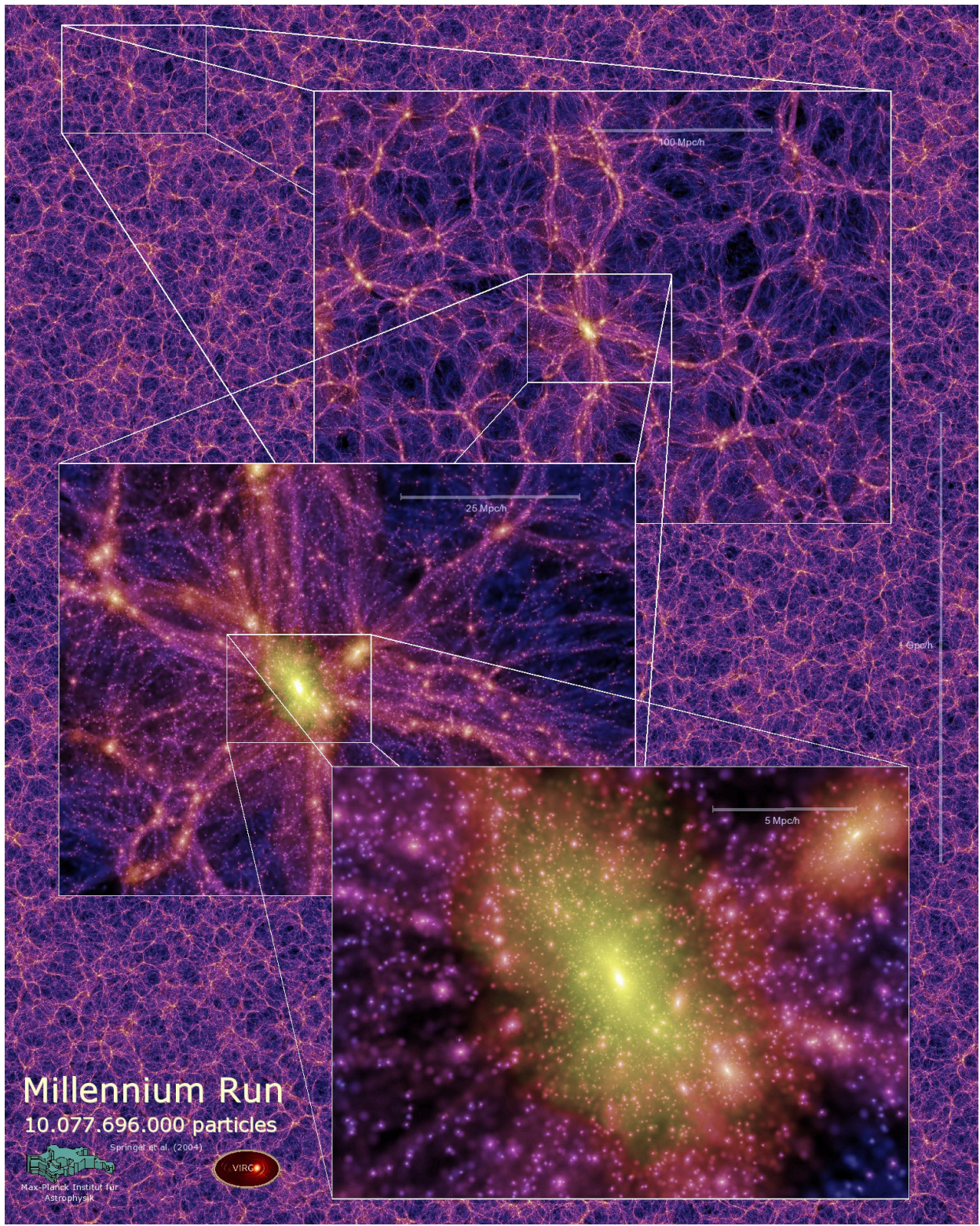

*Figure 3.13* Snapshots from the Millennium Simulation show the growth of large-scale structure as the cosmos expands and cools. The filamentary structure of matter is often referred to as the cosmic web.
*Credit: V. Springel/Millennium Simulation Team/MPA*

> **Citizen Scientists**
>
> In addition to Planet Hunters, mentioned in Chapter 2, **Galaxy Zoo** is also part of the **Zooniverse** project, which provides a platform for citizen scientists to "use the efforts and abilities of volunteers to help researchers deal with the flood of data that confronts them." Citizen scientists use Galaxy Zoo to classify galaxies into different types. Results from the Galaxy Zoo project have been published in astronomy journals, and some unusual and exciting discoveries have been made using this platform. For example, Hanny's Voorwerp (Dutch for "Hanny's object") was found by Hanny van Arkel, a Dutch teacher. This object has been followed up extensively by astrophysicists and is likely a reflection nebula heated by a black hole jet.
>
> 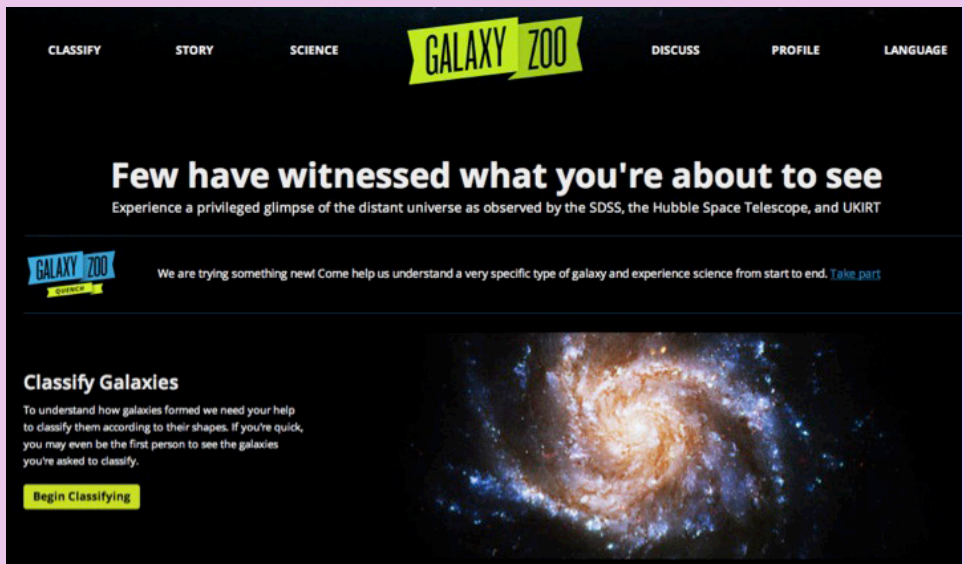
>
> ***Figure 3.14*** *The home page of Galazy Zoo.* **Credit: Zooniverse**

ellipticals, thrive in these groups and clusters. As the superclusters rise, great voids of similar size appear in the cosmos; their few inhabitant galaxies lead isolated lives very different than those in crowded regions. Connecting all these structures are filaments of dark matter that give the cosmic web its name. Only a small fraction of the baryonic matter finds its way into galaxies: most remains attached to the dark matter strands, sheets, and swells of the cosmic web. This is the gas reservoir that galaxies require during their period of maximum growth. In the next decade, the mapping of large-scale structure will be phenomenally advanced by LSST and WFIRST-AFTA. Much of our work in the next several decades will involve characterizing the gas in the cosmic web, mapping its density, temperature, composition, and kinematics, with facilities like the LUVOIR Surveyor and a large-aperture, wide-field X-ray telescope, the X-ray Surveyor. Ultimately, our goal is to have a detailed picture of the intergalactic gas and to understand its connection to circumgalactic gas moving in and out of galaxies as they assemble from the raw materials of the universe.

The combination of JWST, ALMA, VLA and SKA will allow us to directly image light from the first stars, the first clouds of molecular and atomic gas, and dust. Not only will we image the distribution of gas and dust but we will also measure their dynamics—interferometers like ALMA, VLA, and SKA will provide measurements of line widths, velocity dispersion and two-dimensional kinematics so we can measure the possible rotation and stability of these clouds. The next generation of telescopes, including LUVOIR and far-infrared telescopes, are needed to understand the change in the growing stellar endoskeletons and the chemical enrichment within galaxies over cosmic time. With these facilities we will be able to understand precisely what our Milky Way must have been like 4.6 billion years ago, when Earth formed, and we will reconstruct the full assembly history of galaxies.

*First Light and Reionization: The Universe Changes Course*

IMMEDIATELY after the Big Bang, the universe was composed of high-energy elementary particles and light in conditions more extreme than anything found today. As the universe expanded and cooled, massive particles decayed into common elementary particles—protons, neutrons, and electrons—but this cosmic soup remained an "ionized" plasma, much too hot for atoms to form. The universe expanded and cooled until, about 400,000 years after the Big Bang, hydrogen and helium atoms formed. We see this event today as the cosmic microwave background, a relic light of the early universe and a source of copious information on that epoch.

The universe would have forever remained a sea of atoms in the fading glow of primal light were it not for the appearance of stars and massive black holes, both of which flooded space with ultraviolet light energetic enough to strip electrons from their atoms and return the universe to an ionized state. Already, HST has taken us to the very threshold of "first light," allowing us to peek at the brightest protogalaxies in the first billion years. But, to fully explore this critical moment in the history of the universe requires a larger space telescope, one with extraordinary sensitivity to the infrared glow of the distant ancestors of today's galaxies. JWST will provide this next step forward and allow our first peek at the "reionization" process. Furthermore, WFIRST-AFTA will deeply image large areas to find protogalaxies throughout the cosmic dawn epoch, decoding their growth in time despite the patchy fog of neutral gas, which effectively absorbs light from the hot blue stars that dominate early star systems. Spectroscopy of quasars in this epoch, made possible by JWST's extraordinary near-infrared sensitivity, will also reveal the mix of neutral and ionized gas and allow astronomers to follow the process of reionization.

The sea of neutral gas from which all of these protogalaxies and black holes form can be directly imaged at radio wavelengths. In the next 10 to 20 years, ground-based radio telescopes will begin to see the signals and make the first maps of its distribution. Studying this neutral gas back to the time of the first stars—and even beyond, to the period called the "dark ages," when the seeds of the first stars and galaxies were forming—will require sensitive radio telescopes operating at meter and decimeter wavelengths. In 30 years such telescopes may use the moon as a shield from Earth's radio interference to map the reionization process directly through the emission of radio light from the neutral gas. Fully resolving the structure present during the dark ages and the reionization epoch requires an array of radio antennas in this shielded radio-quiet environment, the Cosmic Dawn Mapper of our Visionary Era.

Stars are the key agents that reengineered the universe: they radiate energy into their environments and then alter their environments by synthesizing and distributing heavy elements. Carbon, nitrogen, oxygen, silicon, iron, and other elements are essential to the universe we know, but none of them were created in the Big Bang. Our goal, then, is to chart the first stars and their evolving populations, to see how they individually and collectively changed the structure and composition of the young universe. Stars are the key to increasing complexity in the universe, and as such, are key to the development of life.

The first stars may have been very different from those we know today. They may have been much more massive and destined for blazingly brief lives, but even so, they are likely to have been too faint for us to see individually even with JWST. A star cluster of this first stellar generation—if these stars were born in families—would be detectable with JWST, and the supernova explosion of such a rare star could be bright enough to be discovered with the giant camera of WFIRST-AFTA and then studied in detail with JWST (**Figure 3.15**). Catching such an event and studying it in detail would provide a giant step forward in our understanding of the chemical enrichment that began with these first stars. Finding when and how the first stars were made, determining how massive they were, and learning in what ways they were similar or different to the stars that came later, is a primary goal for JWST. Likewise, images and spectra recording the buildup of galaxy-sized star systems will inform detailed theoretical models of how a hierarchy of structures began to appear within the first billion years of cosmic history. Key details in our understanding of the physical processes at work in these early times will come from higher-resolution spectra provided by the coming generation of 30-m ground-based telescopes, and from unprecedented deep millimeter-wave maps from ALMA.

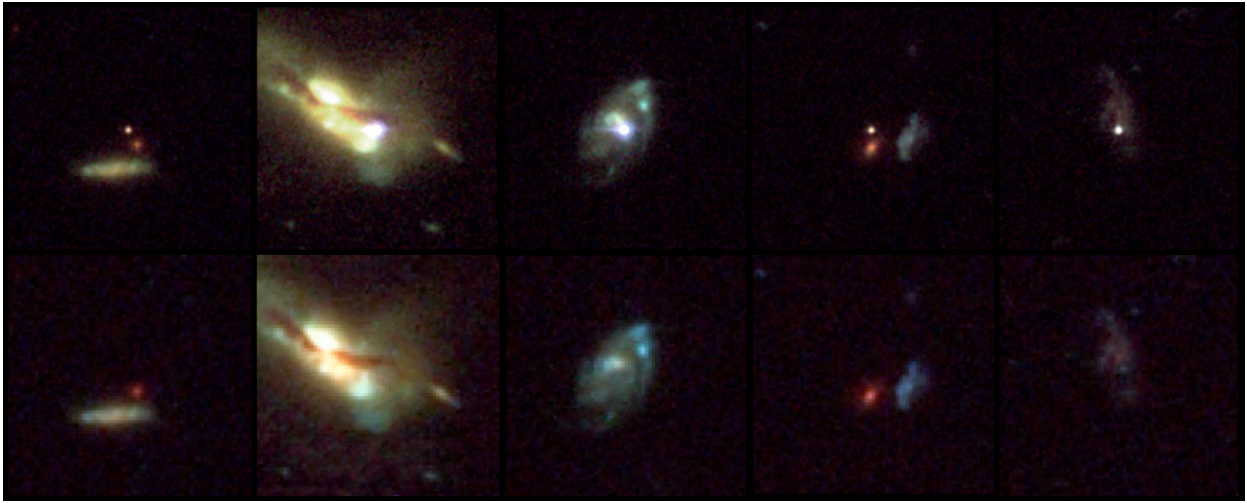

*Figure 3.15* Hubble Space Telescope images of galaxies with supernovae caught in the act of exploding. These galaxies are pictured as they appeared when the universe was roughly half its current age and appear much less organized than galaxies today. Much more powerful telescopes will be required to catch supernova explosions from the universe's first stars. **Credit: NASA/ESA/A. Riess (STScI)**

Over the next five to 15 years, JWST and WFIRST-AFTA will provide crucial information on the era of "first light," but, even with critically needed higher-fidelity theoretical modeling, there is likely no way to predict if JWST will be powerful enough to answer our fundamental questions. Basic uncertainties regarding the nature of the first stars and when they formed, the level to which protogalactic starlight is absorbed by dust and neutral hydrogen, and the dynamic range required to understand the galaxy hierarchy, might result in the need for a much larger near-to-mid-infrared space telescope, one with more than 10 times the light gathering power of JWST. Such a successor to JWST could, in 15 to 30 years, discover some of the first individual stars to form in the universe. Furthermore, as has been shown for present-day galaxies and their immediate predecessors, the ability to study starlight absorbed by dust and re-radiated in the mid-to-far-IR is an essential tool in understanding the life cycle of a galaxy. To repeat such studies for the epoch of galaxy birth requires mid- to far-IR telescopes of much greater size and resolution. In the Formative Era, the Far-IR Surveyor could achieve the kiloparsec-scale spatial resolution required to understand the environments of star formation in early protogalaxies.

While protogalaxies were forming, massive black holes were also growing. It is perhaps not surprising that the most massive black holes we can find today are located in galaxies with the greatest mass of stars, and vice-versa, but present theories are unable to provide a convincing model of how these two different phenomena—the size of a galaxy and the size of its central massive black hole—became linked. In fact, there is some evidence of little connection when galaxies and SMBHs first formed. While there is some evidence that these SMBHs appear in the era of cosmic dawn, there is as yet no adequate theory that explains how or when they first formed.

In the Near-Term Era, JWST and WFIRST-AFTA will be able to chart the appearance over cosmic time of the first SMBHs and determine their contribution to cosmic reionization. However, to fully explore the formation of massive black holes and their growth over the epoch of first light will require reaching to the lesser SMBHs of only millions of solar masses, below the sensitivity of these missions. In the Formative Era, the Gravitational Wave, Far-IR, and LUVOIR Surveyors will explore the full range of this remarkable phenomenon, by finding and studying the first SMBHs over the full mass range found in today's universe.

Our journey to describe and understand the birth and growth of the universe will culminate in the Visionary Era, when we will use revolutionary instruments across the electromagnetic and gravitational wave spectrum to completely map the contents of "selected volumes" of space that represent all of cosmic time.

For each of these epochs, we will describe in detail the distribution of dark matter, gas, and galaxies over a dynamic range of orders-of-magnitude. We will measure the mass, chemical composition, and motions of gas entrained in the cosmic web. We will chart the evolution of the galaxies—their stellar populations, rates of star formation, heavy element production, structure and dynamics, from the birth of the first stars through to today's magnificent spirals—not only in systems like the one we live in, but also in the old, massive ellipticals that stand as great monuments to a young universe of fire.

## 3.4 Activities by Era

### Near-Term Era

- **Stellar Life Cycles and the Evolution of the Elements**

  - *Identify locations of cold dust and gaps in protoplanetary dust disks with ALMA and begin spectral characterization of dust properties in the inner regions with JWST.*

- **Archaeology of the Milky Way and its Neighbors**

  - *Establish a complete inventory of the Milky Way's past, active accretion events, and dwarf galaxy population through sensitive wide-field starcount maps with WFIRST-AFTA and LSST. Identify such fossils in all galaxy types by performing deep surveys of nearby galaxies.*

- **The History of Galaxies**

  - *Hunt for the first seed black holes and measure black hole masses in nearby galaxies with JWST.*

  - *Image light from the first galaxies in the universe with ALMA, JWST, and 30-meter ground-based telescopes.*

### Formative Era

- **Stellar Life Cycles and the Evolution of the Elements**

  - *Map the chemical-dynamical evolution of all nearby planetary systems by resolving the extended structure of protoplanetary and debris disks as a function of age with the LUVOIR Surveyor, and directly detect the distribution and flow of water into the inner regions with the Far-IR Surveyor.*

- **Archaeology of the Milky Way and its Neighbors**

  - *Fully characterize the archaeology of the Milky Way galaxy by establishing sub-Gyr ages (color-magnitude diagrams from the LUVOIR Surveyor) and internal kinematics and space orbits (30-m radial velocities, Gaia distances, and the LUVOIR Surveyor proper motions) for all surviving and disrupted dwarf galaxies. Measure direct star-formation histories with the LUVOIR Surveyor for all galaxy types as we can for Andromeda today.*

  - *Use high-precision astrometry and photometry with the LUVOIR Surveyor to measure the stellar IMF over all mass scales and in different environments of the Milky Way, chart the formation and dissolution of star clusters, build a movie of how the galactic disk rotates and how spiral arms work, and measure the detailed spectra of individual stars in all Milky Way components to tell us the chemical composition of the gas from which they formed.*

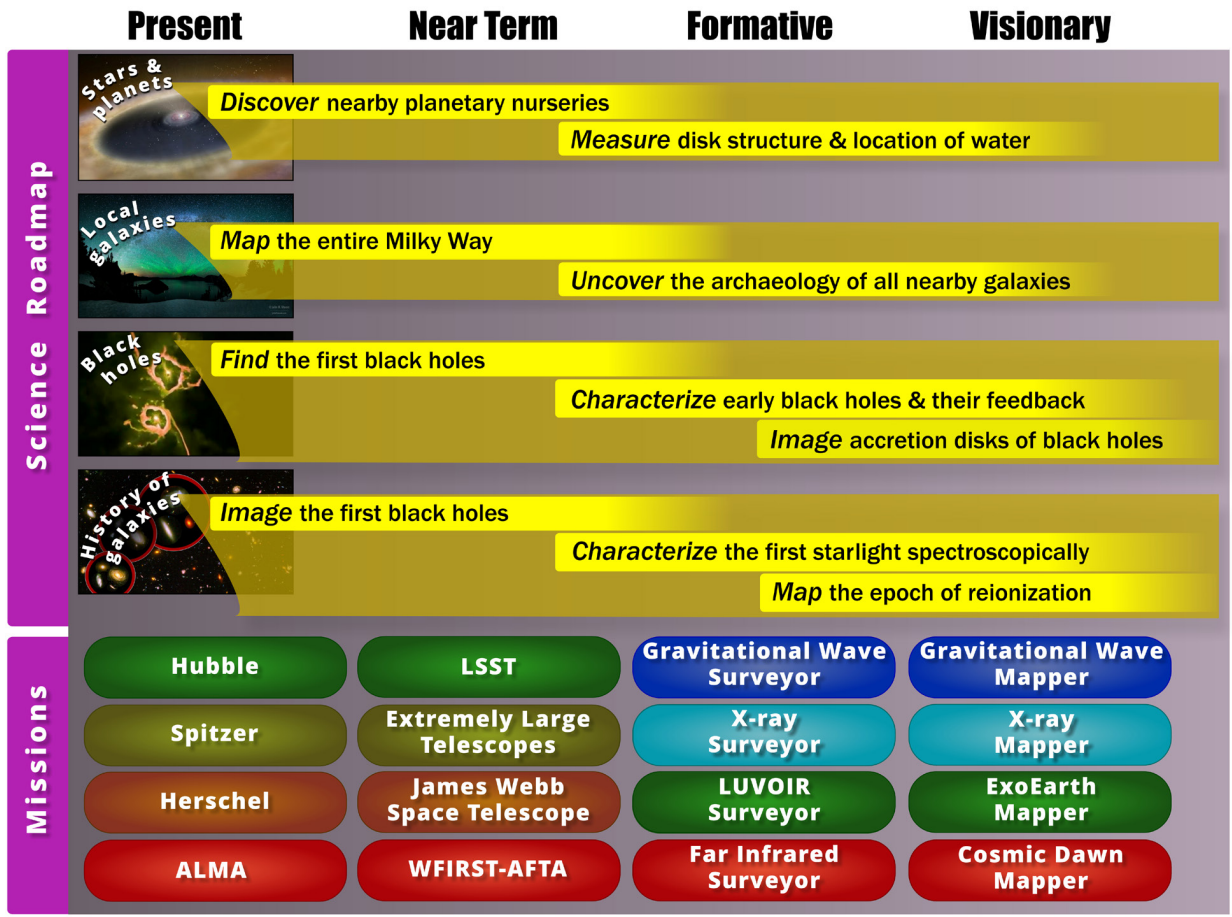

Figure 3.16 Schematic of the Cosmic Origins Roadmap, with science themes along the top and a possible mission sequence across the bottom. Credit: F. Reddy (NASA/GSFC)

- **The History of Galaxies**

    - Characterize black hole masses to the earliest systems and study in detail the physics of feedback and the relation between galaxy growth and black holes with the LUVOIR Surveyor. Measure the spin distribution of black holes in the local universe with the X-ray Surveyor to expose the inner workings of quasars and the feeding of supermassive black holes.

    - Witness and study the sites of violent star formation in early protogalaxies with kiloparsec-scale resolution and directly image forming protoplanetary condensations in new planetary systems with the Far-IR Surveyor.

    - For the first time, record the mergers of massive black holes that accompany galactic collisions and mergers with the Gravitational Wave Surveyor, constrain the growth processes of massive black holes by determining the merger rate, and provide precise measurements of their masses and spins.

    - Measure the properties of the universe's first stars with the LUVOIR Surveyor.

    - Measure the process of reionization in the early universe, through high-spatial-resolution imaging and moderate-to-high-spectral-resolution spectroscopy of the first galaxies and quasars with the LUVOIR Surveyor.

*Visionary Era*
- ***The History of Galaxies***

  - *Completely map the content of selected volumes of space that represent slices through all of cosmic time. Building on the scientific discoveries from the X-ray and LUVOIR Surveyors in the Formative Era, we will describe the dark matter, gas, and galaxies over a dynamic range of orders-of-magnitude and launch a new exploration of the cosmic web with the Black Hole and Gravitational Wave Mappers.*

  - *Measure the spin distribution of black holes across the universe, and quantify the feedback energy released by the first monster black holes at the dawn of the universe with the Black Hole and Gravitational Wave Mappers.*

  - *Witness the growth of black holes by pinpointing the sites of black hole mergers across the universe with the Black Hole and Gravitational Wave Mappers, and directly imaging the accretion disks of nearby feeding black holes with the ExoPlanet Mapper.*

  - *Resolve the structure present during the dark ages and the reionization epoch with the Cosmic Dawn Mapper, an array of radio antennas on the back side of the moon.*

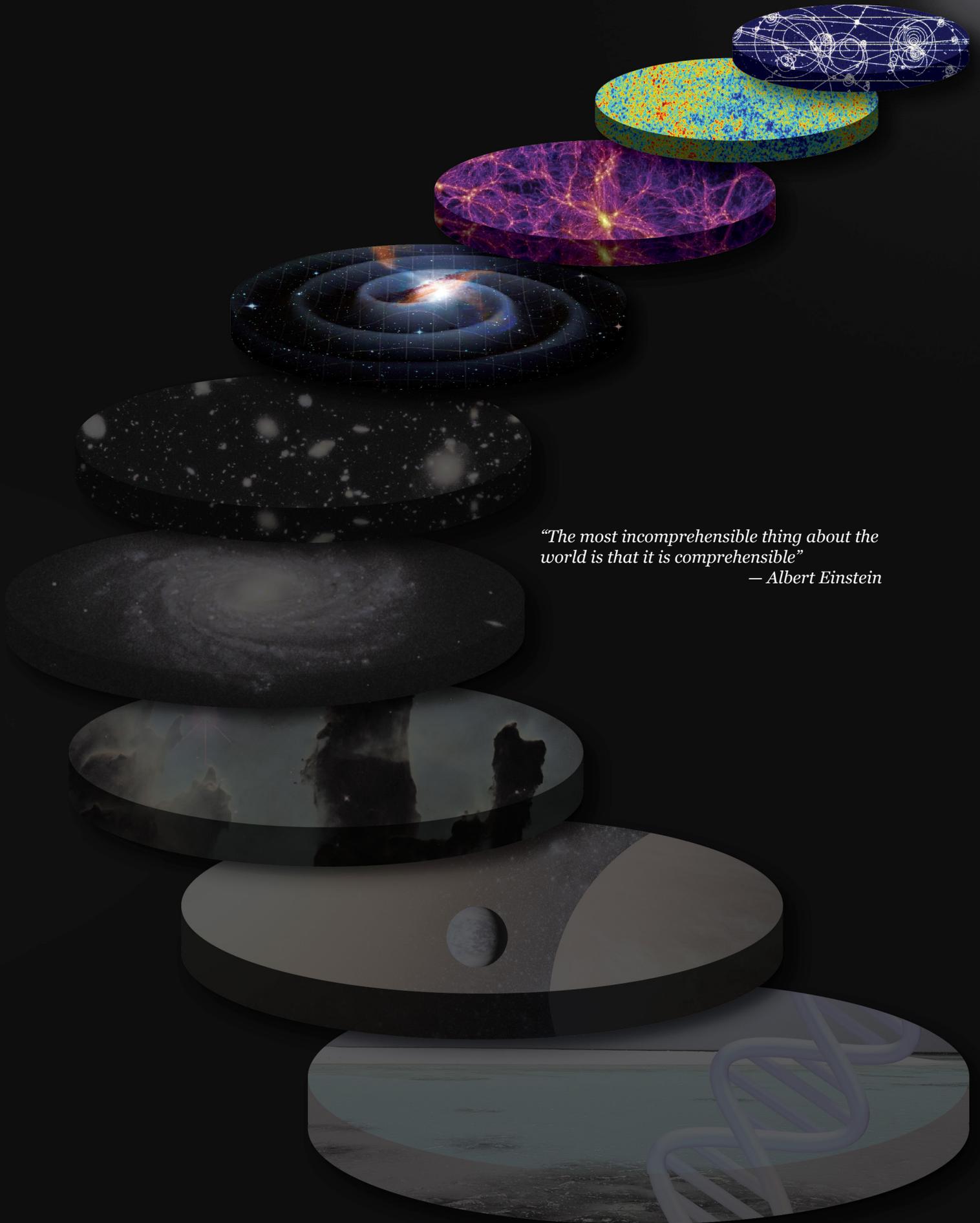

"The most incomprehensible thing about the world is that it is comprehensible"
— Albert Einstein

# 4 How Does Our Universe Work?

Since Newton first connected the drop of an apple with the orbit of the moon to devise a theory of universal gravity, astronomers have sought to understand celestial phenomena by applying principles of fundamental physics and to discover new laws of physics by studying the stars. In recent decades, astrophysicists have learned through astronomical observations that our universe began 13.8 billion years ago in a Big Bang, that the dominant form of matter in the universe is radically different from the protons, neutrons, and electrons that make up the world of everyday experience, and that black holes—objects so exotic that Einstein himself deemed them impossible— are in fact ubiquitous throughout the cosmos.

Today, we seek to probe the first nanoseconds of cosmic history, to solve the extraordinary mystery of accelerating cosmic expansion, to understand the nature and behavior of subatomic particles at the extreme densities of neutron stars, and to map the swirling flows of matter, radiation, and gravity near the event horizons of black holes. In the remnants of exploded stars, the collisions of massive galaxies, and the trillion-degree soup of the early universe, nature creates physical conditions that can never be realized in a terrestrial laboratory. Finally, we are poised to open a completely new window on the universe by "listening" to the rippling waves of space-time emanating from violent gravitational cataclysms.

In the next 30 years we will probe the earliest instants of cosmic history and the properties of dark energy and gravity by mapping the universe with exquisite precision over vast spans of temporal and spatial scales. By measuring the faint signals that reach us from these distant places and ancient times, NASA missions will take us closer to understanding the fundamental laws of nature and may reveal phenomena that lie beyond our current imagination.

## 4.1 The Origin and Fate of the Universe

> Is the universe finite or infinite? How did it begin? How will it end? Modern cosmological observations have provided a powerful empirical basis for addressing these questions, but our answers remain far from complete. The leading hypothesis is that the hot early universe of the Big Bang theory emerged from a still more exotic phase, when the universe grew in size by 100 million billion billion times in less time than it takes light to cross an atomic nucleus. In one of the most counterintuitive scientific discoveries of recent decades, we have learned that the expansion of the universe is speeding up rather than slowing down. The properties of the dark energy thought to drive this accelerating expansion also control the ultimate fate of the universe: whether it will slowly dilute and cool, tear apart in a "big rip" tens of billions of years from now, or collapse again in a fiery cataclysm that could seed a new Big Bang.

### *Inflation and the Cosmic Microwave Background*

The 1965 discovery of the cosmic microwave background (CMB) provided startling evidence for the Big Bang theory and launched the modern era of cosmology. The light of this relic radiation dates from the first half million years of cosmic history, when the universe was hot enough to ionize hydrogen atoms and produce an opaque fog of light-scattering free electrons. The subsequent thousand-fold expansion of the universe has cooled the primordial fireball to a microwave hiss, with a temperature of 2.73 kelvin. The blackbody distribution of CMB photon energies shows that the early universe was in a state of thermal equilibrium, and together with the observed abundances of light elements, it confirms the Big Bang theory's description of cosmic physics back to when the universe was one second old and the temperature was 10 billion degrees. The CMB is almost perfectly isotropic, meaning that its measured temperature in all directions is almost exactly the same. After 25 years of increasingly sensitive searches, NASA's Cosmic

Background Explorer (COBE) achieved the first detection of CMB anisotropies in 1992. These anisotropies reveal the presence of tiny density fluctuations, variations of only a few thousandths of a percent. These are the seeds that subsequently grew into the galaxies, clusters, and superclusters we see in the universe today.

The conventional Big Bang theory accepts, as unexplained initial conditions, the large-scale homogeneity of the cosmos, the existence of primordial fluctuations that seed structure growth, and the enormous size of the observable universe. In the early 1980s the theory of cosmic inflation emerged as a possible explanation for these "just so" facts of the standard Big Bang model. Inflation posits an interval of extremely rapid exponential expansion in the very early universe, perhaps associated with the end of the quantum gravity era or the breaking of a fundamental symmetry between the strong nuclear force and the electroweak force. A tiny patch that had time before inflation to reach a state of thermal equilibrium could grow large enough during inflation to encompass our entire observable universe many times over. At the end of inflation, reheating filled this vastly expanded volume with particles and radiation. Remarkably, the exponential expansion stretched quantum fluctuations from subatomic scales to macroscopic scales, thereby generating the seeds of cosmic structure. All of this, according to the leading models of inflation, occurred in the first trillionth-of-a-trillionth-of-a-trillionth of a second of cosmic history.

Simple models of inflation driven by a primeval energy field predict fluctuations in the CMB that are adiabatic (present equally in all forms of matter and radiation), Gaussian (following a bell curve probability distribution), and nearly but not perfectly scale-invariant (with equal contributions to gravitational potential fluctuations from all scales). These predictions are in beautiful agreement with high-precision measurements of CMB anisotropies from the WMAP and Planck satellites and ground-based experiments (**Figure 4.1**), including the small deviations from scale-invariance that distinguish inflationary predictions from a mathematically generic statistical model. However, while these measurements place the once

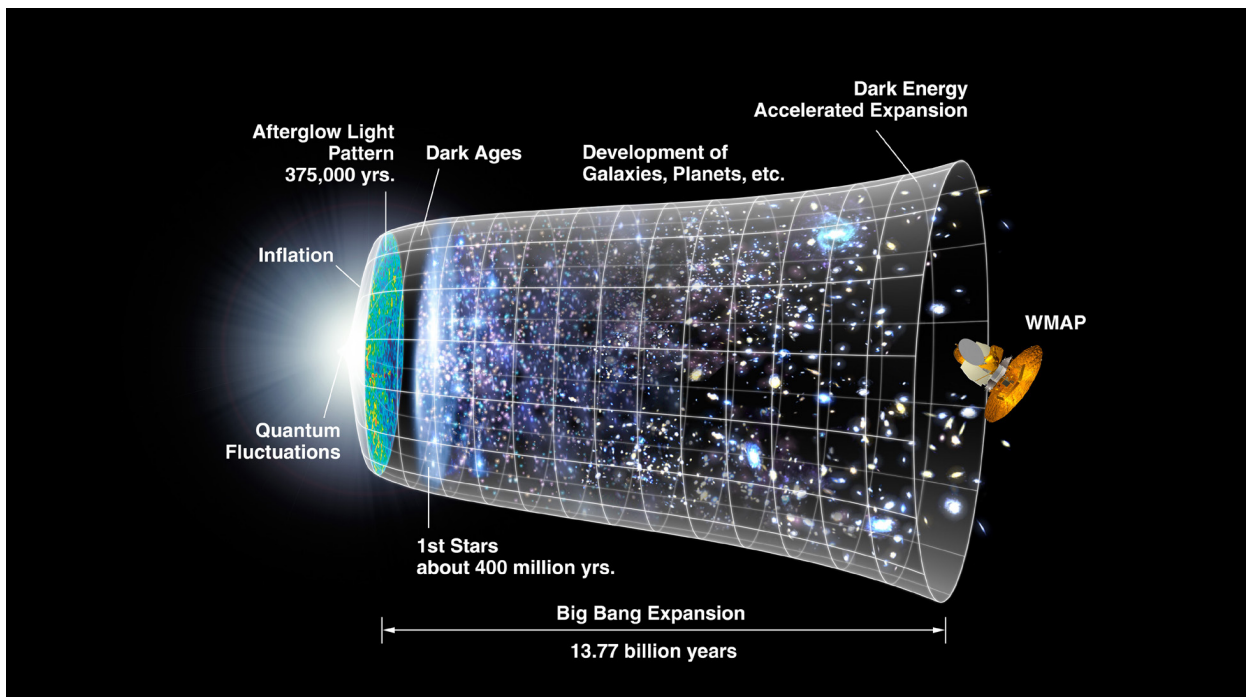

*Figure 4.1* A timeline of our universe. This representation shows the quantum fluctuations (left) that expanded exponentially during the epoch of inflation and became seeds for the growth of cosmic structure. After 375,000 years the universe cooled enough to form neutral hydrogen, leading to the generation of the cosmic microwave background that permeates the universe today. The accelerated expansion of our universe caused by dark energy is a relatively recent event. **Credit: NASA/WMAP Science Team**

speculative theory of inflation on a firm empirical basis, there is still much that we do not understand about how inflation began, the field or fields that drove it, and why it ended. Furthermore, there are competing theories of the very early universe, in which (for example) the initial conditions of the Big Bang emerge not from inflation but from a gas of fundamental strings that oscillate in higher spatial dimensions, or in the contracting phase of a cyclic universe that undergoes repeated periods of expansion and collapse.

In addition to temperature variations, the CMB exhibits fluctuations in polarization because CMB photons scatter off of moving electrons at the epoch of recombination (when electrons and protons first combined into neutral hydrogen atoms). The polarization fluctuations are more difficult to measure than temperature fluctuations because they are an order of magnitude weaker, but they contain distinctive information that can set strong constraints on inflation and alternative scenarios. The density fluctuations that seed structure growth generate polarization patterns with an E-mode geometry, which can be circular or fan like. The exponential expansion of inflation models should also generate a spectrum of primordial gravitational waves, which imprint CMB anisotropies with a B-mode polarization pattern like a vortex (**Figure 4.2**). Ground-based experiments, WMAP, and Planck have measured E-mode polarization fluctuations, and the South Pole Telescope recently detected B-mode fluctuations caused by low-redshift gravitational lensing of the primordial E-mode. However, while the B-mode imprint of primordial gravity waves may be the most distinctive and informative signature of inflation, it is a subtle signal that is extremely difficult to detect.

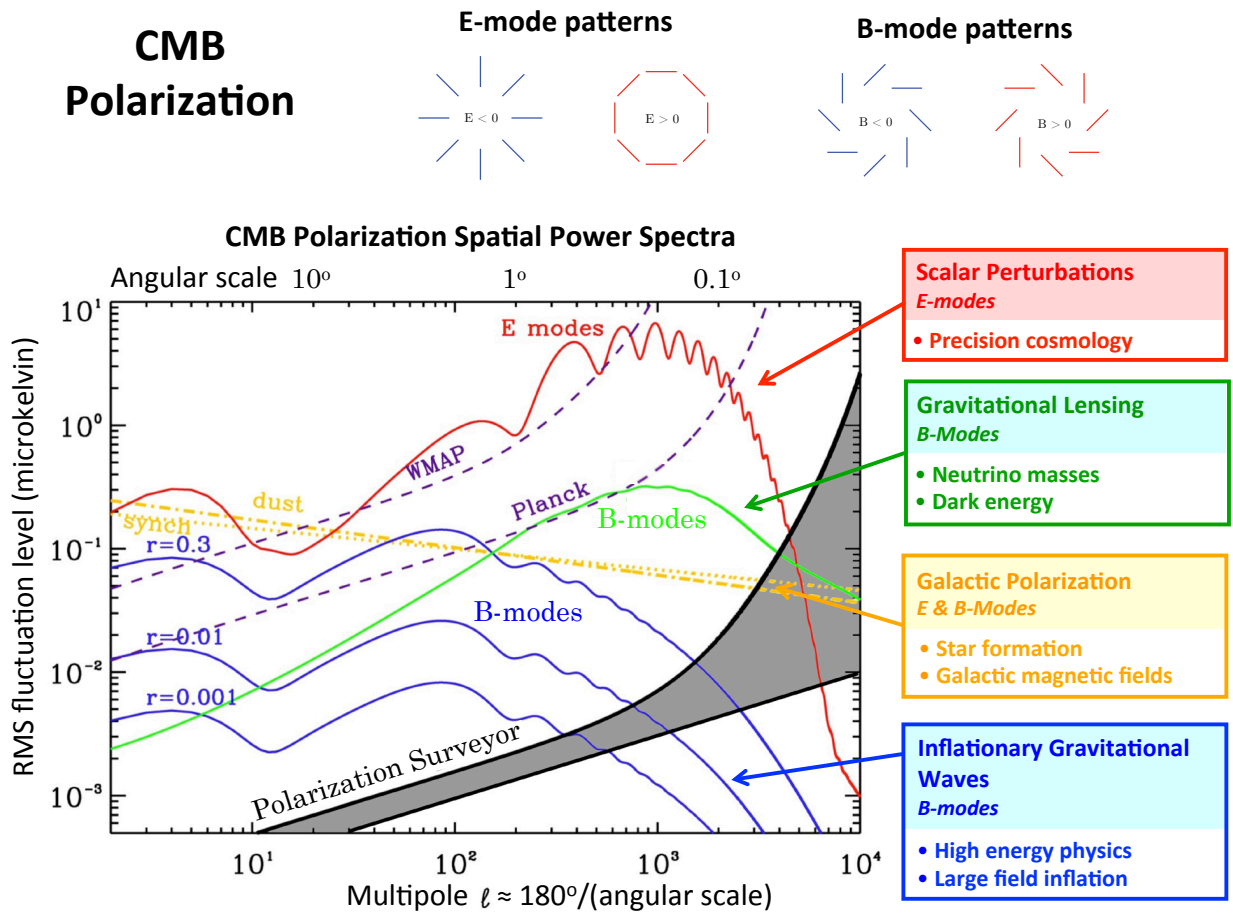

*Figure 4.2* Top: The E-mode polarization patterns do not change sign under mirror symmetry, while B-mode patterns do. This difference enables separation of E- and B-mode polarization in CMB maps. Bottom: The spatial power spectrum of the polarization anisotropies, which is plotted versus the multipole (lower axis) and the angular scale (upper axis). The E-mode spectrum (red) has been measured by WMAP and Planck as suggested by the sensitivity lines (dashed purple) for both satellites. The CMB Polarization Surveyor will characterize the E-mode polarization to cosmic variance limits on all scales up to its resolution and sensitivity limits, which could span the gray region. Observations near the galactic plane will also enable characterization of galactic dust and synchrotron radiation (orange), which will limit the CMB sensitivity in these regions. Measurements of the B-mode spectrum from gravitational lensing (green) will provide constraints on neutrinos and dark energy. The unknown amplitude of the inflationary gravitational wave spectrum (blue) is indicated by the three curves of different amplitudes and is proportional to the unknown energy scale of inflation. The figure is adapted from Bock et al. 2009.

> **Cosmic acceleration is one of the biggest puzzles in fundamental physics.**

The amount of primordial information in the CMB is finite because small-scale fluctuations are blurred by superposition and diffusion, and we have only one sky to observe. The measurements of temperature anisotropy from Planck are already approaching this fundamental limit set by "cosmic variance," the statistical uncertainty inherent in observations of the universe at extreme distances. The natural successor to Planck is a mission, referred to in our Formative Era roadmap as the CMB Polarization Surveyor, which could extract all of the primordial information in CMB temperature and E-mode polarization anisotropies. In addition, a defining goal of such a mission is to detect and characterize the B-mode gravitational wave signal from inflation if it is present at a level detectable above the gravitational lensing B-mode background. Measurement of the gravitational wave signal would provide powerful insights into the energy scale and driving physics of inflation. If it is strong, it would favor "large field" inflation models that naturally give rise to a multiverse, where a large (perhaps infinite) number of Big Bang universes germinate separately in an inflationary sea that separates them at superluminal speed. A weak or undetectable gravitational wave signal would rule out this broad class of models and might challenge the inflationary paradigm itself. This mission would probe fundamental physics at energies a trillion times higher than those reached by the Large Hadron Collider, the most powerful particle accelerator ever built.

Ground and balloon experiments have already driven development of the ultra-sensitive detectors that would be required for such a mission, so in technical terms it is relatively straightforward. This technology needs to be matured, space qualified, and expanded to industrial scale, because the key to sensitivity is to have enormous numbers of detectors that are individually operating at the quantum noise limit. The desired aperture would be similar to or modestly larger than the Planck satellite, so as to fully resolve the angular scale of primordial E-mode anisotropies and characterize the gravitational lensing of the CMB by intervening large-scale structure. The detailed mission requirements will be directly informed by detections or limits set using ground and balloon observatories. In addition to pursuing the inflationary gravitational wave signal, such a mission would provide exquisite sensitivity to any small deviations from adiabatic or Gaussian fluctuations, which, if detected, would provide powerful clues to the physics of inflation or alternative theories. The mission would also provide dramatically improved CMB constraints on the matter, radiation, and energy content of the universe, the history of re-ionization, the masses of neutrinos, and the amplitude of density fluctuations present at recombination. These measurements have profound implications in themselves, and they are crucial for understanding the mystery of cosmic acceleration.

CMB anisotropies map the structure present at recombination, but there is also valuable information encoded in the energy spectrum of the CMB. The COBE FIRAS experiment demonstrated that any deviations from a pure blackbody form are smaller than 0.01-percent, critical evidence that the CMB must have originated in a state of thermal equilibrium in the hot early universe. But events that release energy between light element nucleosynthesis (minutes after the Big Bang) and recombination (a half million years later) can distort the spectrum, as can scattering of CMB photons by hot electrons after the formation of structure. High-precision measurements of the spectral shape could reveal or set limits on a variety of

> **Cosmology and the Arts**
>
> Astronomical explorations provide inspiration for many fields of human endeavor, including the visual and performing arts. The South African artist Karel Nel (pictured below with his work *The Collapse of Time*) spent several years collaborating with the HST COSMOS survey team, culminating in his 2008 exhibition *The Brilliance of Darkness*. His images, created with ancient carboniferous dust and radiant white salt, evoke the COSMOS maps of the distant universe.
>
> The American artist Josiah McElheny, in collaboration with cosmologist David Weinberg, has created a series of monumental sculptures that depict the history and structure of the universe in the visual language of 1960s modernist design. *The Last Scattering Surface* (pictured below right) represents the connection of primordial fluctuations, encoded here by lamps that trace the COBE map of the cosmic microwave background, to present-day galaxies and superclusters.
>
> Exhibited in galleries and museums around the world, works like these communicate scientific discoveries to an audience far beyond the readership of technical journals.
>
> 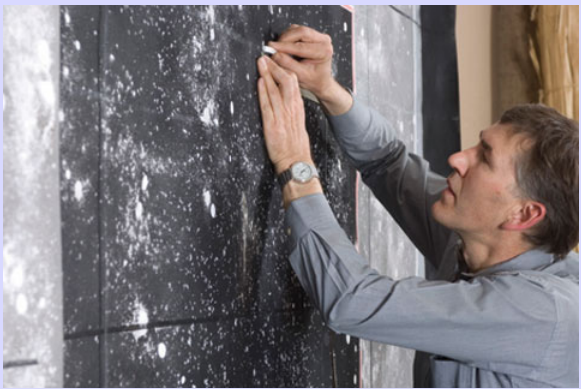 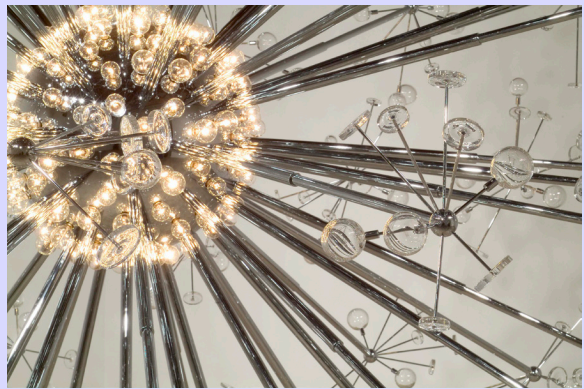
>
> ***The Collapse of Time***, Karel Nel, 2008. Photograph by Graham Borchers, courtesy of the artist and Art First, London.
>
> ***The Last Scattering Surface***, Josiah McElheny, 2006. Image courtesy of the artist.

interesting physical processes, such as annihilation of dark matter particles in the early universe, decay of a sterile neutrino species, or dissipation of acoustic waves generated by small-scale matter fluctuations. Any of these discoveries would mark an important advance in our understanding of the universe, either the forms of matter it contains or the fluctuations produced by inflation. An additional instrument on the CMB Polarization Surveyor or a relatively modest stand-alone mission (Explorer or Probe class) could achieve orders-of-magnitude improvement over the COBE FIRAS measurement using technology that is largely in hand, thus providing a unique probe of cosmic physics in an interval that is otherwise hidden from view.

## *Cosmic Acceleration and Dark Energy*

Even before the discovery of the CMB, a key goal of observational cosmology was to determine whether the average density of matter was sufficient to eventually halt and reverse the expansion of the universe, leading to a "Big Crunch" at some finite time in the future. However, in the most astonishing cosmological discovery of many decades, studies of the 1990s showed that the expansion of the universe is not decelerating but accelerating. This is the cosmic equivalent of tossing an apple in the air and having it accelerate away into outer space. While we know gravity as an attractive force that holds stars, planetary systems, and galaxies together, on the scale of the universe, gravity repels.

Cosmic acceleration is one of the biggest puzzles in fundamental physics. Einstein's theory of gravity, general relativity (GR), predicts that matter, radiation, and all forms of energy we have ever observed in a laboratory will produce attractive gravity and slow the expansion of the universe over time. Within GR, cosmic acceleration can be produced by exotic forms of energy that have negative pressure, known generically as dark energy. This energy could be a cosmological constant whose density is set by the fundamental zero-point energy of the quantum vacuum, but the observed magnitude is extremely far from natural theoretical predictions. It could instead be a dynamical field whose energy density varies in space and time. Alternatively, it could be that GR itself is incorrect, and that the true theory of gravity reproduces the empirical success of GR on solar system and galactic scales but leads to acceleration over cosmic distances. (GR could also break down in the regime of strong, dynamical gravitational fields around black holes and colliding neutron stars, as discussed in Section 4.2.)

Deciding which, if any, of these explanations is correct has become a major focus of contemporary cosmology. The basic approach is to measure the history of cosmic expansion and the history of the growth of matter clustering with the highest achievable precision over a wide span of redshift in order to differentiate the predictions of distinct theoretical scenarios.

Constraining theories of cosmic acceleration is one of the defining goals of the WFIRST-AFTA mission, which will (during 2.5 years of its six-year prime mission) conduct three large surveys to measure the expansion and growth of structure by multiple techniques. The supernova survey will use the wide-field infrared camera to discover and monitor more than 1,500 type Ia supernovae out to redshift $z=1.7$, then measure the relation between distance and redshift by modeling these sources as "standardizable candles." The high-latitude imaging survey, covering 2,000 square degrees in areas of sky that are not obscured by the Milky Way, will measure the shapes of 500 million galaxies at HST-like resolution. Analysis of these images will measure the growth of structure via weak gravitational lensing, the subtle but systematic shearing of background galaxy shapes by the gravitational light-bending distortions of foreground matter. The high-latitude spectroscopic survey, over the same area of sky, will measure the redshifts of 20 million galaxies in the range $1 < z < 2$ that is most difficult to probe from the ground. These will be used to measure the cosmic expansion history via a "standard ruler" imprinted on galaxy clustering by sound waves (known as baryon acoustic oscillations, or BAO) that propagated in the universe before it underwent recombination. The net precision expected for the WFIRST-AFTA expansion and growth measurements is 0.1–0.3% via multiple methods. This represents a one to two orders-of-magnitude improvement on existing data, with much tighter control of systematic uncertainties, crosschecks among independent probes, and a wider range of redshift. The WFIRST-AFTA measurements will complement contemporaneous measurements by LSST and the ESA-led Euclid mission, which use similar methods but differ in their statistical strengths and their primary systematic challenges.

According to our current understanding, the nature of dark energy controls the ultimate fate of the universe, or at least that portion of the universe that we will ever be able to observe. If dark energy is a cosmological constant, then the universe will continue to accelerate, eventually reaching an asymptotic state where it doubles in size every 12 billion years. The growth of large-scale structure will slow to a halt as gravitational attraction falls ever further behind accelerated expansion. The universe will cool as stars burn out and fewer stars are born to replace them, and the volume of space visible from Earth will steadily shrink as galaxies now within our view are accelerated beyond a cosmic event horizon. An exotic form of dark energy known as "phantom energy" would instead drive acceleration at an ever increasing rate, eventually stretching the fabric of space to infinite scale in a finite time, a cosmic fate referred to graphically as the "Big Rip." If the properties of dark energy evolve with time, then its gravity could become attractive rather than repulsive in the distant future, leading to an eventual Big Crunch. If the explanation of cosmic acceleration lies in modified gravity rather than dark energy, then the implications could be weirder still, perhaps demanding that we consider the entire history of our expanding 3-dimensional space from a higher-dimensional perspective.

We do not know which if any of these directions will be favored by future cosmological data. While the high-latitude surveys planned for the WFIRST-AFTA prime mission only cover 5% of the sky, observations in an extended mission could push close to the ultimate cosmic variance limits of the weak lensing and BAO methods and reach the effective limiting accuracy of the supernova method. Whether there are basic cosmology missions to do after WFIRST-AFTA and the CMB Polarization Surveyor depends on what we learn from them. Discoveries from these missions could, for example, highlight the value of mapping structure at redshifts $z > 2$ to pin down the early evolution of dark energy and increase leverage on the spectrum of primordial fluctuations. At redshifts $2 < z < 6$, this structure can be traced with emission-line galaxy surveys.

Several of the missions described elsewhere in the roadmap could also lead to breakthroughs in fundamental cosmology. For example, the Formative Era gravitational wave or X-ray missions could challenge GR in a way that points to a gravitational explanation of cosmic acceleration, or detect gravitational waves originating from energetic processes when the universe was only a picosecond old. The Visionary Era Cosmic Dawn Mapper, a lunar radio array to probe 21-cm radiation from the reionization and pre-reionization epoch, could give extraordinarily precise measurements of early structure because the observable volume at very high redshifts is much larger than that available even to WFIRST-AFTA, Euclid, and LSST. The Visionary Era Gravitational Wave Mapper would measure "standard siren" distances to hundreds of thousands of merging neutron stars and black holes, yielding measurements of cosmic expansion that are 100 to 1,000 times better than those that exist today and well beyond the limit of any other known method.

Mapping the polarization of the CMB and understanding the puzzle of cosmic acceleration could point us in extraordinary directions:

- A universe that is one among many in a multiverse.
- A universe that will end in a Big Rip 20 billion years from now.
- A universe that has expanded and collapsed through many cycles, perhaps infinitely many.
- A universe that has spatial dimensions beyond the three that are visible to our everyday experience.

Today these ideas seem at the speculative edge of science, but the examples of the Big Bang theory and inflation show that cosmological scenarios can move from speculative hypothesis to empirically well-grounded theory by paths that are hard to anticipate. In the century since Einstein proposed the modern theory of gravity and space-time, we have made astonishing advances in understanding the nature of the cosmos. Our picture remains incomplete, but we expect to reveal its deeper layers in the decades to come.

## 4.2 Revealing the Extremes of Nature

> Our universe has extreme places ripe with discovery potential—natural laboratories where the laws of physics are being tested beyond our current capabilities. There the accelerations currently achieved by the most powerful particle physics laboratories on Earth (e.g., CERN) can be dramatically surpassed, allowing collisions of ultra-high-energy subatomic particles. There are unexplored regimes of density and energy where previously unknown particles may be present and the laws of physics, as we now understand them, may be altered. There are supermassive black holes (SMBHs), billions of times the Sun's mass, that seem to have coevolved with their host galaxies.
>
> In the next 30 years, we will probe these unexplored extreme corners and shine a light onto the origin and behavior of matter in the universe. We will measure SMBH properties, masses and spins, as a function of redshift to reconstruct their growth rates and the processes by which they grew. We will study the ripples of space-time created when a small black hole falls into a large black hole in the extreme mass-ratio inspiral events and in compact binary mergers to test Einstein's field equations for gravity. And we will probe the deep interiors of neutron stars by using X-ray observations to determine their mass-radius relation and thus constrain their equation of state, searching for exotic phases of matter that appear under extreme conditions.

### *Black Holes, Accretion Disks, and Powerful Jets*

Despite being the weakest of the four forces of nature, gravity is responsible for some of the brightest emission, the most powerful explosions, and the most massive objects in the universe. The energy released from matter falling into the gravitational field of a massive object, the so-called accretion process, powers quasars that are billions of times more luminous than our Sun with jets that can reach across the scales of galaxies (**Figure 4.3**). However, accretion does not only give rise to these extremes of nature. Its influence also shapes the properties of many astronomical objects and reaches across all disciplines of astronomy. Indeed, accretion plays an important role in the formation of stars from interstellar clouds, of planets around stars, and in the assembly of gas and stars into galaxies.

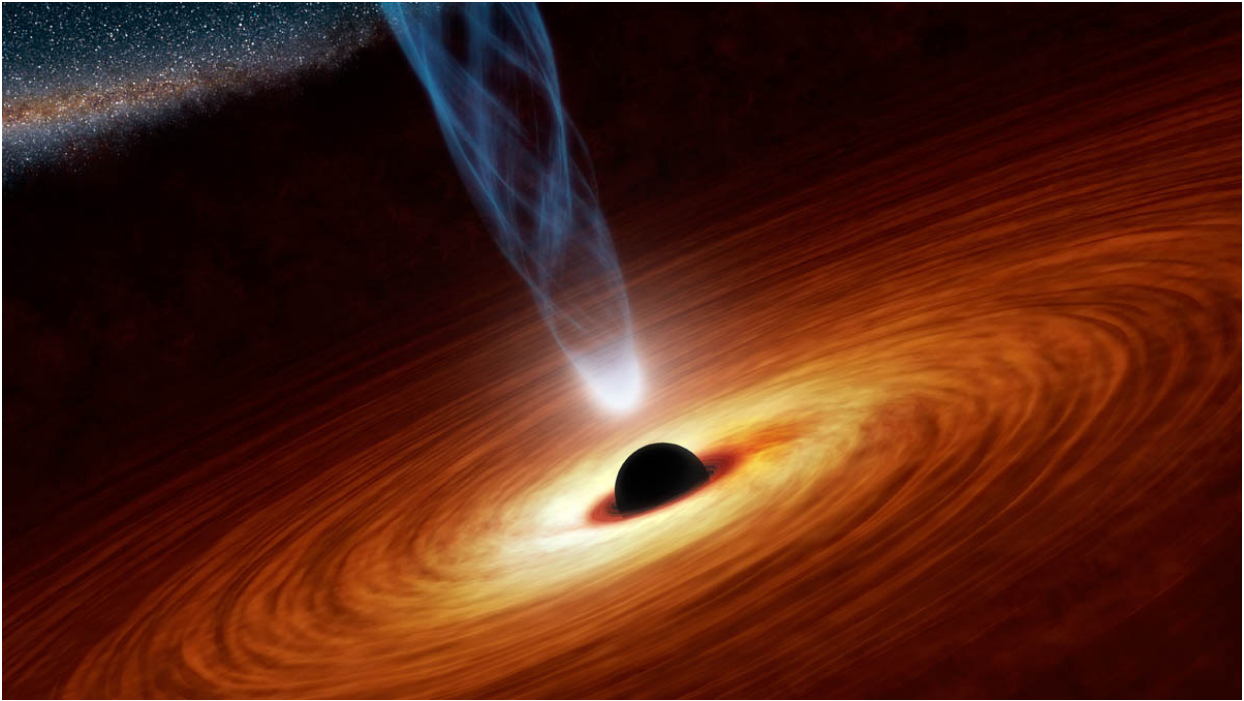

***Figure 4.3*** *An artist's impression of an accreting supermassive black hole and the relativistic jet it drives. Credit: R. Hurt/NASA/JPL-Caltech*

While the role of gravity in driving accretion is well established, numerous important questions remain to be answered: How does matter assemble into accretion disks? What drives the motion of matter onto the central object or away from it in strong outflows? How bright can accretion disks shine and why are some so dim? Black holes (BHs), from stellar mass to the giants that are billions of times the Sun's mass, serve as excellent test beds to understand the process of accretion, its role in the evolution of the universe, and, in turn, to utilize accretion to understand the fundamental laws of nature.

Observations have so far revealed that every massive galaxy has a massive BH in its center, and probably millions of smaller, stellar-mass BHs distributed throughout. In the Milky Way alone, there are tens of confirmed stellar-mass BHs as well as the central monster, Sgr A*, which has a mass of 4 million suns. Looking at our nearby galaxy neighbors, the BH in the nucleus of M87 not only outweighs our own but also has visible powerful jets extending outward from the center that our telescopes can easily resolve.

Given the vastly different temperatures and particle densities present in an accretion disk around a BH and one that is forming planets around a star, it is not clear whether the same process can drive accretion in both cases. Moreover, despite the observational evidence for powerful, fast, highly collimated jets that are launched from some BH systems, the mechanisms powering and aligning these jets remain unclear. Jets may be an important means by which SMBHs truncate the growth of massive galaxies.

Studies of accretion and of BHs are closely intertwined. We see BHs precisely because they accrete, and emission from the accretion flow can be used to understand the BH itself. Accretion also plays a dominant role in the growth of BHs across cosmic time. The accretion of matter onto BH seeds causes them to reach monstrous masses, and it is the release of energy during the accretion process that makes quasars shine even in the early stages of galaxy formation. In the Near-Term Era, JWST will provide an unprecedented look into the accretion process of the first BHs in the universe, their growth, and their interaction with their surroundings. The rate at which BHs grew in the early universe will be directly measured, and this will provide a handle on the efficiency of the accretion process. Moreover, the morphology of central regions of galaxies and their star-formation rate will lead to measurements of mechanical and energetic feedback from the accretion process.

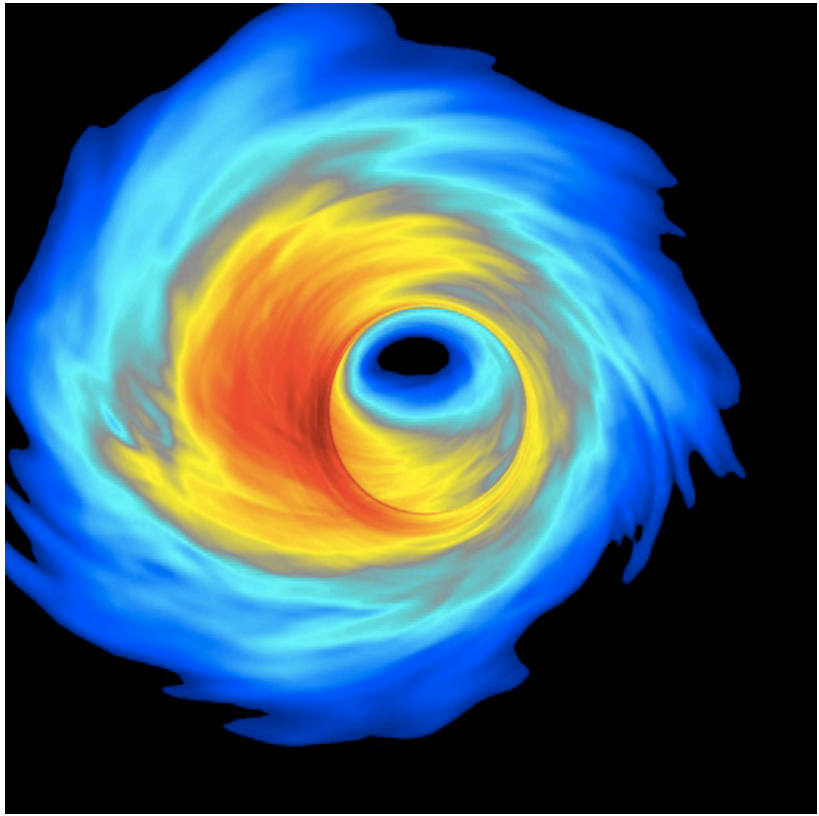

*Figure 4.4* *Simulation of the shadow of a black hole's event horizon.* **Credit: S. Noble (RIT)**

X-ray astronomy is a natural tool to study BH accretion, exemplified by the discovery of the first BH, Cygnus X-1, in 1964. X-ray emission is produced at the inner regions of BH accretion disks (both in stellar-mass and SMBHs) as gas heats to billion-degree temperatures. X-ray observations will enable us to move from the census of BHs and their host galaxies to detailed probing of the inner workings of individual sources.

The strong and rapid variability of the observed X-rays reveals the timescales of turbulence and instabilities, while the X-ray spectrum encodes the temperatures and densities of the accretion disk. Features in the spectrum, such as the K-shell iron lines at 6.4–7.0 keV, are strongly affected by the relativistic Doppler effect, caused both by the extreme velocities of matter near the BH as well as by gravitational redshifts in the deep BH potential. Measuring the shapes of these spectral features, therefore, allows a mapping of the BH space-time and serves as a gravity probe.

In the Formative Era, breakthroughs in our understanding of accretion will come from the next-generation X-ray Surveyor, combining high-resolution spectroscopy with a large collecting area. X-ray absorption line studies of both stellar-mass and SMBHs with energy resolutions of $\Delta E < 4$ eV will characterize the mass, energy and momentum of outflows from the accretion disk, allowing us to determine the impact of BH-driven outflows on their surroundings. With absorption features completely accounted for, the broadened emission features (e.g., the broadened iron line) from the central accretion disk can be characterized robustly. An X-ray Surveyor 10 times larger than current X-ray telescopes (collecting area of 1–3 m$^2$) will enable searches for short-timescale changes in the accretion disk as X-ray flares echo across the inner disk and turbulent hot-spots orbit, successively passing from the approaching to the receding side of the disk and shifting their observed frequency accordingly. Comparative studies of objects with and without jets, as well as monitoring of sources that have unsteady or transient jets, will allow us to understand the disk-jet-wind connection. Taken together, these studies will provide an incisive look into the inner regions of accretion flows and the launching mechanism of powerful jets.

In the Visionary era, interferometry techniques in space can resolve the innermost regions of the accretion disks around tens of SMBHs in nearby galaxies and generate images of the shadows they cast onto their surroundings. Taking a direct picture of a BH in this way will offer an unprecedented look into its properties and into their accretion flows. We will also aim to get our first picture of these powerful engines in X-rays with a BH Mapper, an interferometer with submicroarcsecond resolution. This would enable us to image and perform spatially resolved spectroscopy of the central regions of nearby luminous AGN, by directly mapping the structure of the accretion disk and the base of the jet. And the most exciting results are those we cannot yet anticipate.

## *Probing Black Hole Space-times*

The defining characteristic of a BH is the presence of an event horizon through which matter, energy and light can enter, but nothing can exit. The BHs of Einstein's GR have an event horizon that cloaks the central region of extreme density referred to as the space-time singularity. In addition, GR dictates that astrophysical BHs can be described by two properties alone: their mass and their spin. So GR clearly lays out the gravitational potential and signatures that should be present around each and every BH.

The mass of a BH can be determined by studying its effect on surrounding gas and stars. The precise mapping of the stellar orbits around the very center of our Milky Way galaxy reveals the existence of a 4-million-solar-mass dark object confined within a radius of 20 light-hours. In addition to providing one of the most accurate SMBH masses known, this observation provides one of the most powerful pieces of evidence for the existence of black holes. Studying the stellar motions in the centers of other nearby galaxies leads us to a similar, if less precise, conclusion.

The spin of the BH is a much more challenging quantity to measure since the effects of spin only manifest themselves very close to the BH. At the current time, the only quantitative method for estimating the spin of a BH is via X-ray spectroscopy, either through a characterization of the temperature of the inner accretion disk or the gravitational redshift of spectral features (both of which increase as the BH spins faster in the same sense as the accretion disk rotation). After a decade of intensive study with RXTE, Chandra, XMM-Newton, Suzaku, and now NuSTAR, we have spins for approximately a dozen stellar-mass BHs and 20 SMBHs in some of the nearest and brightest AGN. While it is exciting to have a first look at spin in BH systems, the small number of measurements and the remaining possibility of systematic errors limit the astrophysical conclusions that can be drawn.

The Formative Era will see an explosion in our ability to probe BH spins. The quantum leap in X-ray spectroscopy enabled by the next-generation large X-ray Surveyor will yield BH spin measurements in hundreds of AGN, and allow us to push spin studies beyond the local universe. The systematic errors in these measurements will be dramatically reduced due to the mission's high energy-resolution, allowing a much more thorough characterization of the X-ray spectrum. In the same time frame, the deployment of a space-based Gravitational Wave Surveyor will allow us to make extremely precise measurements of the masses and spins of merging SMBHs out to very high redshifts (essentially the formation epoch of million-solar-mass BHs), giving us a completely new window on the BH population.

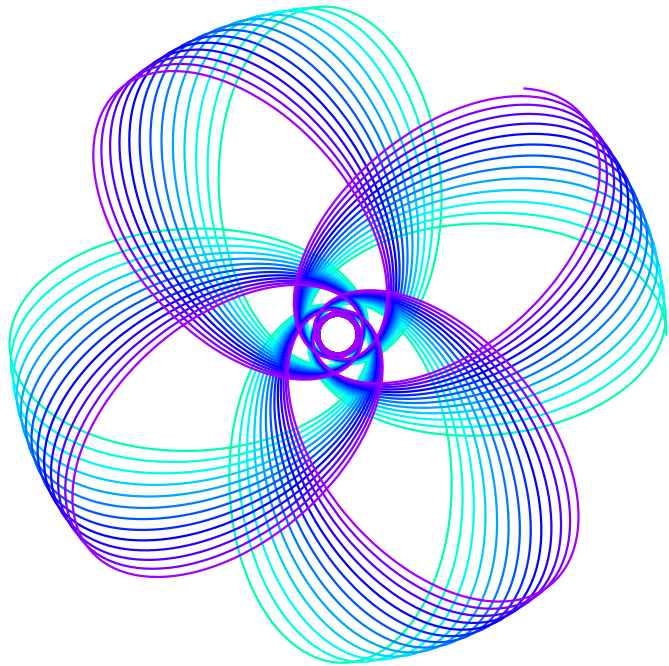

*Figure 4.5* Extreme mass-ratio inspirals. The path traced out by a stellar-mass black hole as it orbits a far more massive galactic black hole produces a complex "spirograph" pattern that is encoded in the gravitational wave signal. Here each color corresponds to one complete orbit. Detecting these signals will allow us to produce detailed maps of the black hole space-time. Different theories of gravity predict a unique fingerprint for these black hole orbits, and by comparing the signals to these predictions we can perform high-precision tests of Einstein's theory. **Credit: G. Perez-Giz (New York University)/J. Levin (Columbia University)**

In fact, a space-based gravitational wave detector permits a detailed 3D mapping of the space-time structure around a BH by studying the so-called extreme mass-ratio inspirals (EMRI), systems where a stellar-mass BH spirals into a million-solar-mass BH (**Figure 4.5**). These systems will emit gravitational waves in the mHz band, which can only be accessed with a space-based detector; a ground-based detector would be swamped by seismic noise and the gravitational disturbances from trucks, clouds and other nearby moving objects. The gravitational waves produced by an EMRI not only encode a map of the space-time around the massive BH, but also encode other subtle effects such as the absorption of some of the waves by the SMBH's event horizon, and the small tides raised on its event horizon by the infalling BH. These effects, while small in magnitude, are clearly recorded in the gravitational wave signal, and would constitute direct evidence for the existence of an event horizon. Even more powerful evidence would come from the abrupt termination of the signal as the smaller BH is swallowed. Thus we would know if BHs truly exist, and if they conform to the predictions of Einstein's theory of gravity.

> **A question of central interest is whether Einstein's theory of gravity is really the correct law of nature on astrophysical scales.**

A question of central interest is whether Einstein's theory of GR is really the correct law of nature on astrophysical scales. Theorists have discussed other types of "BH-like" objects with characteristics very different from those predicted by standard GR. For example, string theory has motivated the possibility of "superspinners," objects rotating so fast that they do not possess an event horizon. The discovery of superspinners (or any other form of non-Einsteinian

"BH-like" objects) would be profound—it would provide the first piece of observational evidence for physics beyond the standard models of particle physics or cosmology. If such objects accrete gas, many of their properties would be very similar to normal BHs. However, in principle, the exotic "stringy ball" at the heart of the object can be observed directly. While more theoretical work is needed on such exotic objects, the predicted X-ray spectrum and variability of a superspinner would be distinctly different from that of a BH. An even more dramatic difference would be seen during a merger event involving a superspinner: gravitational waveforms would keep increasing in frequency rather than being "chopped off" by the event horizon.

Einstein's theory of gravity has survived a wide range of high-precision experimental and observational tests, but these tests have been limited to the weak, static gravitational fields found in the solar system (fields of order $10^{-8}$ in dimensionless units) and low-velocity compact binaries (fields of order $10^{-6}$). But what happens in the strong, dynamical fields that pertain on cosmological scales and in the vicinity of binary BH systems? Black hole mergers are expected to generate powerful bursts of gravitational waves that travel unimpeded across the universe. Einstein's theory predicts that gravitational waves come in two polarizations and travel at the speed of light. By directly detecting gravitational waves we will be able to test these basic predictions for the first time, and in addition, we can use the information brought to us by the waves to carry out a host of other high-precision tests of gravity in the dynamical strong-field regime.

Some modifications to Einstein's theory of gravity yield "floating orbits" in which orbital decay halts altogether by a new interaction; others predict that the graviton, the quantum particle that transmits the gravitational force, has a small mass and up to six different polarizations. Direct observations of gravitational waves can thus identify or stringently constrain new physics. Multiple observations at different positions and redshifts would allow us to test the fundamental assumption that the laws of gravity are independent of location and age of the universe.

## *Discovering Fundamental Laws with Neutron Stars*

Our universe offers us many natural laboratories where the laws of physics can be tested beyond our current capabilities. Neutron stars, which are the small dense remnants of massive stars, are a prime example. What makes neutron stars unique is the truly extreme density present in their cores, where matter can take on new forms, previously undetected particles may be present, and the laws of physics can differ from those tested on normal atomic nuclei. When densities are compared, neither the early universe, when the relic background radiation formed, nor particle colliders on Earth, such as the Large Hadron Collider at CERN, can match the interiors of neutron stars.

As stellar cores reach higher and higher densities, distinct possibilities arise for the composition and interactions of particles composing the core. For example, atomic nuclei could decompose into the quarks that make up protons and neutrons (**Figure 4.6**). No experiment has ever shown the presence of free quarks, not bound into nucleons, so observing this phenomenon in neutron stars would be the first indication of this unprecedented behavior. Another possibility is that when the energy of the interactions between nucleons rises with increasing density, new particle interactions can take place that completely change the character of the matter. For example, interactions between particles such as pions and kaons resemble those in superconductors and lead to much lower pressures. In contrast, if neutrons and protons prevail, the pressure in the stellar interior keeps rising. Finally, another possible outcome is the restoration of mathematical symmetries in some of the physical laws, such as those present in the very early universe that were subsequently broken. The restoration of symmetries may have many consequences that would reshape our understanding of nuclei.

Neutron stars are abundant, with about a billion estimated to be present in our galaxy alone. They are observable in different manifestations across the entire electromagnetic spectrum, from radio wavelengths to gamma rays, depending on the type of radiation their surfaces or energetic magnetic fields produce.

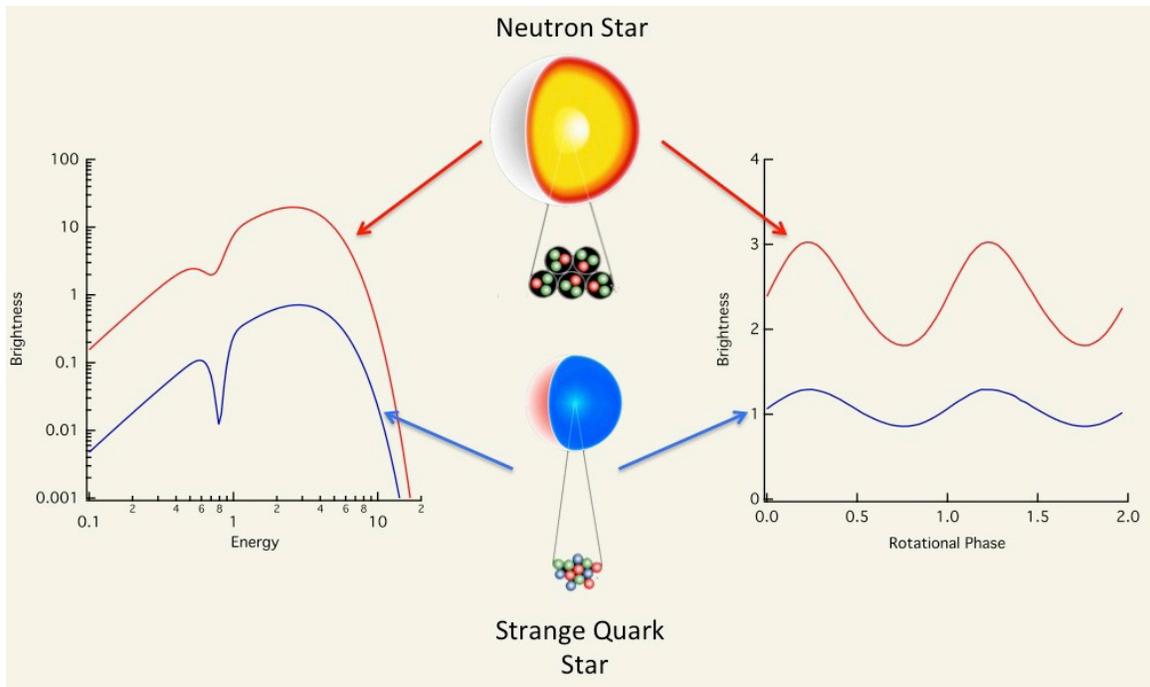

*Figure 4.6* *A neutron star composed of neutrons all the way down to its core has a different size and mass than one made up of free quarks (middle). The size difference can be detected in the shape and depth of spectral lines (left) and in the amplitude and shape of periodic changes in the star's brightness (right).*
***Credit: NASA Chandra X-ray Observatory and F. Ozel (Univ. of Arizona)***

NASA plays a key role in the study of neutron stars because, for a large fraction of these observations, space-based capabilities are required due to the fact that the surfaces of neutron stars reach million-degree temperatures and emit primarily in X-rays and ultraviolet light. This is true for isolated neutron stars, whose emission is powered by rotation or their own interior heat, as well as for neutron stars that accrete matter donated by a companion star. In the latter case, matter freshly accreted onto the surface allows the neutron star to shine in X-rays, which can be observed with space telescopes.

One goal of current NASA missions like the Chandra X-ray Observatory and Fermi Gamma-ray Space Telescope is to study and characterize these emissions. Snapshots taken at X-ray and gamma-ray wavelengths every few months to years reveal the evolution of bursts and energetic emissions produced from neutron star surfaces and magnetospheres. These observations can also trace how neutron stars change as new material accretes from binary companions and accumulates on their surfaces. However, to study the new interactions and new forms of matter that may arise in extreme densities, it is necessary to probe the deep interiors of these compact objects, which are not directly observable.

There are no in-situ experiments or observations that can directly reveal the properties of a neutron star's core, so how can we probe matter at these extreme conditions? Luckily, the microscopic particles and interactions present in the interior become mapped with high fidelity to the large-scale, visible properties of neutron stars. The interactions that shape the interior of the neutron star and give rise to a particular composition of particles also determine the mass, size, and even the temperature of the entire star. Measurements of these macroscopic properties serve as excellent probes of new physics. In the example in **Figure 4.6**, a stellar core composed of neutrons is larger and more massive than one containing free quarks.

We can probe neutron star interiors, utilizing the direct relation between the size of the neutron star and its equation of state. Interactions that generate larger pressure at a given density give rise to larger stars. A measurement of the stellar radius at the 10% level can distinguish an interior made up of quarks from one composed of nucleons. Current and future observations, therefore, are aimed at high-precision measurements of the neutron star radii.

We cannot rely on any technological capabilities to obtain a resolved image of a neutron star, even in the Visionary Era, owing to their extremely small sizes, (approximately 20 km across) and large distances from Earth. Nevertheless, in the Near-Term and Formative Eras, missions relying on different methodologies can provide fruitful probes into new physics in neutron stars. One powerful technique leading to the measurement of neutron star properties analyzes periodic pulsations in light from its surfaces. Neutron stars spin rapidly, so any bright spot appears to an observer as a periodic pulse. These pulsations encode information about the star's mass, size, and surface temperature. For example, neutron stars with larger radii for the same mass, produce pulsations with larger amplitudes (**Figure 4.6**). Measuring pulsations with high time resolution and decoding their nature serves as a direct probe of the stellar interior. In the near future, an approved NASA mission called NICER—Neutron Star Interior Composition ExploreR, slated to be deployed on the Space Station—will exploit this technique by obtaining high timing resolution data from pulsing neutron stars in soft X-rays.

A second method relies on very high-resolution X-ray spectra obtained from light emitted by the surface. In these spectra, atomic lines can be observed from matter that is present in the thin stellar atmosphere. The star's extreme gravity shifts each of these lines to lower energies by the same amount. In addition, the lines are broadened by Doppler effects due to the star's rapid spin (**Figure 4.6**). The redshift measurement reveals the mass-to-radius ratio of the star, while the line width allows a high-precision measurement of the radius. However, detecting these features in individual spectra requires a large (> 1 m$^2$) collecting area because neutron star surfaces are typically dim and their spectral lines relatively weak. This measurement will be possible with the Formative Era X-ray Surveyor.

Spectra also can be used to determine the surface temperatures of neutron stars. Combined with the observed brightness, this allows a determination of the object's size, which is a powerful diagnostic of the star's interior. A telescope used for these observations needs to be able to separate light from the neutron star from light originating from nearby sources. It therefore needs to have a subarcsecond spatial resolution to extract the surface emission from background objects in crowded fields of the galaxy.

The latter two measurements will be possible in the Formative Era with an X-ray Surveyor that possesses a large collecting area as well as high energy and timing resolution. The combination of the high spatial resolution with the ~3 m$^2$ collecting area will allow the spatial features to be detected, and the surface temperatures and emitting areas to be measured with high precision.

The hot neutron star created by the supernova explosion subsequently radiates away its residual heat through its surface, and the cooling rate is sensitive to the composition of its core. For example, a neutron star with a nucleonic core and an iron surface cools down to a 100,000 kelvin a million years after the explosion, while a neutron star with a pion condensate core takes only 10,000 years to reach that temperature. Therefore, observations of neutron star surface temperatures at different ages can be a direct probe of the neutron star equation of state.

In the Formative Era, a LUVOIR Surveyor that reaches 35th magnitude in U or V bands can detect the cooling emission from neutron star surfaces within a few kpc from Earth and reconstruct their cooling history for approximately a million years by combining observations of neutron stars at different ages. This will provide a probe to the extreme physics of matter in the cores of neutron stars that is complementary to X-ray and gravitational wave observations.

## 4.3 Listening to the Cosmos

> For all the advances we have made in astronomy in the past 100 years - opening up the electromagnetic spectrum from gamma rays to radio waves and resolving objects far dimmer than can be seen by the naked eye—it is humbling to realize that we have yet to detect the most powerful explosive events in the universe, caused by the collision of two black holes. What else might we be missing by being unable to "hear" the gravitational vibrations of the universe?

Gravitational waves are produced by the movement of matter or energy—when you wave your hands you generate extremely weak gravitational waves. But to produce waves that we have any hope of detecting across the vast reaches of space requires large amounts of high-density material moving at a good fraction of the speed of light. This makes gravitational waves excellent probes of extreme astrophysical environments.

The gravitational wave symphony of the universe is expected to span a wide range of frequencies from a diverse collection of astrophysical sources. The coming decade will see the initiation of a new age in astronomy, as gravitational wave telescopes make their first recordings of this symphony. The high

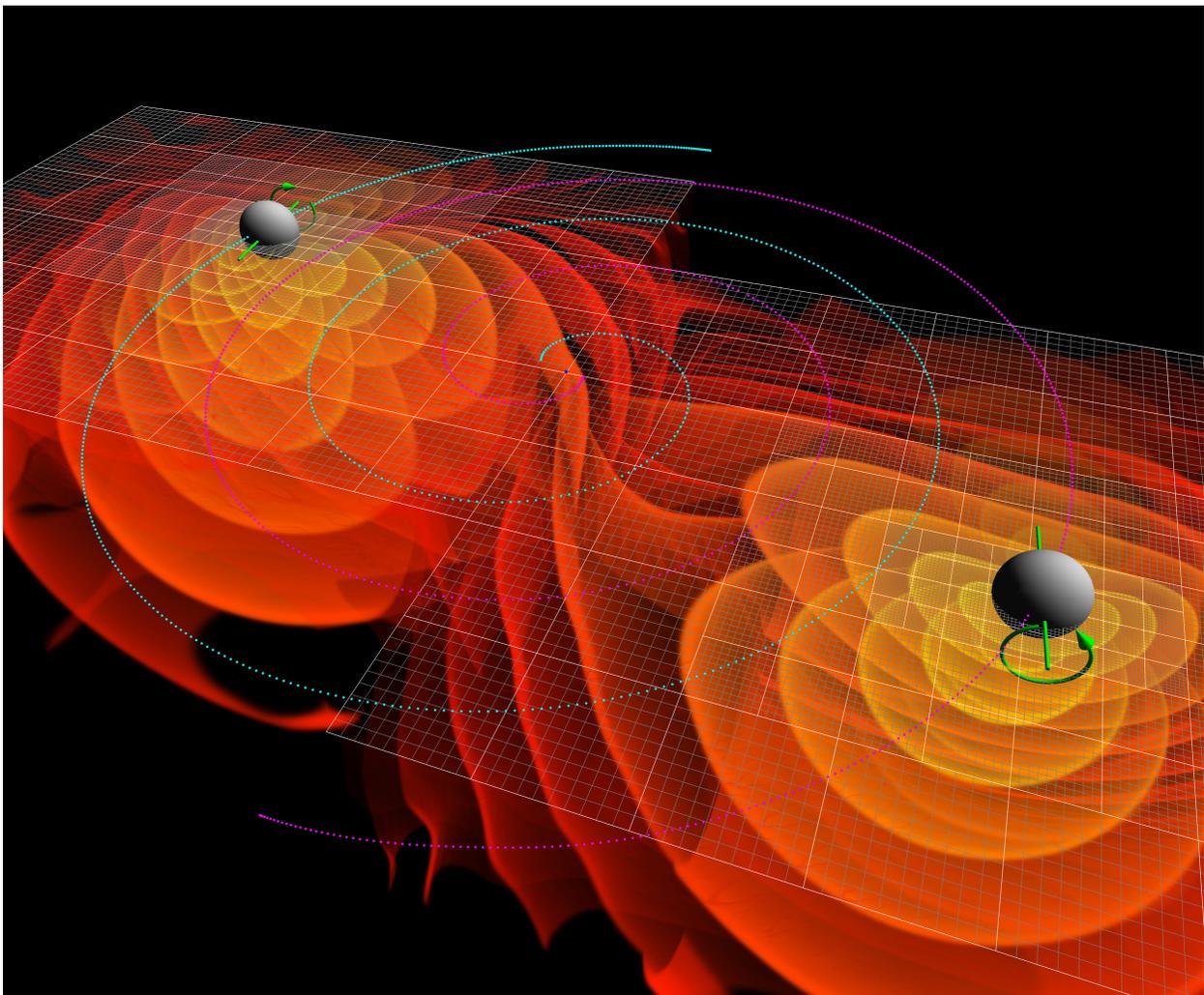

*Figure 4.7* This snapshot from a numerical simulation shows the gravitational waves produced by a pair of merging black holes. **Credit: Chris Henze (NASA Ames), from a simulation by J. Centrella, B. Kelly, J. Van Meter, and J. Baker (NASA GSFC)**

frequency portion of the spectrum will be made accessible by the advanced LIGO/Virgo ground-based laser interferometers, while the very low frequency portion of the spectrum will be detected in the timing signals from an array of radio pulsars. The signals recorded by these instruments have the potential to revolutionize our understanding of the universe. However, ground-based detectors and pulsar arrays are unable to detect signals in the mid-range of the gravitational wave spectrum, where many of the most interesting and powerful signals are to be found. To draw an analogy with sounds and hearing, pulsar arrays are tuned to the infra-sound of an elephant's call while LIGO is tuned to the ultra-sonic chirps of bats and dolphins, but neither can hear ordinary speech or music. To explore the source-rich mid-range frequencies requires a space-based array located far from the gravitational disturbances of Earth. A Gravitational Wave Surveyor, operating in the millihertz frequency range, would open a vast new realm of discovery with the potential to radically transform our understanding of the universe.

Gravitational waves and light provide complementary information about astrophysical events. With light we can image the outer layers of an object and learn about its chemical and thermal properties, but we usually cannot see what is happening below the surface. With gravitational waves we are able to probe deep into the interior and track the flows of mass and energy, but we cannot produce an image. Objects such as our Sun are strong sources of light but weak sources of gravitational waves, while mergers of two isolated black holes (**Figure 4.7**) are strong sources of gravitational waves but emit no light. There are other systems, such as the merger of two neutron stars, or black hole mergers embedded in a gas cloud, that are powerful sources of light and gravitational waves. Such multimessenger systems will open up new ways of exploring the universe and probing fundamental physics.

To date we have only indirect evidence for gravitational waves from the orbital decay of neutron star and white dwarf binaries, but we are able to anticipate many other sources. Likely candidates for space-based detection include binary systems of SMBHs, and the capture of stellar remnants by galactic SMBHs. In the mHz band covered by the Gravitational Wave Surveyor, it is likely that the signals from white dwarf binaries in our Milky Way will be so numerous that they become the limiting "noise" source at low frequencies. These detectors will be able to record massive BH mergers ($10^8$–$10^3_{\text{solar masses}}$) throughout the observable universe. Less certain, but paradigm-changing possibilities include signals from a network of cosmic (super)strings or other topological defects associated with phase transition in the 100 GeV to 1,000 TeV range, a mere trillionth of a second after the Big Bang. Other possibilities include brane-world effects, TeV-scale reheating scenarios, and exotic inflationary scenarios. In addition, there may be sources no one has anticipated. What we do know is that they must involve large concentrations of high-density material moving at close to the speed of light. Certainly the most surprising result would be that the gravitational wave universe holds no surprises!

Looking ahead to the Visionary Era, a more sensitive, multi-element Gravitational Wave Mapper would be able to detect and localize hundreds of thousands of neutron star and stellar-mass BH binaries. The critical requirement for such a mission is a significantly increased sensitivity over the GW Surveyor in the 10 to 100 mHz range, i.e., above the maximum frequency produced by white dwarf binaries. This will result in dramatically reducing the astrophysical foreground noise that otherwise limits the sensitivity to distant sources and faint backgrounds. A second requirement is few-arcsecond angular resolution to identify the host galaxies of most sources directly from the gravitational wave signal. This capability will enable measurements of "standard siren" distances to hundreds of thousands of galaxies, yielding a high-precision map of the space-time of our universe. The host galaxies of massive black BH mergers could be identified, allowing detailed multi-messenger studies of galaxy mergers across cosmic time. Even more exciting is the possibility of detecting the faint ripples of gravitational waves produced during the Big Bang, allowing us for the first time to see past the recombination curtain and to directly witness the moment of creation.

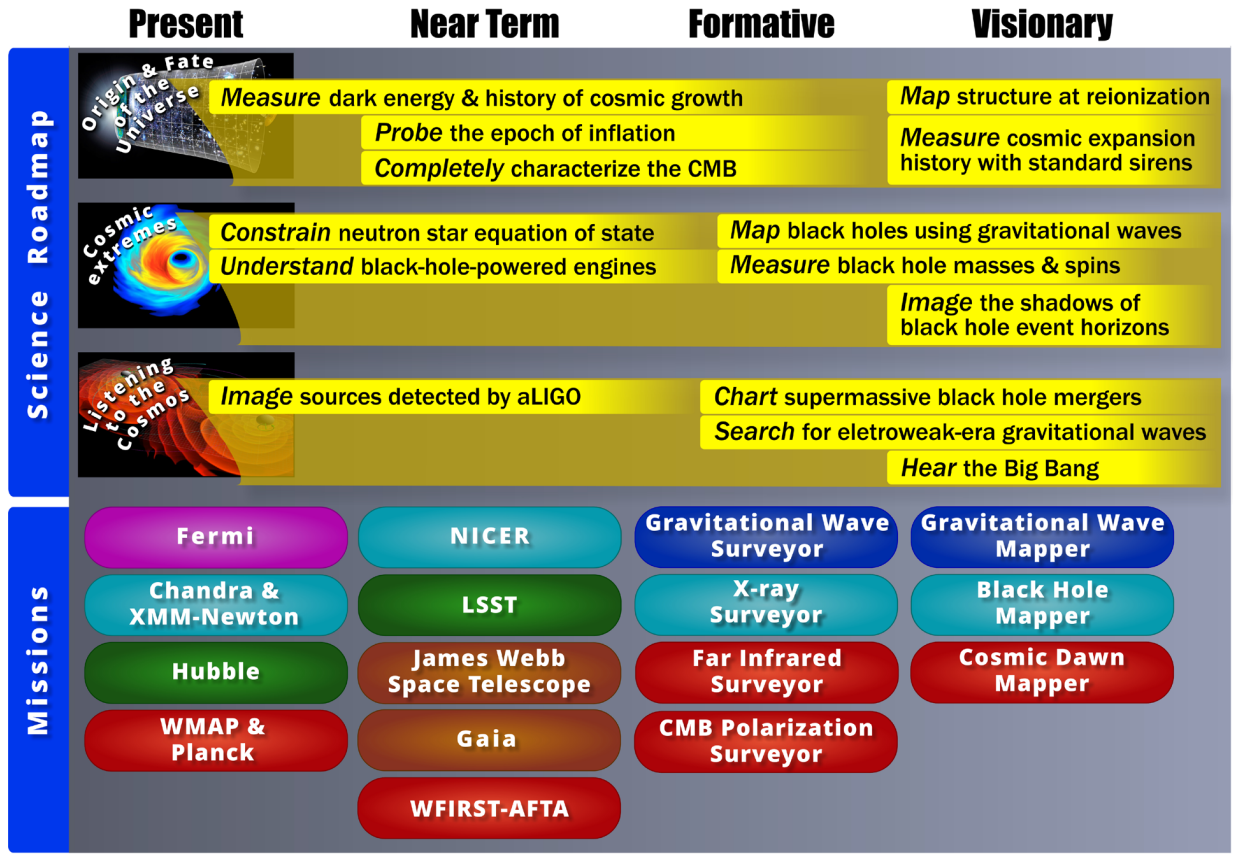

*Figure 4.8* Schematic of the Physics of the Cosmos Roadmap, with science themes along the top and a possible mission sequence across the bottom. **Credit: F. Reddy (NASA/GSFC)**

## 4.4 Activities by Era

### Near-Term Era

- **The Origin and Fate of the Universe**

  - Complete characterization of CMB anisotropies from Planck, WMAP, and ground-based experiments. Create integrated picture with other cosmological data.

  - Improve precision measurements of the Hubble constant with HST, JWST, and Gaia.

  - Measure the history of cosmic expansion and structure growth at 0.1% level precision using WFIRST-AFTA, to probe the origin of cosmic acceleration.

- **Revealing the Extremes of the Universe**

  - Use X-ray observations from Chandra and NICER to measure the masses, radii, and surface temperatures of neutron stars and constrain the equation of state of nuclear matter.

*Formative Era*
- ***The Origin and Fate of the Universe***

  - *Measure the polarization of the CMB to cosmic variance limits with the CMB Polarization Surveyor. Search for the signature of primordial gravity waves from inflation.*

  - *Measure the spectrum of the CMB with precision several orders of magnitude higher than COBE FIRAS, from a moderate-scale mission or an instrument on CMB Polarization Surveyor.*

- ***Revealing the Extremes of the Universe***

  - *Probe the workings of nature's most powerful engines, accretion onto BHs and their associated jets, using X-ray observations.*

  - *Make high-precision determinations of neutron star properties with spectroscopic measurements from the X-ray Surveyor.*

  - *Measure BH spins and test basic GR predictions for BH space-times with X-ray Surveyor observations.*

  - *Map BH space-times with Gravity Wave Surveyor measurements of extreme mass-ratio inspirals, and test GR predictions for the physics of BH mergers and the polarizations of gravitational waves.*

- ***Listening to the Universe***

  - *Measure SMBH mergers throughout the observable universe and white dwarf binaries throughout the galaxy.*

  - *Search for unanticipated sources of gravitational waves from combined Gravitational Wave Surveyor observations and electromagnetic measurements.*

*Visionary Era*
- ***The Origin and Fate of the Universe***

  - *Measure early matter clustering at high precision using redshifted 21-cm observations of the reionization and pre-reionization epoch, from the Cosmic Dawn Mapper.*

  - *Measure the cosmic expansion history at ultra-high precision using standard siren distances to hundreds of thousands of merging neutron stars and BHs, from the Gravitational Wave Mapper.*

- ***Revealing the Extremes of the Universe***

  - *Image the innermost regions of SMBH accretion disks and see the X-ray shadow of the event horizon, from the Black Hole Mapper.*

- ***Listening to the Universe***

  - *Directly detect the background of gravitational waves from the inflationary epoch or high-energy phase transitions of the early universe.*

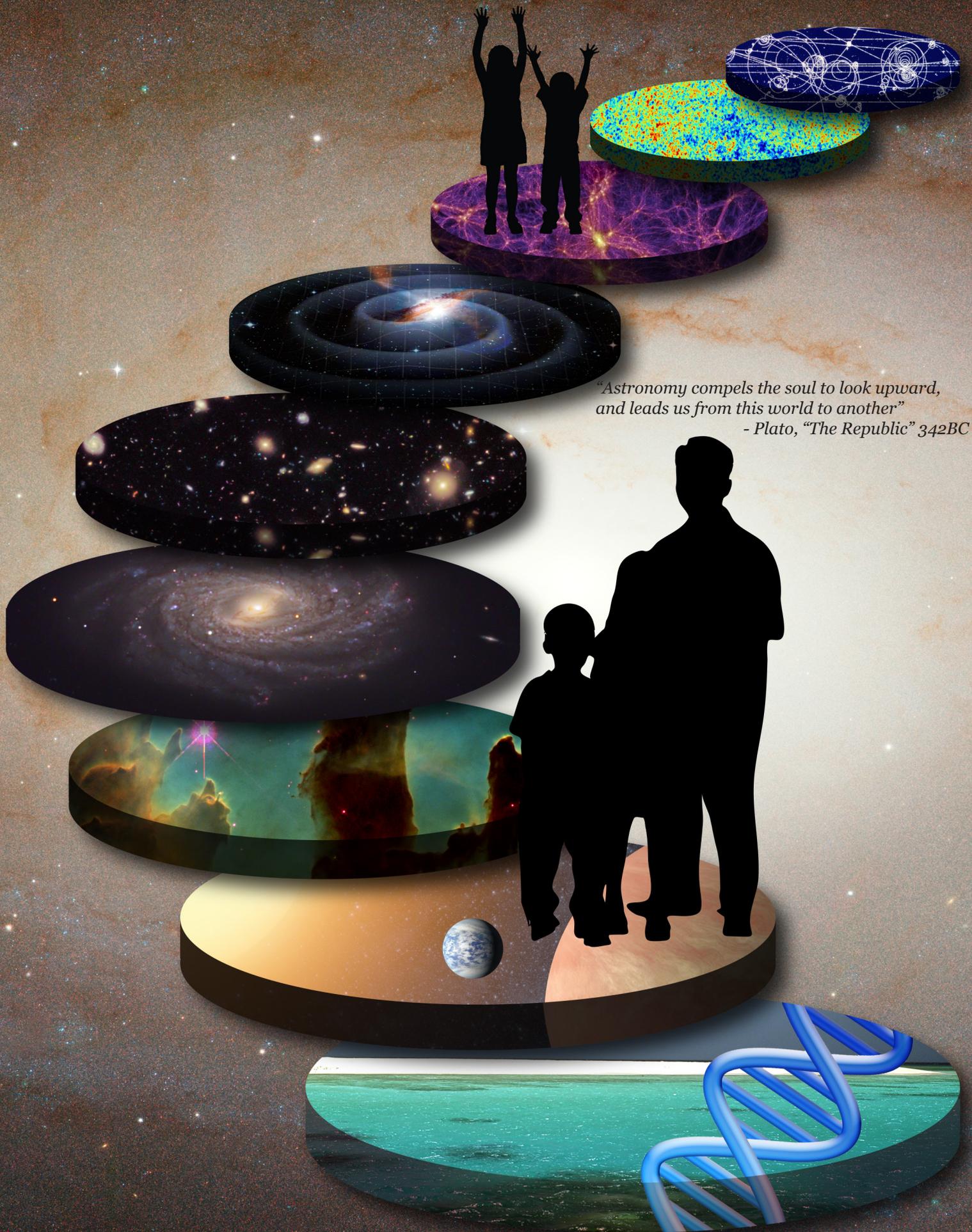

"Astronomy compels the soul to look upward, and leads us from this world to another"
- Plato, "The Republic" 342BC

# 5 PUBLIC ENGAGEMENT: CONNECTING THROUGH ASTRONOMY

We as a species are at a unique turning point in history—a place where we find ourselves poised to answer some of humanity's biggest questions. These are central to who we are; they have been asked since humans have been able to ask questions at all and pervade our society, as evidenced by the public's fascination with astronomy (**Figure 5.1**). For this reason, the astronomer's quest to answer the universe's most enduring questions is a quest for all humanity, and every effort must be made to connect with and engage the public and future explorers with astrophysics research. NASA's charter is to expand human knowledge and preserve the role of the United States as a leader in space science and technology, *for the benefit of all mankind* and to *provide for the widest practicable and appropriate dissemination of information concerning its activities and the results thereof* (Space Act of 1958). The future of astronomy public engagement offers great opportunities to better align to national priorities and extend our network of collaborations. Ensuring that NASA data are easily accessible for public engagement and that scientific discoveries are effectively communicated to the widest possible audience is extremely important.

One of the primary reasons we are approaching this scientific turning point at this juncture in history is the technology advances that have recently been made, and will be made in the near future. Not only is this true in an engineering sense, but it also applies in the realm of communication and outreach. Technology has given us an opportunity to break down many of the barriers that have long existed between scientists and the public, both in terms of ideas and knowledge communicated, and of access to actual data.

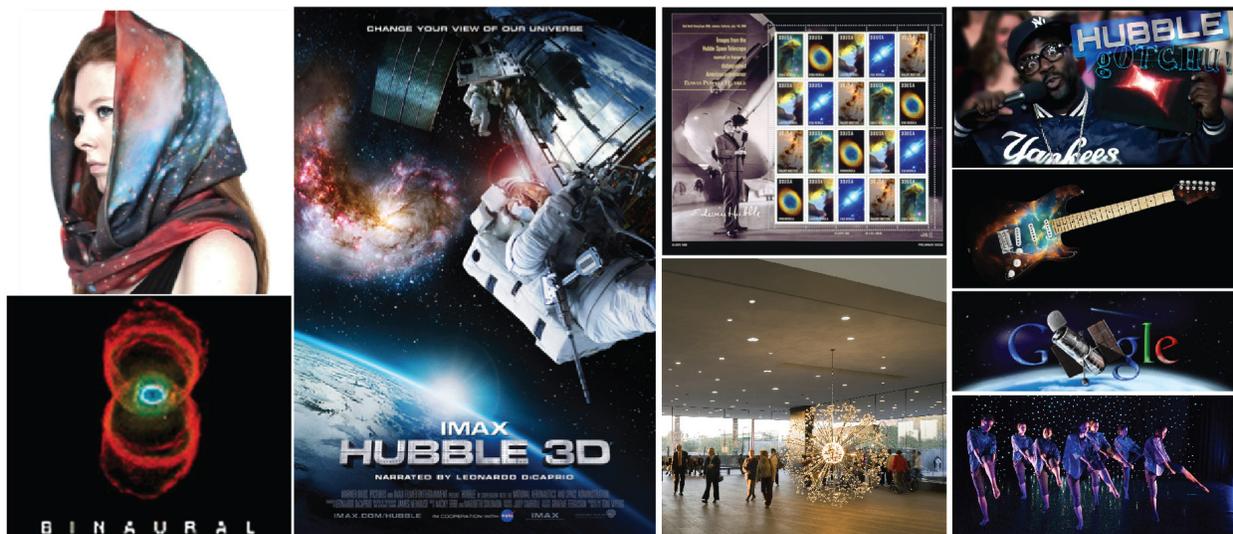

*Figure 5.1* Society's love of astronomy is evident throughout popular culture. From left: Hubble imagery on clothing by Shadowplaynyc (top) and the cover of the Pearl Jam album, "Binaural"; a 3D IMAX® film about Hubble's history, discoveries, and servicing; postage stamps honoring a decade of Hubble (top) and sculptures inspired by the cosmic microwave background (see p. 59, photograph by David Weinberg); one of several "Hubble Gotchu!" segments on "Late Night with Jimmy Fallon" (top, courtesy of NBC), a "Hubble skinned" guitar (photo by Nathanial Burton-Braford), a Google Doodle celebrating Hubble, and "Stargazing," an astronomy-themed fusion of dance, shadow play, and music performed by the alight dance theater in 2013 (photo by Brian Allard).

## 5.1 The continuum of astronomy learners

TECHNOLOGICAL advances in communication are enabling the public to participate in astronomy in a variety of ways. The job of scientists and science communicators is to provide avenues for people to share in the discovery process in whatever ways they choose. This relatively new continuum of participation in astronomy not only allows a wider population to be reached, but lets a broader spectrum of astronomers participate in outreach—in whatever way is best for them. In the past, it was typical for people to learn about astronomy discoveries either in a formal setting—in classrooms—or via "traditional" media such as press releases and television documentaries. These avenues are undoubtedly still important to the portfolio of outreach, but new technologies are rapidly providing new ways to communicate astronomy to the public, and to invite them to participate directly in discovery.

### *The cornerstone of communicating astronomy: unique content*

THE cornerstone of communicating astronomy is *unique content* that the public is eager to learn about. For astrophysics specifically, successful communication of science ideas often hinges on visualization of NASA data. As more telescopes come online in the next 30 years, providing a wealth of new data, we must be prepared to visualize these vast datasets in such a way that benefits not only the scientists involved in the projects, but also the public who has shown such fascination with previous similar efforts. With such substantial audiences in the realm of social media, the science community should ensure that these visible channels stay populated with accurate and engaging multimedia data-based products that inform and excite the public.

Similarly, accessibility to NASA data via online archives has given the public an opportunity to actively participate in data analysis alongside professional astronomers. This new field of citizen science exists in many disciplines, but astrophysics is uniquely poised to build on the public's inherent fascination with astronomy and to engage people in authentic experiences with NASA-unique data. Projects like Galaxy Zoo and Planet Hunters (with over 855,000 registered users in Zooniverse as of this writing; see Chapters 2 and 3) have led the way in astrophysics citizen science, providing online user-friendly interfaces through which anyone can classify galaxies or look for signatures of planets in actual data. Similarly, programs such as the NASA/IPAC Teacher Archive Research Program have provided ways for educators to become involved in ongoing astronomy research using NASA data archives. Future programs should build on the community's best-practices in order to make more NASA data accessible to the public and teachers in this highly participatory way.

## 5.2 Audiences: From online to one-on-one

THE advent of social media and the accessibility of information via the internet has completely changed science communication: *scientists are more often now interacting directly with the public* in forums like Facebook and Twitter, and NASA specifically has been a leader in engaging huge audiences online (with over 5 million Twitter followers as of this writing). NASA's success with social media is largely due to the unique content provided via the extremely popular **NASA.gov** website (**Figure 5.2**), which is host to all of NASA's missions and science. NASA.gov has averaged 11.5 million visits per month from February through September 2013, with 5.25 million unique visitors per month for the same period.

Knowledge gained via the work of professional astronomers not only may be communicated to the public—adults and children alike—via the methods outlined above, but also can be carried down to the next generation of scientists via the more traditional paths of *formal and informal education*. Resources provided to formal and informal educators through avenues such as the NASA/STScI HST education programs are critical in translating astronomy research into products relevant for the classroom. For example, STScI formal education programs (mostly HST-related with some new JWST content) are utilized by half a million teachers and 6 million students per year, with informal education programs reaching 9 million

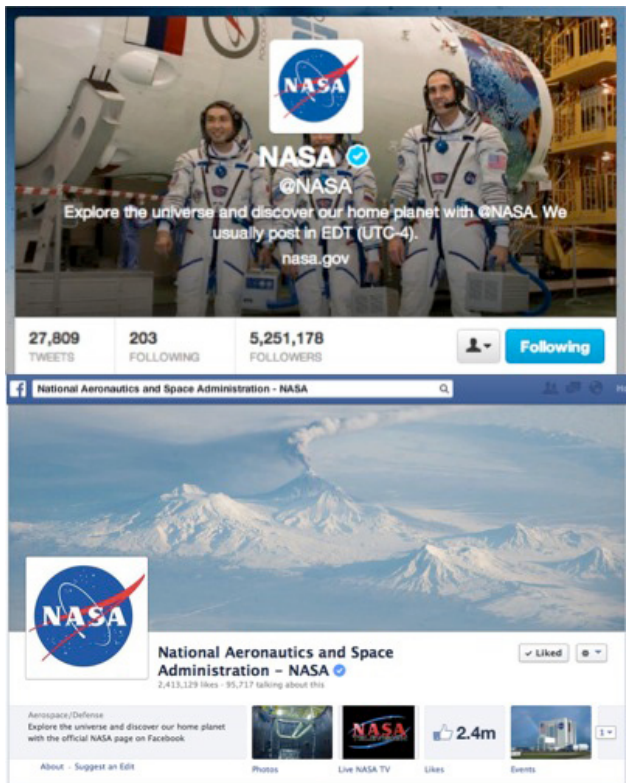

*Figure 5.2* NASA is an award-winning leader in web presence and social media, with the most Twitter followers of any federal agency (over 5 million as of this writing). **Credits: Twitter and Facebook**

people per year (detailed metrics and evaluation at **http://outreachoffice.stsci.edu/education-metrics/**). Continued support for teachers and informal educators is important to ensure that accurate and timely astrophysics knowledge is shared in ways that most benefits both teacher and student.

NASA's Science Mission Directorate programs (including Astrophysics) have a strong history of rigorous evaluation standards, and it is critical that future education programs rely on similar strict standards in order to ensure that teachers and students are presented with products that best meet their needs. This evaluation methodology must continue not only for formal and informal education programs, but also in public engagement efforts to ensure that impact is maximized while duplication of effort is eliminated. Throughout development of formal and informal education programs, close ties to the missions and scientists making the discoveries are critical to ensure that the information relayed is accurate. Astronomy is already considered to be a "gateway science" which inherently inspires and attracts students at an early age; this should be leveraged in the future as we work to attract students into the STEM workforce. The work that formal and informal primary and secondary educators do to develop this workforce is extremely important, as are the higher education STEM fellowships and opportunities.

In addition to the vast audiences that can be reached via online interactions and the very important work that educators do for astronomy, there is no doubt that *in-person experiences with scientists are invaluable to the public*, especially students who are trying to decide on career paths. While it is important to ensure that astrophysics discoveries are distributed widely to the largest audiences possible and accurately translated in our nations' classrooms, these more individualized encounters should be prioritized as well. Even though such endeavors have been successful in the past, there are many ways to increase collaboration and partnerships such that these authentic encounters happen more often, and in more impactful ways. For example, effectiveness could be increased by setting up a coordinated and visible program with an online database of localized astronomers who are willing to do school visits, give public talks, and partner in astronomy-themed events in different locales. Such a program could make astronomers available to public schools, museums, planetariums, lecture series, and other ongoing events. As many scientists express the desire to participate in such outreach, but often feel under-equipped, a coordinated program to train them in effective communication with a variety of audiences could enhance their capabilities and confidence to engage with the public.

In this same vein, the effectiveness of general public-engagement astronomy programs and large events cannot be understated. During the International Year of Astronomy in 2009, "…at least 815 million people in 148 countries participated in the world's largest science event in decades" (the full report is available at **http://www.astronomy2009.org/resources/documents/detail/iya2009_final_report/**). Today, interactions such as Google+ Hangouts and Twitter Q&As gather the public into participatory

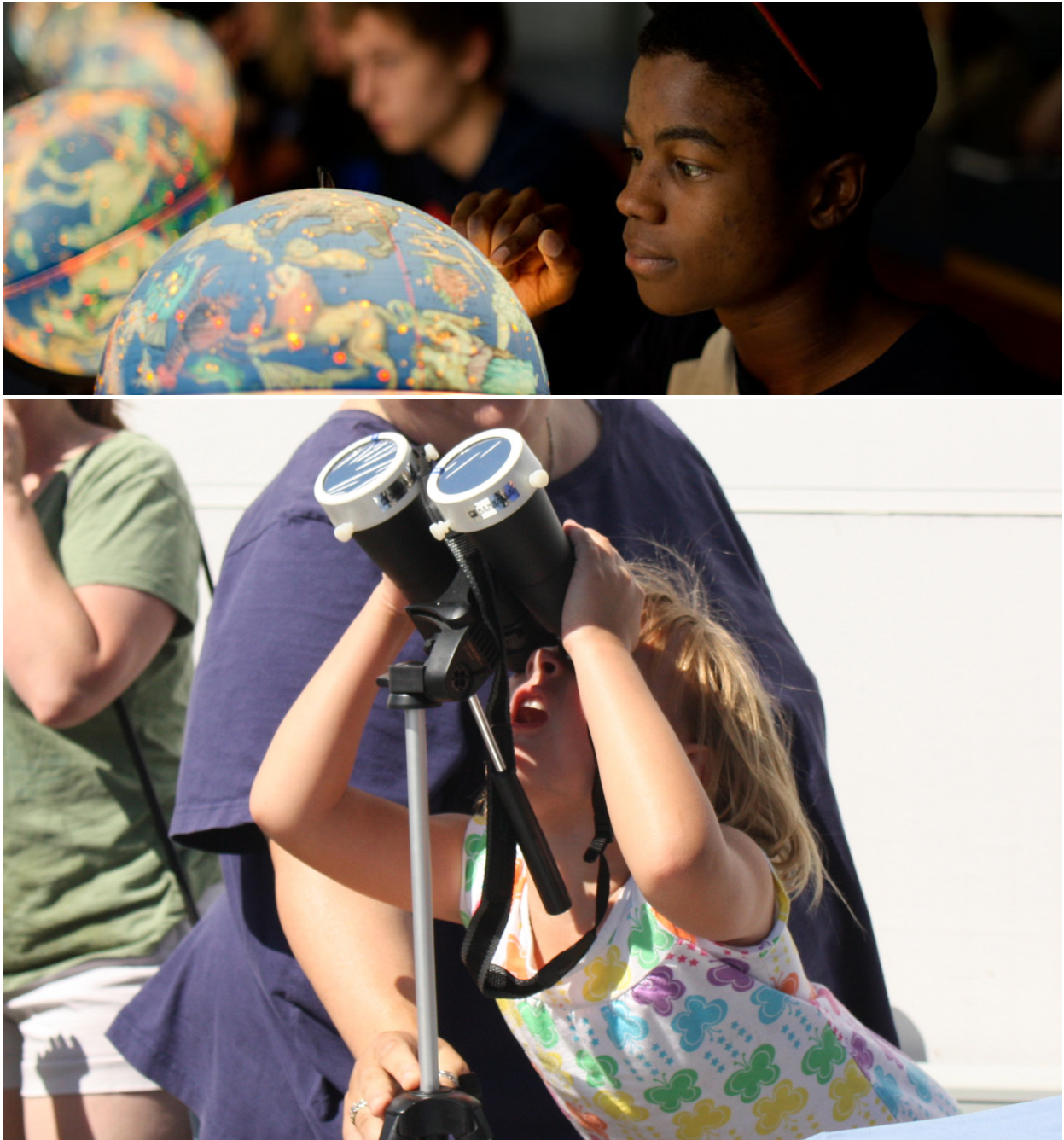

*Figure 5.3* *Astronomy inherently inspires students and serves as a gateway for entering STEM careers. Future efforts to connect existing programs on increasing diversity—and to build on their successes—will help to ensure an inclusive future for astronomy.* **Credits: GSU (top) and N. Batalha (NASA/Ames)**

interaction online, and NASA Socials gather large numbers of NASA social media followers in a central location for in-person behind-the-scenes experiences at launches and other events. Statistics such as these demonstrate the extent to which the public is ready to learn and engage when astronomy content is offered.

*Diversity and inclusion: Critical for STEM success*
A focused and coordinated effort to increase engagement with underrepresented populations will ensure a future of astronomy that is inclusive and enriched through diversity. Many successful efforts are already in place, including undergraduate astronomy programs at colleges and universities targeting women and minorities, committees within professional societies such as the AAS and APS dedicated to discussing these issues, films dedicated to the subject, and NASA-wide programs focused on increasing diversity in all STEM fields. A high-level effort connecting all of these various programs for astrophysics would leverage all of the already existing resources and expand opportunities for underrepresented minorities.

The central goal for the future is to make NASA's exciting discoveries accessible to everyone—people of different ages, backgrounds, and experiences—using an array of communication, outreach, and education tools to best reach them. As NASA science is publicly funded, this is not only a commitment; it is an obligation. Engaging people *where they are* using a variety of techniques will be crucial for success. We should rely on current best practices for education, public engagement, and science communication in order to expand our reach and impact, and connect into the lives of people where they happen to be. As we participate as scientists in the Daring Visions outlined in this roadmap, we are committed to sharing our discoveries with those outside the academic and laboratory walls, and to actively engaging people in these Enduring Quests to understand our universe.

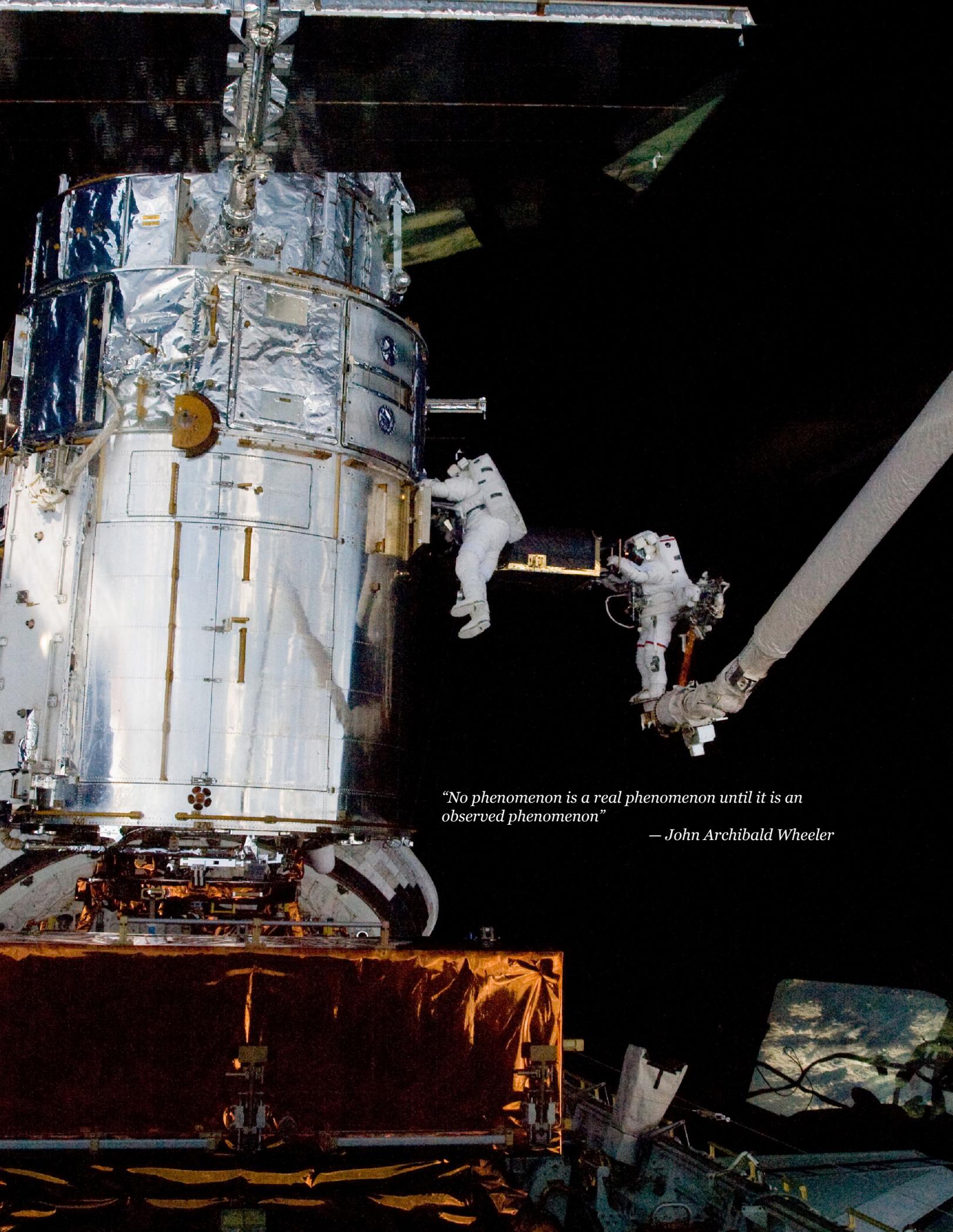

"No phenomenon is a real phenomenon until it is an observed phenomenon"

— *John Archibald Wheeler*

# 6 Realizing the Vision: Notional Missions and Technologies

Realizing the science vision will require observational capabilities beyond what is currently available. Seeking other life-bearing worlds will require observing faint small planets located near much brighter stars—a significant observational challenge yet to be overcome. Understanding the origin of galaxies, stars and planets requires observations of faint distant sources across the electromagnetic spectrum at a sensitivity and angular resolution well beyond the levels available today. Unlocking the physical principles driving the universe will be made possible by observing gravitational waves and photons produced in regions of the universe where physical conditions are extremely different from those found in our local environment. Capturing these signatures will likewise require entirely new observational capabilities.

In this section, we describe a set of notional missions, primarily to highlight key areas where technology advances are needed. These notional missions are driven by fundamental observational requirements for parameters such as wavelength, angular or spectral resolution, and sensitivity. We intentionally keep the description of these missions at a high level and provide minimum detail in terms of the specific hardware. Both the notional missions and enabling technologies described in this Chapter are ordered within each Era (Formative Era, corresponding roughly to the next 20–30 years, followed by the Visionary Era), by increasing wavelength, with gravitational wave missions arbitrarily listed first.

Although medium and small-scale notional missions are not listed in this chapter, they will undoubtedly play a key role in realizing the vision described here. Thanks to their shorter development timescales, small and medium-scale missions provide an opportunity for more rapid adaptation to new technologies and new science questions, and often serve as scientific and technology precursors to larger missions/programs. Such missions also permit focused and well-defined science investigations to be pursued, helping to maintain scientific progress between (and even during) large missions. Parts of the roadmap science vision will be best achieved by such focused small-scale efforts, while others may be obtained using probe-scale missions, such as:

- Measuring blackbody spectrum distortions in the cosmic microwave background
- Mapping the universe's hydrogen clouds using 21-cm radio wavelengths via a lunar orbiter observing from the far side of the moon
- Monitoring energetic transients with X- and gamma-ray telescopes
- Measuring X- and gamma-ray polarization

However, over the 30-year timescale of this roadmap, it would be futile to specify what such small/medium efforts should be. Doing so would also undermine the flexibility advantage of smaller missions, which can be formulated and developed more rapidly. This chapter therefore focuses on larger missions, which must answer elements of the science vision for which the required measurement quality (angular resolution, sensitivity) undoubtedly requires a large-scale effort.

## 6.1 Formative Era

### *Gravitational Wave Surveyor*

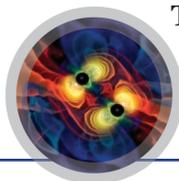
The technologies required to observe the universe with gravitational waves are very different from those used to explore the electromagnetic spectrum. However, the basic measurement principle is simple. A passing gravitational wave causes space to expand and contract by a small amount in the plane transverse to the propagation direction. By carefully measuring the change in distance between free masses, it is possible to record the wave pattern, much like a microphone records the small displacements of a diaphragm caused by sound waves. Laser interferometry can be used to track the distance between a constellation of free-floating "proof masses" to incredible precision. Each proof mass is surrounded by a spacecraft that provides shielding from outside disturbances, such as solar radiation and the solar wind, and serves as a platform to provide power, communications and housing for the laser metrology system. The measurement sensitivity at high frequencies is limited by the laser power, the size of the telescopes used to focus the laser signals, and the distance between the spacecraft (waves with periods shorter than the light travel time between the detectors suffer from partial signal cancellation). The measurement sensitivity at low frequencies is limited by residual parasitic forces on the proof mass caused by capacitive or gravitational coupling to the spacecraft.

**Technological Needs**

Key technologies include precision micro-thrusters to keep the spacecraft centered on the free-floating proof masses, frequency-stabilized lasers, high-rigidity telescope assemblies and optical benches, precision gravitational reference sensors and high-cadence phasemeters.

Extensive design studies and technology development in the U.S. and Europe over the past 20 years have led to a mature design concept for a gravitational wave surveyor with a peak sensitivity in the millihertz frequency range that is capable of detecting massive BH mergers throughout the observable universe. Such an instrument could also detect stellar-remnant captures by galactic BHs out to redshifts $z \sim 1$ and thousands of compact binaries in our galaxy. The LISA Pathfinder, a European-led technology development mission with U.S. participation, is currently undergoing final assembly and is scheduled to launch in mid-2015. Ground testing of the Pathfinder disturbance reduction system shows performance levels that are significantly better than the design goals. These translate to a gravitational wave sensitivity for a million-kilometer-scale array that is limited by the foreground signals from compact white-dwarf binaries in our galaxy. In other words, the detector's sensitivity limit is set by an interesting astrophysical source of gravitational waves and not by instrument noise.

The Gravitational Wave Surveyor would be made up of three spacecraft flying in a triangular configuration with laser links tracking the distance between each pair of spacecraft, allowing for the synthesis of three independent interferometry signals—two that measure the two polarizations of the waves, and a third "null channel" that is insensitive to low-frequency gravitational waves and can be used to distinguish between unexpected signals and instrument noise. By precisely measuring the wave patterns produced by colliding BHs, the surveyor will provide accurate determinations of their masses and spins. The direction and distance to the signals will be less well measured (several square degrees on the sky and tens of percent in distance for typical systems), making this first-generation surveyor more akin to a spectroscopic telescope. Full imaging capability, in terms of precise localization of the sources, will require multiple arrays.

### *CMB Polarization Surveyor*

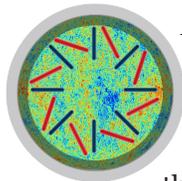

A number of concepts for next-generation space CMB missions have been explored, from lower-cost missions that could be pursued in the near term using technology that is largely proven, to more ambitious larger-scale missions that aim to extract all of the primordial information available in the CMB. The wavelength range of the surveyor will span much of the CMB blackbody spectrum, which peaks between 1–2 millimeters. Angular resolution is a key parameter since it determines the telescope size and drives mission cost. CMB B-mode polarization from inflationary gravity waves is expected to have a maximum amplitude on $\sim 2°$ angular scales, which requires only a rather modest telescope. However, access to finer angular scales is needed in order to characterize both the E-mode polarization to cosmic variance limits (see also **Figure 4.2**) and the B-mode polarization produced by gravitational lensing of the CMB to separate this effect from the inflationary signal. Such considerations lead to a 1- to 4-meter aperture telescope that will achieve few-arcminute resolution with high-throughput optics.

**Technological Needs**

The large gain in mapping speed that is needed relative to the Planck mission will be achieved through the use of large arrays of sensitive millimeter-wavelength detectors. Arrays of superconducting detectors have reached the fundamental quantum noise limit; therefore, achieving ≥ 100 times improvement in Planck mapping speed will require at least ~ $10^4$ detectors (~100 times more detectors than Planck, since mapping speed approximately scales with detector number). Arrays of ~ $10^3$ detectors are currently being used for ground-based and balloon-borne CMB instruments. The first array of ~ $10^4$ superconducting detectors is now being used for submillimeter observations on the James Clerk Maxwell Telescope in Hawaii and is of comparable size to detector arrays being planned for next-generation ground-based CMB experiments.

The performance and technological readiness of millimeter wavelength detector arrays is steadily advancing, as highlighted by the recent first flight of multiplexed superconducting detector arrays on a NASA balloon. The most mature superconducting detector array technology is transition-edge sensor (TES) detectors, which are measured using superconducting quantum interference device (SQUID) cryogenic amplifiers. Similar detectors and SQUID readout systems are being developed for X-ray observatories, and both fields will benefit from common technology advances. Continued investment in array technologies will improve sensitivity, fidelity, uniformity, operability, and multiplexability, and will also reduce their system-level impacts, e.g., mass and power. Parallel development in component technologies is also warranted to maximize mapping speed and retire risk for the CMB Polarization Surveyor. Critical component technologies include: detector array readout electronics, large cryogenic optics systems (ideally cooled below 4 kelvin to minimize noise), anti-reflection coatings, polarization modulators, optical filters, and subkelvin cryogenic systems for the detector arrays.

## Far-IR (FIR) Surveyor

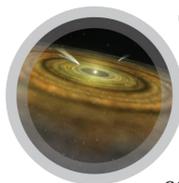

The scientific potential of the FIR spectral region was already understood 50 years ago and has been highlighted in each of the last four decadal surveys. In 1996, results from the NASA COBE mission confirmed the remarkable prediction that the universe is just as bright in the FIR as it is in the near-IR/optical band. Importantly, the FIR band allows access to the earliest stages of star and planet formation that are obscured by dust in the visible and near-IR bands. Furthermore, it contains fundamental spectral features of key species such as carbon (atomic and ionized) and water.

The 3.5-m ESA/NASA Herschel Space Observatory, launched in 2009, represents the first major FIR astrophysics space mission (the MIPS instrument on the 0.85-m Spitzer Space Telescope also provided FIR imaging out to 160 mm). The wide-field galaxy surveys performed by Herschel revealed the spectacular richness of the FIR sky and provided crucial information on the history of dust-obscured star formation over cosmic time. Future missions can surpass Herschel's capabilities in a number of ways. First, Herschel used passive cooling to achieve a telescope temperature of 70 kelvin; a large gain (> $10^3$) in sensitivity may be achieved by actively cooling the telescope to 4 kelvin. Spectroscopy represents a second area where large gains are possible, through the use of larger and more sensitive detector arrays combined with new instrument designs to enable multiobject spectroscopy, imaging spectroscopy, and/or tomographic spectral line mapping. FIR line mapping of star-forming regions with high spectral resolution using heterodyne detection is also relatively unexplored. Third, Herschel's angular resolution is limited by diffraction to around 15″ at 200 mm wavelength. This value, combined with the high surface density of FIR galaxies, means that Herschel's sensitivity in deep cosmic surveys is limited by spatial confusion: multiple FIR galaxies are blurred together at Herschel's resolution. Spatial confusion could be overcome by the use of a larger single-aperture telescope and/or the use of imaging spectroscopy to separate galaxies by redshift.

Ultimately, a multi-aperture interferometer will be required to achieve detailed subarcsecond FIR images of star-forming regions and galaxies.

**Technological Needs**

*Telescopes:* Herschel's 3.5-m diameter makes it the largest space telescope ever flown. Herschel used a monolithic primary mirror made by brazing together twelve silicon carbide segments. Significant progress is possible on a relatively short timescale using modestly larger apertures (4–6 m) cooled to 4 kelvin and equipped with wide-field imagers and spectrometers. Further in the future, a large single-aperture (10–20 m) FIR telescope would need to use a segmented approach. Compared to the shorter wavelength bands (IR through X-ray), the non-cryogenic technical requirements for FIR telescopes are rather modest. Thus, the FIR band may provide a lower-risk opportunity to develop and demonstrate advanced concepts for segment deployment, phasing, and/or robotic assembly while simultaneously offering a strong science return.

*Interferometry:* A 10-m diffraction-limited single-aperture telescope has an angular resolution of 2.5″ at $\lambda$ = 100 μm. A variety of sparse-aperture interferometer concepts have been explored for reaching subarcsecond resolution in the FIR, typically involving telescopes with 1–4 m apertures cooled to 4 kelvin. The telescopes may be positioned using mechanical means such as booms or tethers, or through formation flying. As for large single-aperture telescopes, the technical requirements for interferometry in the FIR are not as demanding as for shorter wavelength bands, so FIR interferometry may again be a logical starting point that provides a useful training ground while delivering crucial science.

*Detector arrays:* Detector technology has historically been a major limiting factor for FIR astrophysics, especially at wavelengths beyond the reach of doped silicon detectors ($\lambda$ > 40 μm). Fortunately, this situation is changing rapidly thanks to advances in superconducting detector technologies similar to those being developed for millimeter-wavelength CMB experiments and for energy-resolved single photon detection at optical and X-ray wavelengths. These advances are leading to arrays that are both far larger and much more sensitive than were used on Herschel. Larger formats, higher sensitivity, better multiplexing, reduced cosmic ray susceptibility, and lower system impacts (mass, power, operating temperature, etc.) are all important development priorities. Continued laboratory development and full system demonstrations on ground-based and suborbital platforms (SOFIA and balloons) are needed.

*Component technologies:* The priorities here include improved subkelvin focal-plane coolers, space-qualified 4 kelvin mechanical coolers, detector readout electronics, and wide-field or multi-beam spectrometers.

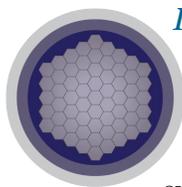

## Large UV/Optical/IR (LUVOIR) Surveyor

Ever since Galileo first turned his telescope toward the sky, optical telescopes have carried an indelible legacy, transforming our scientific understanding of the universe and our place within it. Technological advancements since then have allowed us to continue expanding the regions of the electromagnetic spectrum that we are able to probe with today's telescopes, and to study ever more exotic phenomena. But the fact remains that the vast majority of the baryonic matter in the universe is contained within stars and clouds of gas. Many of the science questions we have about stars, clusters of stars, galaxies, and even clusters of galaxies, can only be answered by studying these structures in the wavelength regimes where their signatures are the strongest—the near-infrared, optical, and ultraviolet wavelengths.

Today, HST is the most famous telescope operating at these wavelengths. The order-of-magnitude increase in sensitivity and angular resolution afforded by the space-based location of HST contributed to such exciting new discoveries as the existence of dark energy and the accelerating universe, the knowledge that most massive galaxies host a central supermassive BH, and images of a planet orbiting around a star other than the Sun. Another order-of-magnitude increase in sensitivity and resolution will provide a step forward in answering remaining questions that are beyond the capability of HST to address. For example, limitations in sensitivity and angular resolution have ensured that only the very brightest individual stars can be studied in galaxies other than the Milky Way, and so our knowledge of galaxy star-formation histories is limited only to the galaxy we live in. Limitations in ultraviolet sensitivity have severely handicapped our ability to explore the cycles of gas flowing into and out of galaxies. The ability to image the immediate surrounding of nearby bright stars with a large sensitive telescope equipped with a coronagraph to suppress stellar glare would allow unprecedented study of exoplanets similar to Earth, and thus potentially habitable. Last but not least, an 8-16 m aperture would dramatically enhance detection of Earth-sized planets to statistically significant numbers, and allow in-depth spectroscopic characterization of the most favorable targets. Furthermore, an order-of-magnitude increase in our capabilities will pave the way for exciting new discoveries that have never been imagined.

Such a prime space-based LUVOIR observatory could have a wide range of instruments capable of performing a variety of imaging and spectroscopic measurements, including:

- wide-field imaging across a broad spectral range
- high-contrast imaging and spectroscopy of circumstellar environments
- single-object spectroscopy across a broad spectral and resolution range
- multi-object spectroscopy of thousands of objects
- diffraction-limited spatially-resolved spectroscopy
- astrometry of nearby bright stars (to infer exoplanet masses) and of more numerous fainter stars (galaxy dynamics)

The following requirements have evolved from the science goals described above.

*Telescope diameter*: Scientific return in several areas is a steep function of telescope diameter, providing both a larger collecting area and higher angular resolution. The telescope should therefore be as large as technologically realistic within the Formative Era. We notionally consider an 8–16 m telescope, somewhat larger than current ground-based telescopes and significantly larger than HST (10 m would be ~100× more sensitive than HST in the background-limited regime), and also significantly more capable than the JWST. The diffraction limit of a 16-m telescope would, for example, provide angular resolution of 8 milliarcseconds in the optical at $\lambda = 6000$ Å, allowing structures with physical sizes of only a parsec to be imaged out to distances of 25 Mpc. It would also allow spectral characterization of Earth-like planets at a distance 10 pc out to ~3 microns with realistic assumptions on coronagraph performance.

*Wavelength coverage:* While the full IR-to-UV spectrum from ~10 microns to 91 nm (Lyman limit) is rich in information content, wavelength coverage will be strongly constrained by technology considerations such as coatings, cooling and detectors. A telescope and instruments supporting imaging and spectroscopy from near-IR to near-UV would be very compelling for a wide range of measurements. The near-IR to optical is, e.g., particularly rich in spectral signatures for characterization of exoplanets. Extending wavelength coverage outside the near-IR to near-UV will be contingent on technology and programmatic considerations, such as the deployment of other missions more narrowly aimed at spectral ranges outside the near-IR to near-UV.

### Technological Needs

Several new technologies need to be developed to support such a mission. These include:

- ***Optics deployment and co-phasing***: An 8–16 m telescope will require a segmented approach, and advanced options for optics deployment, such as robotic assembly may prove attractive. The wavefront accuracy and stability requirement is particularly challenging for exoplanet imaging, and may drive aperture format. For example, a single large segment may be surrounded with smaller segments, and the most demanding observations (high-contrast imaging at short wavelength) may only use the central large segment.

- **Coatings**: Achieving high reflectivity from UV to near-IR is very challenging and will require improvements in coatings. Availability of such coatings may drive the telescope wavelength range.

- **Detectors**: Advances in detector technologies can have a major impact of the performance and scientific reach of the telescope. Large-format high-sensitivity detectors will be required across a wide spectral range (IR to UV) to take full advantage of the telescope's wide field of view and unprecedented angular resolution. Direct imaging of exoplanets would greatly benefit from low-noise photon-counting detectors. Energy-resolving detector arrays such as the microwave kinetic inductance detectors (MKIDs) would be a game-changing capability, considerably simplifying the instruments' optical design and allowing a significant gain in scientific return: they could allow simultaneous spectroscopic imaging of exoplanets and wavefront control for a high-contrast imaging coronagraphic instrument, and could enable multi-object spectroscopy to a scale not possible with conventional detectors

- **Starlight suppression systems**: To overcome the ~$10^{10}$ contrast between an Earth-like planet and its host star, advanced optical systems that suppress the starlight within the instrument will be critical. In addition to the optical design challenges associated with this goal, wavefront control will be essential to maintain the required subnm wavefront stability within a coronagraph instrument. To enable high contrast imaging, a low-vibration telescope will need to be designed, requiring technology development in vibration isolation, low-vibration reaction wheels, and active sensing/control of vibrations. If such efforts prove to be too challenging or costly, a fallback option is an external occulter, although this approach becomes extremely challenging (occulter size, distance, amount of energy required for repointing) for large telescope diameters.

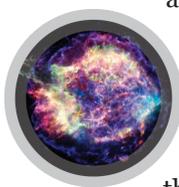

## *X-ray Surveyor*

X-ray astronomy has developed in a spectacular fashion over the past 50 years, culminating in NASA's Chandra and ESA's XMM-Newton space observatories, both launched in 1999. Imaging in the hard X-ray band (6–80 keV) has also been opened up recently with the 2012 launch of NASA's NuSTAR mission. X-ray observations probe extreme astrophysical environments large and small, from the hot intergalactic gas comprising and surrounding clusters of galaxies to the accretion regions surrounding BHs. The angular resolution needed to spatially resolve these regions spans a huge range, from tens of arcseconds for clusters down to several microarcseconds for BHs. Spectroscopic capability is also key as it allows detailed studies of plasma states and interactions, motion, substructure, and turbulence.

The optics used to focus X-rays are a central technology challenge for X-ray astronomy. To date, all X-ray focusing telescopes for astronomy—with the exception of some solar telescopes—have relied on grazing-incidence reflective optics, which use a pair of nested reflective shell assemblies to redirect the paths of X-ray photons by angles of order a degree or less in order to bring them to a focus. This small deflection angle results in a large ratio between the effective collecting area and the physical area of the optics and makes large collecting areas difficult to achieve. The angular resolution of the telescope is set by a combination of small-scale surface errors and large-scale figure errors, which are influenced by the design and by the fabrication processes. For example, Chandra has a 1 keV collecting area of 0.08 m² and achieves an angular resolution of 0.5″ HPD (0.2″ FWHM); meanwhile XMM's three mirror assemblies have a combined collecting area of 0.4 m², or 5× more than Chandra, but with an angular resolution of 15″ HPD (6″ FWHM) that is 30× worse than Chandra.

The required collecting area and angular resolution obviously depend strongly on the science goals. A large gain in collecting area over Chandra and XMM, combined with an angular resolution under 1 arcsecond, represents an excellent choice for the next general-purpose observatory, especially when coupled to energy-resolving detectors that provide high-throughput spectroscopy. A mission with ≥ 3 m² of effective collecting area and at this level of imaging with state-of-the-art detectors would provide nearly two orders of magnitude improvement on Chandra and be technically feasible given suitable developments over a 15-year period. The mission should provide high-resolution spectroscopy over a broad energy band with a resolving power of a few thousand and a field of view of at least 5 arcminutes. Some key challenges to realizing such a mission are given below.

**Technological Needs**

*Optics*
- ***Mirrors***: The large effective-area requirement of the X-ray Surveyor precludes the use of the thick Zerodur shells, meticulously figured and polished, as used in Chandra. Therefore, a technology is needed that can produce a large number of very thin shells relatively inexpensively and, as these will likely not meet the axial figure requirement for subarcsecond imaging, a technique is needed to correct these figures. Several technologies are under development that could provide this correction. These include active figure control via piezoelectric or magnetostrictive methods, or coating techniques to selectively deposit material to smooth out figure imperfections.

  An alternative approach is to utilize a very rigid structure, as with the silicon pore optics development in Europe. Significant challenges remain with this approach, which is currently at the ~ 15-arcsecond resolution level.

- ***Coatings***: The typical iridium coating used on many X-ray mirrors is highly stressed and this presents a problem for very-thin-shell optics, which are easily distorted. The challenge here is to deposit low-stress coatings or correct for stresses after the coatings have been applied while still meeting the requirements of low surface roughness and high bulk density.

- ***Assembly and alignment***: Major challenges for thin shell optics are mounting, alignment and bonding. Thin shells have little rigidity, particularly if a segmented approach is chosen, and are easily perturbed during mounting and assembly. While post-attachment figure correction may be possible, this remains a key development requirement in realizing subarcsecond optics. If assembled in a modular form, primary-to-secondary alignment is critical, but intramodule coalignment should be straightforward.

- ***Metrology, calibration, and verification***: Thin-shell mirrors will have significant distortion under Earth's gravity, when oriented horizontally, and so a vertical test facility is desirable. For adjustable optics, a challenge will be to optimize the figure both during ground calibration and again after launch.

*Detectors*

The ideal focal plane detector would combine the pixel size of a CCD, to support subarcsecond optics, with the spectroscopic resolution of a superconducting microcalorimeter at higher energies (few keV) and the resolution of gratings at lower energies. This may be achievable with future microcalorimeter arrays, or may require a compliment of microcalorimeter and grating instruments.

- ***Microcalorimeter arrays***: Current microcalorimeter pixel sizes will need to be reduced by close to an order of magnitude and increased in pixel number by several orders of magnitude. This will require progress in the development of large arrays of small pixels with high energy resolution, high yield, and uniformity. Array sizes approaching $10^5$ pixels will be required. Work will be required in fabrication techniques, sensor layout, absorber design, heat-sinking, shielding, wiring, and integration into appropriate cryogenic cooling systems.

- **Microcalorimeter readout**: Substantial advances in multiplexers for these large arrays of microcalorimeters will need to be demonstrated. Either GHz frequency-division multiplexing or code-division multiplexing, or perhaps a combination of these, would be needed.

- **Gratings:** Development of lightweight, high-efficiency and high-dispersion diffraction gratings with high-line densities, efficiencies, and yield are needed. Work will be required to improve fabrication processes and validate grating performance.

## 6.2 The Visionary Era

The notional missions listed in this section are beyond the ~ 30-year time frame of the Formative Era, but they require technology to be developed in the next 30 years and help establish a long-term vision for NASA in which earlier and smaller missions will play an essential role.

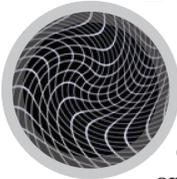

### Gravitational Wave Mapper

It is always difficult to predict the direction of an entirely new field of astronomy before the first detections are made in that observational window. Increasing the sensitivity is a natural path that opens new discovery space, but the game-changer will be the advent of multi-element arrays of interferometers. With two or more widely separated detectors the angular resolution can be significantly improved, particularly if the peak sensitivity is moved to higher frequencies (~ 0.1 Hz), and the overall sensitivity of the instrument is increased by using more powerful lasers, larger telescopes and improved gravitational reference sensors. With two colocated arrays, it becomes possible to detect and characterize the stochastic background of gravitational waves produced by violent events in the early universe and thus to provide unique constraints on physical processes that occur at times far earlier than can be probed with light.

The transition to multidetector arrays will transform gravitational wave astronomy into an imaging science with the ability to localize sources to individual galaxies and enable the routine discovery of electromagnetic counterparts. Such multispectrum observations will provide a new way of mapping the cosmos, and potentially reveal the exotic physics that powers the accelerated expansion of the universe. The enhanced sensitivity and consistency tests provided by a multidetector Gravitational Wave Mapper will allow us to explore the unseen universe, and fully realize the discovery potential of this new branch of astronomy.

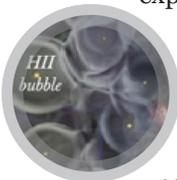

### Cosmic Dawn Mapper

One of the most challenging periods to study in the history of the universe is the so-called dark ages, which began when the universe cooled enough to form neutral hydrogen and helium atoms 400,000 years after the Big Bang and ended when protogalaxies and black holes produced enough light to ionize the universe around a billion years after the Big Bang. The dark ages and the epoch or reionization, when the first stars (cosmic dawn), black holes, and galaxies were formed, can be studied by measuring light given off by neutral hydrogen at a wavelength of 21 cm. Because we are observing light produced long ago, this 21-cm light is red-shifted to wavelengths between 2 and 20 meters, which fall into wavelengths of artificial and atmospheric radio signals on Earth. To fully map the neutral gas in this period, an array of thousands of radio antennas separated from meters to tens of meters would be needed on the far side of the moon, shielded from all the radio noise produced on Earth. The angular resolution and wide frequency range would be required to disentangle this cosmic signal from the very bright radio foreground emanating in our own galaxy, with the goal of making a three-dimensional map of the neutral gas from the epoch of reionization all the way back to the dark ages. There are many challenging aspects of such a mission in areas of antenna and radio-receiver design and deployment, such as, e.g., robotics, computation, radiation hardness, and distributed power.

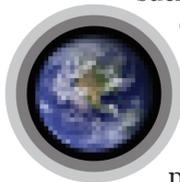

### Exo-Earth Mapper

A large optical/near-IR space-based interferometer would spatially resolve nearby habitable planets, delivering multicolor images and even spectra over the face of a potentially or known life-bearing planet. Ultimately, this kind of information will be crucial for analyzing and understanding evidence for life, since, for example, finding biosignatures that are identified with land features, or a chlorophyll-like feature ("red edge"), could be decisive evidence for advanced life—beyond the single-cell phase that occupied a large fraction of Earth's history. To fully exploit this capability, the facility would have to be sufficiently large to produce such measurements in only a few hours of integration, since rotation of the planet will (over longer observation times) dilute such signatures. While such a challenging mission is clearly beyond the 30-year timescale, it appears more feasible than travel (manned or unmanned) to other habitable planets, and is, therefore, the most credible option to map the surfaces of habitable planets.

There are three fundamental parameters for a multitelescope interferometry facility: the maximum separation between the telescopes, the total light collecting area of the telescopes combined, and the number of individual telescopes. For the notional architecture described here, we assume a goal of a 30 × 30 element map at optical wavelengths (0.3 to 1 micron) of an Earth located 10 parsecs (33 light-years) away. To achieve the needed spatial resolution at all wavelengths, the maximum separation between the telescope units must be ~ 370 kilometers. A total collecting area of around 500 square meters will provide the sensitivity required for R ~ 100 spectroscopy of every spatial element within a day of exposure time. The number of individual telescopes needed depends on the exact details of the observations and observing strategy, but not more than 20 units will be necessary. In this case, each telescope unit would need to have a diameter of ~ 6 meters (larger if fewer telescopes are used). Other architectures capable of achieving our science goals could be envisaged, but the above provide a sense of the technical challenges involved.

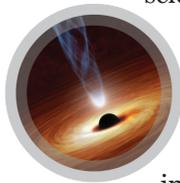

### Black Hole Mapper

A visionary mission for X-ray astrophysics could either entail a major leap in collecting area, to study the cosmic dawn era with high-throughput and high-energy resolution, or a major leap in imaging resolution, opening the door to, among other science, true imaging of the environment around a BH event horizon. At the end of the Formative Era, the community should be informed by the insight gained from the X-ray Surveyor, as to which of these options (or even a completely different one) will yield the greatest science at the visionary stage. The Black Hole Mapper is presented here as a visionary notional mission, to complement the X-ray Surveyor. We stress, however, that improvements in imaging resolution, telescope area, and detector energy resolution are all key aspects of future missions. All these technologies need, therefore, to continue being developed, to enable selection in the future of a visionary mission for the best science.

An X-ray interferometer with (sub)microarcsecond resolution would allow us to take the next giant step in our understanding of supermassive BHs. Such an observatory would allow us to see the shadow of the event horizon in our galactic center and in the giant elliptical galaxy M87, strongly complementing studies in the mm-band and testing one of the most fundamental predictions of the general relativistic theory of black holes. Of equal importance, an X-ray interferometer will produce velocity-resolved maps of the innermost accretion disk, jet, and central hole/shadow in nearby active galactic nuclei. Such observations provide the ultimate view of how these powerful astrophysical engines work.

X-ray interferometry poses significant technical challenges. To image event horizons will require a space array of optics about a kilometer in diameter with focal plane detectors spaced thousands of kilometers farther away. The elements of the array and the lateral position of the detector must be held (or at least monitored) to about the size of a detector pixel (~15 microns) across this large stretch of space. The stability and figure of the optical elements must similarly be held to interferometric standards (~5 nm). Even more formidable is the pointing requirement, wherein we need information on the direction of the array with respect to the celestial sphere to below a microarcsecond. This can be accomplished with either an ultrahigh-stability gyroscopic reference or monitoring of nearby point sources in the sky, possibly interferometrically as well. While the list of technologies is daunting, there is no single show-stopper. A targeted program of technology development could likely greatly ease the difficulties.

## 6.3 Cross-cutting, Game-changing Technologies

In addition to the mission-specific technologies listed above, several technologies reach across multiple notional missions and can either facilitate or enable a large part of the astrophysics roadmap. The following are some examples of cross-cutting technologies and ideas of potential game-changing capabilities. We expect that new technologies will come along before the Visionary Era, which could dramatically expand the range of possibilities for missions.

### *New Technology Mirrors, On-orbit Fabrication and Assembly Technologies*

Except for the enormous sophistication of the HST, Spitzer, and JWST, our methods of building space telescopes have not progressed much beyond building and testing a ground-based telescope and rocketing it into space. This is particularly problematic because a telescope designed for zero-gravity is extremely difficult to test at 1 g, where it will never operate, to say nothing of contamination issues, and the vulnerability of a precisely assembled hardware to survive a violent ride into space. The key to bigger and better space telescopes may rely, instead, on assembling and testing telescopes on-orbit, from subcomponents produced on Earth, and perhaps in the visionary period, from actually producing many components in space using so-called 'smart materials,' advanced robotics, and possibly astronauts.

> **Several of the notional missions listed in our roadmap rely on interferometry to answer key science questions, from radio to X-rays.**

A particular hurdle to our ambitious science goals is the production of 10-m or larger apertures of 1/10-wave quality for the FIR-to-optical regime. The next generations of space telescopes will have light-collecting surfaces that are "active," meaning they will achieve their required high-accuracy figures only after they are part of a complete system and ready for operation. Beyond this, even more advanced schemes, for example, combining in space lightweight 2D frames and low-mass flexible, reflecting membranes, would enable huge collecting areas at a small fraction of the cost of present technologies. Such breakthroughs should be developed for FIR space telescopes and refined in subsequent development to be suitable for optical and even UV space telescopes.

3D printing was invented ~30 years ago. Today, 3D printers are being used to manufacture a wide range of products, from machine parts to human transplants and even houses. Current 3D printers rely on gravity; however, a zero-gravity 3D printer is now being developed for the space station. In the future, 3D printers on the moon could use lunar material to fabricate nonconductive structures; 3D printing of ice from water has the potential to achieve nanometer-scale precision. A lunar fabrication facility could e.g., directly support the development of a large array of 21-cm telescopes on the far side of the moon or enable great leaps forward and cost savings by using lunar material for some of the largest and most costly-to-launch system components for future space telescopes.

*Interferometry*

Interferometric techniques allow multiple telescope beams to be coherently combined to reach angular resolutions not attainable with single aperture telescopes. Several of the notional missions listed in our roadmap rely on interferometry to answer key science questions, from radio to X-rays. All notional missions in the Visionary Era are interferometers, and technology maturation of interferometric techniques is thus highly relevant to realizing the science vision, especially on longer timescales. Interferometry requires the following techniques/challenges to be mastered:

- precision laser metrology
- formation flying
- beam combination, possibly with delay lines
- aperture synthesis techniques: beam combiner optimization, data analysis techniques

Interferometry has historically progressed from longer wavelengths, where technological challenges are less extreme, to shorter wavelengths. Most radio telescopes are now interferometers, and technological advances are steadily enabling interferometry at shorter wavelengths. Astronomical applications of interferometry will also benefit from advances in computing power and detectors.

## 6.4 Science Summary

|  | Formative Era | | | | | Visionary Era | | | |
|---|---|---|---|---|---|---|---|---|---|
|  | GW Surveyor | CMB-pol Surveyor | FIR Surveyor | LUVOIR Surveyor | X-ray Surveyor | GW Mapper | Cosmic Dawn Mapper | ExoEarth Mapper | Black Hole Mapper |
| Demographics of planetary systems |  |  | ▒ | █ |  |  |  | ▒ |  |
| Characterizing other worlds |  |  | ▒ | █ |  |  |  | █ |  |
| Our nearest neighbors and the search for life |  |  |  | █ |  |  |  | █ |  |
| The origins of stars and planets |  |  | █ | █ |  |  |  | █ |  |
| The Milky Way and its neighbors | ▒ | ▒ | █ | █ | █ | ▒ |  | ▒ | █ |
| The history of galaxies | █ |  | █ | █ | █ | █ | █ | ▒ | █ |
| The origin and fate of the universe | ▒ | █ |  | ▒ | ▒ | █ | █ |  |  |
| Extremes of matter and energy | █ | ▒ |  |  | █ | █ |  |  | █ |
| Ripples of space-time | █ | ▒ |  |  |  | █ |  |  |  |

█ Primary Goals
▒ Secondary Goals

## 6.5 Technology Summary

| | Formative Era | | | | | Visionary Era | | | |
|---|---|---|---|---|---|---|---|---|---|
| | GW Surveyor | CMB-pol Surveyor | FIR Surveyor | LUVOIR Surveyor | X-ray Surveyor | GW Mapper | Cosmic Dawn Mapper | ExoEarth Mapper | Black Hole Mapper |
| Formation flying | | | | ◐ | | ● | | ● | ● |
| Interferometry: precision metrology | ● | | ◐ | | | ● | ● | ● | ● |
| X-ray interferometry | | | | | | | | | ● |
| High-contrast imaging techniques | | | | ● | | | | ● | |
| Optics deployment and assembly | | | ● | ● | ◐ | | ● | ● | |
| Broadband coatings | | ● | | ● | | | | | |
| X-ray optics | | | | | ● | | | | ● |
| Large-format detector arrays | | ● | ● | ● | ● | | | | ● |
| New detector capabilities | | | ● | ◐ | ● | | | | ● |
| Cryogenics | | ● | ● | ◐ | ● | | | | |

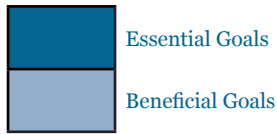

■ Essential Goals
■ Beneficial Goals

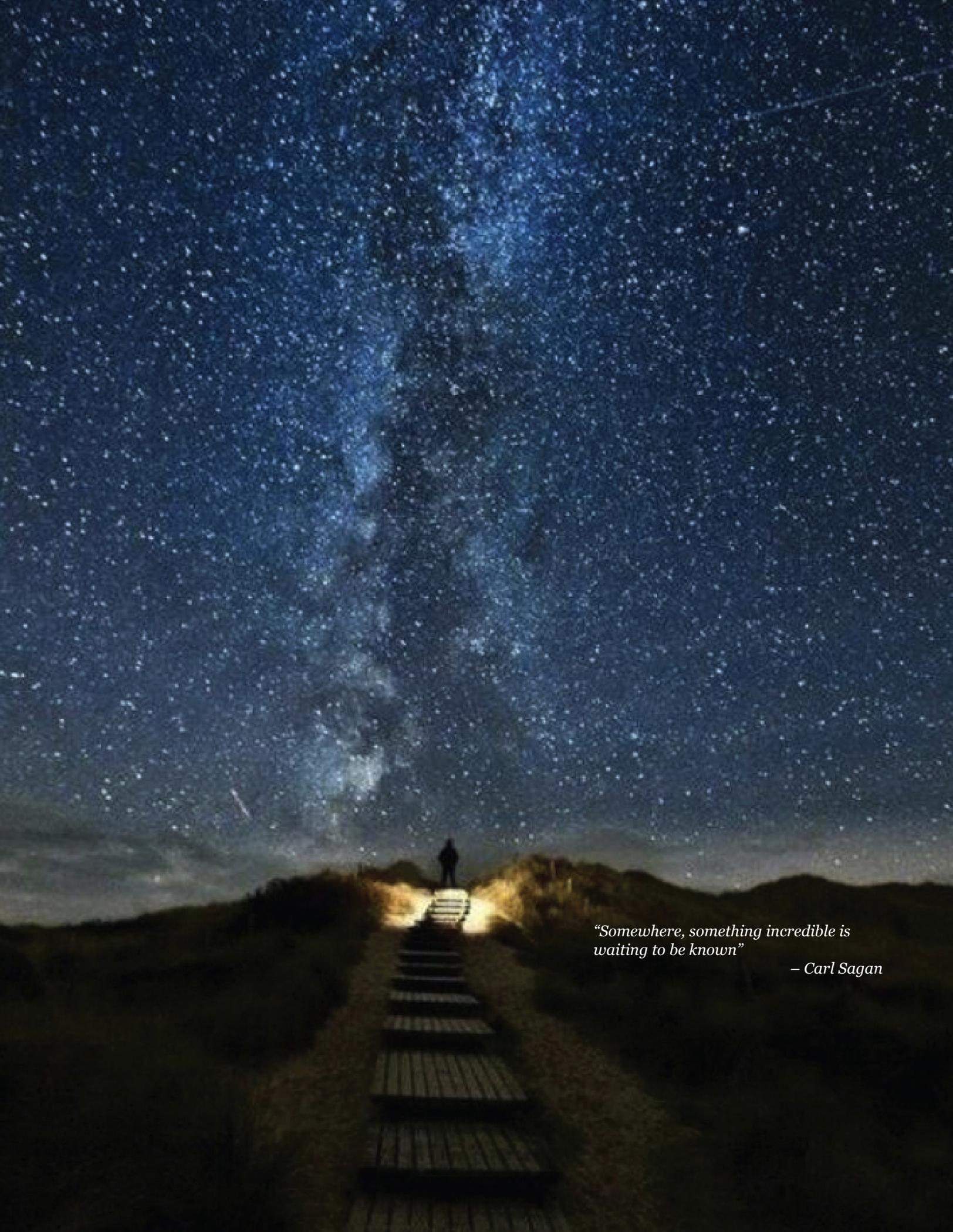

"Somewhere, something incredible is waiting to be known"
— Carl Sagan

# 7 Daring Visions

### Are we alone?

### How did we get here?

### How does the universe work?

Over the last half-century, we have made startling progress toward answering these enduring human questions on the basis of empirical data and physical theory. NASA's contributions—from Chandra, COBE, HST, Kepler, WMAP, solar system probes, lunar and Mars rovers, and dozens of others missions—already rank among the greatest achievements of the agency and of the U.S. science community. NASA Astrophysics is now poised to embark on its most exciting era, building on remarkable scientific and technological advances to vastly expand our understanding of the cosmos, and of humanity's place within it.

Where will the next 30 years of NASA astrophysics missions take us? The answer depends equally on our own ambition and creativity and on the surprises that nature has in store beyond the borders of our current knowledge. The preceding sections of this roadmap outline a route forward and provide a platform for Daring Visions:

*We will analyze the light from hundreds of planets around nearby stars,* searching for traces of oxygen and complex molecules that suggest an atmosphere shaped by life. We will map the surfaces of other Earths, revealing a jigsaw puzzle of continents and oceans that tell of tectonic processes, like the ones that form our own mountains and river valleys. As the continents of a distant world rotate past our view, subtle shifts of light may show that they are carpeted with plant life. We may even observe the change of seasons on a planet that orbits a sun many light-years away.

*We will image the disks of gas and dust around young stars in fine detail,* using both visible and infrared light. We will trace the signs of newborn planets that carve structure into their surroundings or stir belts of asteroids into grinding collisions. By piecing together images of hundreds of individual systems, each of which represents a snapshot frozen in time, we will reveal the history of planet formation. We may see the vapor released from billions of swarming comets and know that they are cratering moons and filling the oceans of rocky inner planets.

*We will measure the spins of hundreds of black holes* using X-ray telescopes and gravitational wave observatories, probing both the remnants of long-dead stars in the Milky Way and the supermassive monsters that lie at the hearts of other galaxies. From these spins we will learn about the collapse of stellar cores that have exhausted their nuclear fuel, about the chaotic flows of superheated gas that allow black holes to grow and quasars to shine, and about the fates of black holes after the gravitational mergers of their parent galaxies. We expect these observations to show incontrovertible evidence of event horizons, the space-time curtains that forever shield a black hole's interior from any external view. But we could find that Einstein's theory of relativity breaks down at the border of causality, and that in gravitational extremes nature creates an object that is even stranger than a black hole.

*We will characterize in detail the cradles of star formation in the Milky Way*, extend our sampling of the coupled gas-star systems in nearby galaxies, and trace the star-formation history of the universe as a whole, back to the epoch of the first stars. We will follow the cycle of gas from the primordial intergalactic medium to the regions of star formation and back out through stellar explosions. We will reveal how elements were forged in the cauldrons of stellar interiors.

*We will observe the cosmic dawn from radio antennas on the far side of the moon*, where they are shadowed from the electronic chatter that is inescapable on Earth's surface. These radio maps will trace an intricate structure of ribbons, tunnels, and bubbles in the 200-million-year-old universe as the light of the first stars burns through the fog of absorbing hydrogen that fills intergalactic space.

*We will listen to the cosmic symphony of gravity waves* using ears spanning a hundred million kilometers and designed with subatomic sensitivity. Within this symphony we will pick out the crescendos of neutron stars that crash together in galaxies billions of light-years away, mapping the rise and fall of massive stars and measuring the accelerating expansion of the universe with unprecedented accuracy. In the background hum of this symphony, we may detect the quantum noise of the inflationary epoch or the reverberations of colliding domain walls that mark the cracks of primordial symmetries of nature.

These visions flow from synthesizing current knowledge across many fields of astrophysics and extrapolating to a possible future, based on attainable progressions of technological development and scientific discovery. Realizing these opportunities will require innovation, perseverance, and sustained national investment at the frontiers of science and engineering.

Past NASA investments have created the extraordinary missions that placed humans on the moon, marked our presence at every planet in the solar system, and extended our view to the far reaches of the universe. They have produced spinoff technologies that permeate everyday life, in materials, transportation, agriculture, security, and medicine. NASA research constantly presses outward on the limits of the possible, posing challenges that inspire the most creative minds and the most innovative companies. This forward-looking spirit, seeking out and rising to the sharpest of challenges, sparks the creation of revolutionary technologies. Inventions that we have not even dreamt of today will, if history is any guide, be driving forces of Earth's economies a century from now.

Generations of explorers before us have mapped the continents and the oceans and studied the heavens to unlock the secrets of nature. Only in the NASA era have we begun to visit other worlds and to explore the deeper cosmos from observatories above Earth's atmosphere. In the continuing quest to trace the history of the universe and understand its workings, NASA will inspire people from all backgrounds and all walks of life.

The vision presented in the previous chapters is an expression of the work of hundreds of dedicated astronomers over the last several decades, distilled and developed by the roadmap team. The team was asked to dream big and far out into the future, limited only by our imagination and our extrapolations of technological progress, and create a compelling vision for the NASA Astrophysics research programs including very ambitious and innovative notional missions—indeed, Daring Visions.

# ACRONYMS

| | |
|---|---|
| AAS | American Astronomical Society |
| aLIGO | Advanced Laser Interferometer Gravitational Wave Observatory |
| ALMA | Atacama Large Millimeter/submillimeter Array |
| AMS | Alpha Magnetic Spectrometer |
| APS | American Physical Society |
| AU | astronomical unit |
| BAO | baryon acoustic oscillations |
| BH | black hole |
| CCD | charge-coupled device |
| CERN | European Organization for Nuclear Research (Conseil Européen pour la Recherche Nucléaire) |
| CMB | cosmic microwave background |
| COBE | Cosmic Background Explorer |
| DOE | Department of Energy |
| ELT | Extremely Large Telescope |
| EMRI | extreme mass-ratio inspirals |
| ESA | European Space Agency |
| FIR | far-infrared |
| MIPS | Multiband Imaging Photometer for Spitzer |
| FIRAS | Far Infrared Absolute Spectrophotometer |
| FWHM | full width at half maximum |
| GCM | global circulation model |
| GR | general relativity |
| GW | gravitational wave |
| HPD | half power density |
| HST | Hubble Space Telescope |
| IMF | initial mass function |
| IPAC | Infrared Processing and Analysis Center |
| IR | infrared |
| ISM | interstellar medium |
| JEM-EUSO | Japanese Experiment Module Extreme Universe Space Observatory |
| JWST | James Webb Space Telescope |
| WFIRST-AFTA | Wide-Field InfraRed Survey Telescope   Astrophysics Focused Telescope Assets |
| LBTI | Large Binocular Telescope Interferometer |
| LIGO | Laser Interferometer Gravitational Wave Observatory |
| LISA | Laser Interferometer Space Antenna |
| LSST | Large Synoptic Survey Telescope |
| LUVOIR | large ultraviolet-optical-infrared |
| MKIDS | microwave kinetic inductance detectors |
| NASA | National Aeronautics and Space Agency |
| NICER | Neutron Star Interior Composition Explorer |
| NSF | National Science Foundation |
| NuSTAR | Nuclear Spectroscopic Telescope Array |
| RV | radial velocity |
| SDSS | Sloan Digital Sky Survey |
| SKA | Square Kilometer Array |

# Acronyms

SMBH....................supermassive black hole
SN..........................supernova
SOFIA....................Stratospheric Observatory for Infrared Astronomy
SQUID...................superconducting quantum interference device
STEM ....................science, technology, engineering, and mathematics
STScI.....................Space Telescope Science Institute
TES........................transition-edge sensor
TESS......................Transiting Exoplanet Survey Satellite
UV .........................ultraviolet
VLA .......................Very Large Array
WMAP...................Wilkinson Microwave Anisotropy Probe
XMM-Newton.......X-ray Multi-Mirror Mission

# Roadmap Team

Chryssa Kouveliotou (NASA/MSFC), Chair
Eric Agol (University of Washington)
Natalie Batalha (NASA/Ames)
Jacob Bean (University of Chicago)
Misty Bentz (Georgia State University)
Neil Cornish (Montana State University)
Alan Dressler (The Observatories of the Carnegie Institution for Science)
Scott Gaudi (Ohio State University)
Olivier Guyon (University of Arizona/Subaru Telescope)
Dieter Hartmann (Clemson University)
Enectali Figueroa-Feliciano (MIT)
Jason Kalirai (STScI/Johns Hopkins University)
Michael Niemack (Cornell University)
Feryal Ozel (University of Arizona)
Christopher Reynolds (University of Maryland)
Aki Roberge (NASA/GSFC)
Kartik Sheth (National Radio Astronomy Observatory/University of Virginia)
Amber Straughn (NASA/GSFC)
David Weinberg (Ohio State University)
Jonas Zmuidzinas (Caltech/JPL)
Brad Peterson (Ohio State University), APS Chair
Joan Centrella (NASA Headquarters), APS Executive Secretary

Roadmap Support

Francis Reddy (NASA/GSFC, Syneren Technologies Corporation)
Patricia Tyler (NASA/GSFC, Syneren Technologies Corporation)

# Acknowledgements

The visions presented in this document benefited from many interactions with members of the scientific community. We drew inspiration from submissions of white papers solicited by the roadmap team, a subset of which were presented in a **Town Hall**. The team is especially grateful to the following colleagues who delivered invited presentations to the roadmap team on specific topics outside the Town Hall: Rachel Bean, Roger Blandford, Jack Burns, Victoria Meadows, Sara Seager, Jim Ulvestadt, and Martin Weisskopf. We also solicited presentations from the Chief Scientists (CS) and the Chief Technologists (CT) of the three Astrophysics Division Programs: Dominic Benford (CS: COR), Ann Hornschemeier (CS: PCOS), Wes Traub (CS: ExEP), Mark Clampin (CT: COR+PCOS), and Peter Lawson (CT: ExEP). Members of the roadmap team consulted individual colleagues for specific scientific input. In particular we would like to thank Sean Brittain, Jacqueline Hewitt, Steve O'Dell, Brian Ramsey, Denise Smith, Harvey Tananbaum, and Michelle Thaller.

A Red Team consisting of Jean Cottam, Debra Elmegreen, Amy Kaminski, Tom Koshut, and Joel Parriott reviewed the draft manuscript and provided extremely useful comments leading to significant improvements of this report. Additional critical comments and suggestions were provided by the Astrophysics Subcommittee (APS) members and from the Director of the Astrophysics Division, Paul Hertz. The unwavering support of the APS Executive Secretary, Joan Centrella, was instrumental to our success in delivering on time and on target. Special thanks go to the APS Chair, Brad Peterson, who joined many of our abundant telecon sessions to provide guidance.

While many images used in this report were obtained from outside sources, the document is greatly enhanced by the beautiful original graphics developed for us by Pat Tyler, who captured our visions by drawing upon her vast expertise and an extensive reservoir of NASA images. Francis Reddy improved the entire manuscript with his skillful editing and enhanced our graphical presentations with insightful suggestions.

We would also like to acknowledge the PCOS/COR Program office at Goddard and the ExEP Program Office at JPL for supporting our face-to-face meetings and several other activities.

Last but not least we are thankful for the invitation to envisage this journey, and we hope that our roadmap helps inspire the U.S. scientific community to turn these visions into reality.

# Acknowledgements

# Astrophysics Roadmap Team Charter

An Astrophysics Road Map will be developed during 2013.

This Road Map will:

- present a compelling, 30-year vision;
- take the Astrophysics 2011 decadal survey as the starting point and build upon it;
- be science based, with notional missions;
- be developed by task force of the Astrophysics Subcommittee (APS);
- take into account community input solicited Town Hall meetings and other potential calls for input;
- be delivered to APS.

The roadmap report should:

- have a compelling, over-arching theme;
- contain multiple paths (science areas) forward towards a long-range vision;
- consider cross-cutting opportunities as well as the larger context of ground-based and international astrophysics;
- be built on science investigations, leading to notional missions that achieve the science;
- consider the technology needed to achieve goals;
- identify challenges (e.g., science challenges, technology challenges, …);
- consider a variety of mission sizes to achieve the science;
- consider way-stations at 10 and 20 years out, as well the full vision at 30 years out.

Note that the roadmap

- is not a mini-decadal survey with recommendations and priorities;
- is not an implementation plan;
- is a long-range vision document with options, possibilities and visionary futures.

**Schedule**

The Roadmap Task Force shall:

- report to the APS Chair at least monthly or more often as the team deems desirable, and to the entire APS at regular meetings;
- deliver an interim report, with high-level themes, to the APS in time for approval by the APS by August 30, 2013;
- deliver their final report to the APS in time for approval by the APS, by December 16, 2013;
- disband once their final report has been approved and accepted by the APS.